# Integration of a Synthetic Molecular Motor Into a Rotary DNA Nanostructure: A Framework for Single-Molecule Actuation

Seham Helmi[1,2*], Erik Benson[3], Jan Christoph Thiele[1,2], Emilie Moulin[4], Nicolas Giuseppone[4] and Andrew J Turberfield[2,3*]

[1] University of Oxford, Department of Chemistry, Mansfield Road, Oxford OX1 3TA United Kingdom

[2] Kavli Institute for Nanoscience Discovery, University of Oxford, Department of Biochemistry, Parks Road, Oxford OX1 3QU United Kingdom

[3] Science for Life Laboratory (SciLifeLab), Department of Microbiology, Tumor and Cell Biology, Karolinska Institutet, 171 77 Solna, Sweden

[4] SAMS Research Group, Université de Strasbourg, CNRS, Institut Charles Sadron UPR22, ,67000 Strasbourg, France, and Institut Universitaire de France (IUF), Paris, France

## Abstract:

Synthetic molecular motors are an appealing means to control motion at the nanoscale, but understanding their behaviour as single-molecule actuators and integrating them into larger, functional systems remain technical challenges. Translating molecular actuation into coordinated device-level behaviour requires precise placement and orientation of the motors: DNA origami provides a powerful platform for positioning molecules with nanometre precision. Here, we demonstrate integration of a light-driven, rotary molecular motor into a DNA-based nanoscale actuator through site-specific, four-point conjugation. The motor is labelled with four distinct oligonucleotides, two on each side, using DNA-templated chemistry. This modular approach enables stable, oriented incorporation of the motor into a DNA assembly through DNA hybridization. Upon photoactivation with UV light, the motor transduces photon energy into rotary motion. By coupling the motor to a fluorescently labelled DNA rotor arm we amplify its movement and enable real-time observation using total internal reflection fluorescence microscopy. A subset of assembled devices exhibits light-induced conformational transitions and directional motion consistent with the expected photochemical mechanism. These results establish a programmable framework for integration of light-driven molecular motors into synthetic nanomachines and tools for the study of their behaviour.

## Summary:

A DNA-templated functionalization strategy enables precise integration of synthetic rotary motors into nanoscale rotary devices, allowing real-time, single-molecule optical tracking of UV-driven motion.

## Introduction:

The development of optically driven molecular rotors marked a breakthrough in nanoscale systems engineering [1–6]. Upon photoexcitation, overcrowded alkenes undergo unidirectional *cis*–*trans* isomerization coupled to a thermal helix inversion (THI) driven by steric hindrance across the double bond. Full 360° rotation is completed by a second phoroisomerization / THI sequence. This mechanism has inspired the design of assemblies with increasingly complex and controllable behaviours, including molecular whirligigs [7], light-responsive gels [8], supramolecular polymers [9] and motor-doped active liquid crystals [10,11].

Studying the dynamics of individual motors reveals behaviours that are often obscured in ensemble measurements [12–16]. However, the small scale of the molecular conformational changes, typically sub-nanometre, presents significant challenges for direct optical observation. Krajnik et al. [17] used defocused imaging to track individual, fluorescently labelled molecular rotors and revealed an unexpected sequence of motor steps; however, mechanistic characterization of molecular rotors remains largely limited to ensemble spectroscopic measurements [18,19]. Synthetic molecular motors have been incorporated into functional materials [6–8,10,11,20], but their integration into nanoscale devices has been restricted by the difficulty of achieving precise control over their position, orientation and interactions within the surrounding superstructure.

DNA nanotechnology [21,22], and particularly DNA origami [23,24], offers a robust and modular framework for integrating functional molecules in larger nanosystems[25–28] . DNA origami enables the construction of a wide variety of mechanical devices such as pivots, hinges, sliders, and rotary systems [29–35], which can be actuated by a range of stimuli, including pH changes [36,37], ionic strength variations and temperature shifts [38,39], strand displacement reactions [40–42], electric fields [43,44], and interactions with proteins[45] including natural motor proteins[33,44].

Recent work has shown that DNA origami structures can be used not only to build mechanical systems but also to probe motion and force at the nanoscale, including the linear [46] and rotary [33] motions of protein motors, force-dependent DNA–protein interactions [47], and receptor–ligand mechanics in immune cells [48]. The programmability of DNA origami allows extrinsic components to be precisely positioned at locations identified by unique DNA base sequences, provided they can be site-specifically tagged with oligonucleotides or DNA-binding moieties. Here, we introduce a strategy for integrating light-driven rotary molecular motors into DNA origami nanostructures with defined spatial orientation. Using a two-template DNA-templated conjugation method, each motor is functionalized with four distinct oligonucleotide adapters, enabling its insertion with spatial and orientational control by hybridization within the DNA framework. By coupling the motor to a fluorescently labelled lever arm extending from the DNA origami device, we amplify its motion so that it can be tracked optically. This platform enables monitoring of the photo-induced conformational changes in synthetic rotary motors at the single-molecule level.

## Results:

### Motor-DNA Conjugation:

In this study we use a second-generation light-driven molecular rotor [7–9] comprising two counter-rotating components shown in red and blue in Fig. 1a. The blue part (stator), to be attached to a stationary base structure, is functionalized with two triethylene glycol (TEG) chains, each terminating in an alkyne end group; the red part (rotor) carries two carboxyl end groups. Upon UV irradiation at 366 nm [8], the rotor is driven out of equilibrium and rotates unidirectionally via repeated *cis-trans* photoisomerization of the central double bond followed by thermal helix inversion [7,8].

To enable integration of the motor into a DNA nanostructure, we conjugated its alkyne and carboxyl groups to four distinct DNA adapters. These oligonucleotide adapters are designed to hybridize with specific domains on the scaffold strand of the DNA origami nanostructure, allowing precise control over motor placement and orientation. Our approach builds on our previously reported DNA-templated conjugation method [49], in which homobifunctional molecules are conjugated to two distinct DNA adapters held in close proximity by hybridization to a third, template strand, thereby selectively increasing the local concentration of their reactive groups. The templated conjugation reactions used to functionalize the stator and rotor components of the motor are shown in Fig.1b. For the stator part, azide-modified DNA adapters A and B were hybridized to a common template strand and conjugated via copper-catalysed azide-alkyne cycloaddition (CuAAC) (Fig. 1b, SI Methods 1.1). Similarly, in a separate reaction, the carboxyl groups of the rotor were conjugated to amine-modified DNA adapters C and D through templated amide bond formation (Fig. 1c, SI Methods 1.2). To facilitate identification of reaction products during gel analysis, the adapters used in each reaction (A/B and C/D) were designed with different lengths and labelled with distinct fluorophores (3′-Cy3 for B and D, 5′-Cy5 for A and D), positioned opposite the reactive groups. Reaction products were analysed by denaturing PAGE and imaged in Cy3 and Cy5 fluorescence channels. The appearance of a yellow band in merged gel images (Fig. 1d and 1e, Figure S1 and S2) confirms colocalization of the two fluorophores, indicating successful conjugation of adapter pairs to the motor's stator (Fig. 1d, lane 4) and rotor (Fig. 1e, lane 3) components, yielding the designed motor-AB and motor-CD intermediate constructs, respectively.

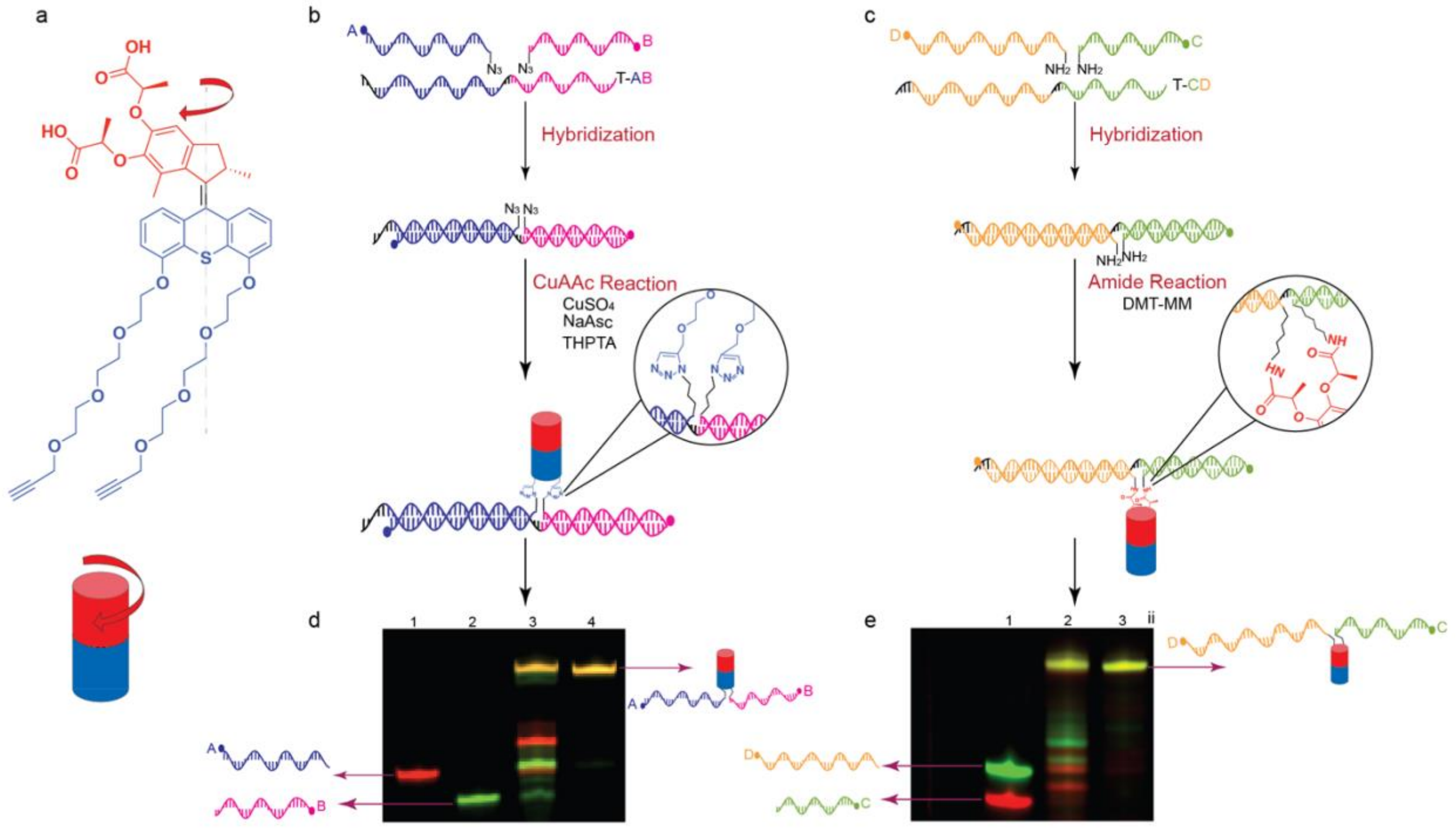

*Figure 1: Template-directed conjugation of the molecular rotor to DNA adapters. (a) The rotary motor. The stator part is coloured blue and the rotor part red. Templated conjugation scheme for: (b) CuAAC reaction and (c) amide functionalization of the stator and rotor parts, respectively. In each reaction, a template (T-AB or T-CD) is used to colocalize, two distinct DNA adapters, A and B or Cand D, respectively. A short gap (2 template nucleotides) between the reactive end groups of the adapters accommodates the molecular rotor. Adapters A and C are labelled with a Cy5 fluorophore; adapters B and D are labelled with Cy3. (d) and (e) show merged images from a 20% denaturing PAGE gel, scanned in Cy3 (green) and Cy5 (red) channels, displaying the products of templated CuAAC (d) and amide (e) reactions of the rotary molecule in lanes 3 and 2, respectively. motor-AB, and motor-CD, conjugates are identified by a yellow band due to the colocalization of both fluorescent dyes. Lane 4 in (c) and lane 3 in (d) contain the HPLC-purified motor-AB and motor-CD conjugates.*

The orthogonality of the chemistries used to functionalize the stator and rotor components of the motor necessitates a two-step conjugation strategy. We tested both reaction sequences: CuAAC followed by amide bond formation and the reverse. In each case, the first conjugation yielded heterofunctionalized intermediates, motor-AB or motor-CD, which could be purified by high-performance liquid chromatography (HPLC, SI Methods 10, Figure S9) (Fig. 1d, lane 4; and Fig. 1e, lane 3, respectively). These purified intermediates were then used in a second reaction to attach the remaining pair of DNA adapters. Although the initial conjugation step proceeded effectively for both reaction orders, the second step failed in each case (SI Methods 2). We attribute this failure to steric hindrance by the DNA already conjugated to the motor, which may restrict access to the remaining reactive groups. Attempts to overcome this limitation, such as omitting the template during the second reaction, either for the motor-DNA intermediate or the reacting adapters, did not improve yield (SI Methods 2, Figure S3 and S4). To address this challenge, we developed a dual-template strategy designed to relieve the steric constraints and facilitate the conjugation (Figs 2a, S5). Starting from a purified motor-AB conjugate, we introduced two new template strands, each designed to hybridize with one of the existing adapters (A or B) and one of the missing ones (C or D). A five-nucleotide gap between the binding sites was included to accommodate the motor core. Following annealing of templates T-AD and T-BC to the motor-AB construct (Fig. 2a, hybridization 1), oligonucleotides C and D were added to complete the four-point conjugation (Fig. 2a, hybridization 2). In this configuration, the motor is suspended between two DNA templates, which both alleviates steric hindrance and increases the local concentration of the remaining adapters, promoting efficient attachment.

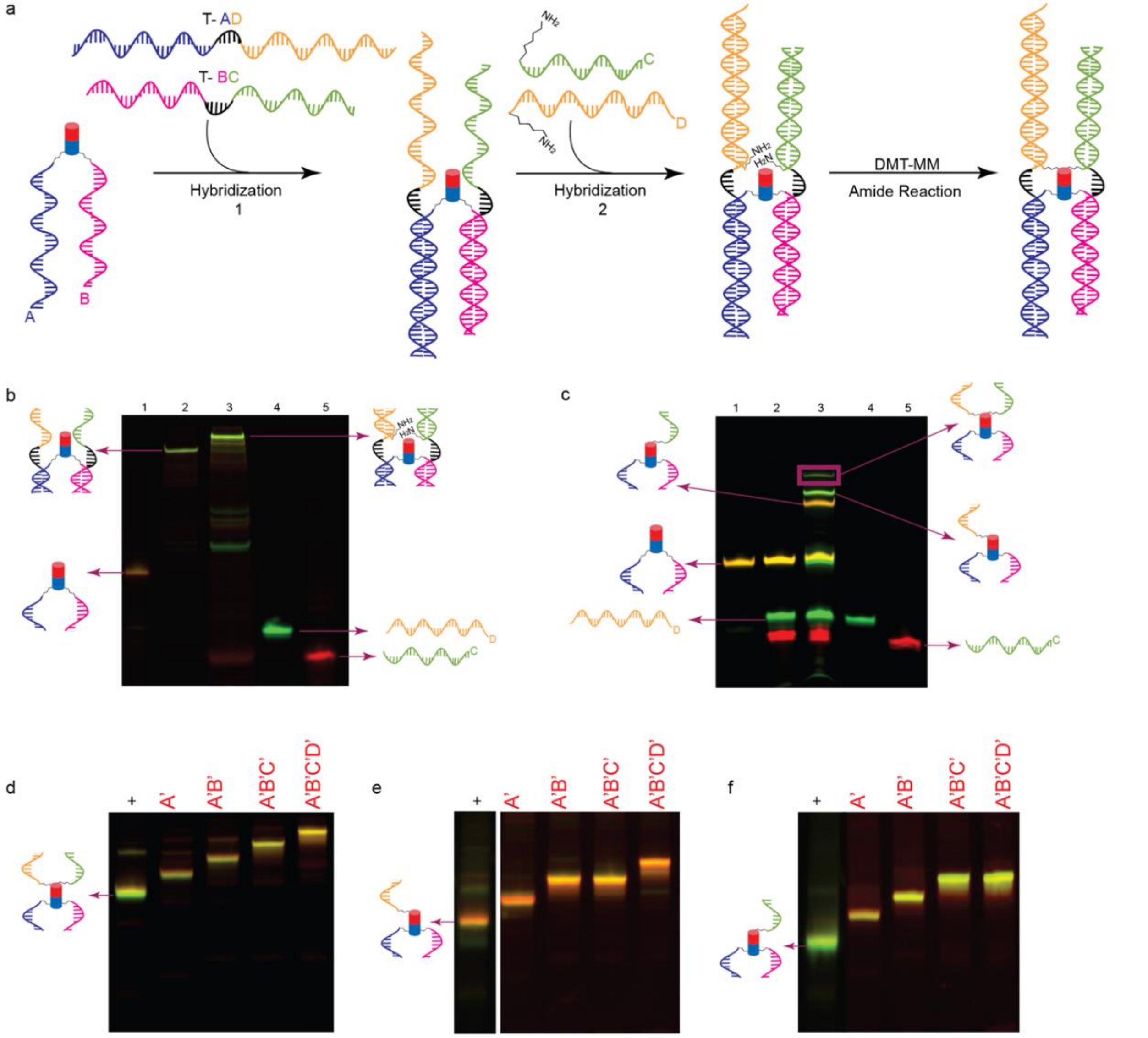

***Figure 2: Two-template conjugation scheme.*** *(a) Simultaneous use of two templates to colocalize adapters C and D with purified motor-A,B. (b) 10% native PAGE gel showing purified motor-AB (lane 1), hybridized to templates T-AD and T-BC in lane 2. Lane 3 shows the result of adding oligos C and D to the motor–template complex from lane 2. Lanes 4 and 5 are controls containing only oligos D and C, respectively. (c) 20% denaturing PAGE gel showing the results of an amide bond formation reaction with the two-template complex (lane 3), yielding three high-molecular-weight bands. Lanes 1,2,4, and 5 show motor-AB, two-template reaction mixture before the amide reaction, adapter D, and adapter C, respectively. (d, e, f) To confirm the attribution of these bands, their contents were gel-purified and tested against the complement of each DNA adapter (A', B', C', and D'). (d) A 10% native PAGE confirms that the top band from (c, lane 3) corresponds to the motor conjugated to all four adapters (A, B, C, and D). (e, f) Similar confirmation that the middle and lower high-molecular-weight bands from (c, lane 3) correspond to the motor conjugated to three adapters: A, B, and D (gel e) or A, B, and C (gel f). All gels were scanned using Cy3 and Cy5 fluorescence channels.*

Following amide bond formation, denaturing PAGE analysis confirmed the formation of three higher-molecular-weight bands (Fig. 2c, lane 3). The lowest-mobility band corresponds to the motor conjugated to all four adapters, A, B, C, and D, as designed. The two faster-migrating bands correspond to motor conjugates with three of the four adapters: either A, B, and D or A, B, and C. To confirm the identities of these bands, each was gel-purified and tested for hybridization against the complementary strands A′, B′, C′, and D′ (Figs. 2d-f, Figure S6). The slowest-migrating band hybridized to all four complements, confirming the successful formation of the fully functionalised motor-A,B,C,D product. In contrast, the other two bands showed no gel shift when tested against either C′ or D′, verifying partial conjugation to the other three adapters.

To complete the partially functionalised constructs, motor-ABC and motor-ABD were separately reacted with the missing fourth adapter using the original template T-CD (Figure S7-8 a and b), which facilitated hybridisation between the conjugated and missing adapters (Figure S7.c). Final analysis after the second round of amide bond formation confirmed successful addition of the fourth adapter in each case (Figure S7.d). These results demonstrate that the two-template conjugation strategy effectively overcomes steric hindrance encountered in the original approach, and underscore the importance of spatial design and flexibility in the construction of multifunctional molecular assemblies.

**Assembly of the Motorized DNA Origami Device:**

We used the optically driven motor to link two DNA origami nanostructures: a baseplate (60 × 64 × 5.4 nm) and a 220 nm rotor arm (Fig. 3a). The baseplate consists of two parallel layers of DNA helices, while the rotor arm is constructed as either a 4- or 10-helix bundle (HB). In the case of the 4HB design, one end features a broader rectangular region. Both rotor arm designs carry identical staple extensions designed to hybridize with Cy3-labelled oligonucleotides (reporter oligos) for optical tracking – 66 and 48 reporter sites for the 4HB and 10-HB rotor arms, respectively. The baseplate is anchored to a streptavidin-coated cover slip via 10 biotinylated oligonucleotides (extended staples) from its underside.

The motor is the only element connecting the rotor arm to the baseplate. The A and B adapters on the stationary part of the motor were designed to hybridize to two scaffold domains (16 and 20 nt) near the centre of the top layer of the baseplate. For the 4HB rotor arm, adapters C and D were complementary to two contiguous scaffold domains (16 and 25 nt) within a central region of the 4HB rotor arm scaffold, located 110 nm from either end (Fig. 3b). With the 10HB rotor arm, adapters C and D formed 16-nt duplexes with two extended staples at the mid-section of the 10HB rotor. Where adapters were hybridized directly to the scaffold they were fully complementary to their respective scaffold domains, with no unpaired or flexible regions. The purified motor-ABCD conjugate was added to the rotor arm assembly mix (scaffold, staples, Cy3 reporter oligo), at a ratio of 0.5:1 ratio (motor to arm scaffold) before assembly of the arm by annealing its components (SI Methods 12.1). The motor-arm complex was then combined with a pre-assembled baseplate, also at a 0.5:1 ratio (arm to baseplate). Assembly was verified by agarose gel electrophoresis. A distinct dimer band visible in the Cy3 channel (Fig. 3c, lane 5, Figure S8) confirmed successful linkage of the rotor arm to the base via the motor. TEM imaging further validated the assembly of the rotary device as designed (Fig. 3d, Figure S10 a and b).

**Optical Tracking:**

To observe the motion of individual DNA origami devices at the single-molecule level, we integrated a 365 nm UV LED into a total internal reflection fluorescence (TIRF) microscope, enabling simultaneous UV illumination, to drive motor rotation, and tracking of motion via Cy3 fluorescence (Fig. 3e). We imaged fields of view containing tens to hundreds of devices at 100 Hz and tracked selected devices for up to 60 seconds. Using localization software[50], we identified clusters of fluorescent signals corresponding to individual rotor arms in each frame, and used all localizations to create maps of the time-dependent rotor position for each device (Fig. 3f), see SI “Scripts and Single-Particle Analysis” section.

Devices with 4-helix bundle arms displayed unexpected switching behaviour, consistent with ~180° rotation of the rotor arm. In some cases, we observed this motion only at low frequency and only upon UV illumination, while in others it occurred more frequently before the UV illumination (Figure S14). This motion may result from preferential alignment with the helices in the top layer of the baseplate.

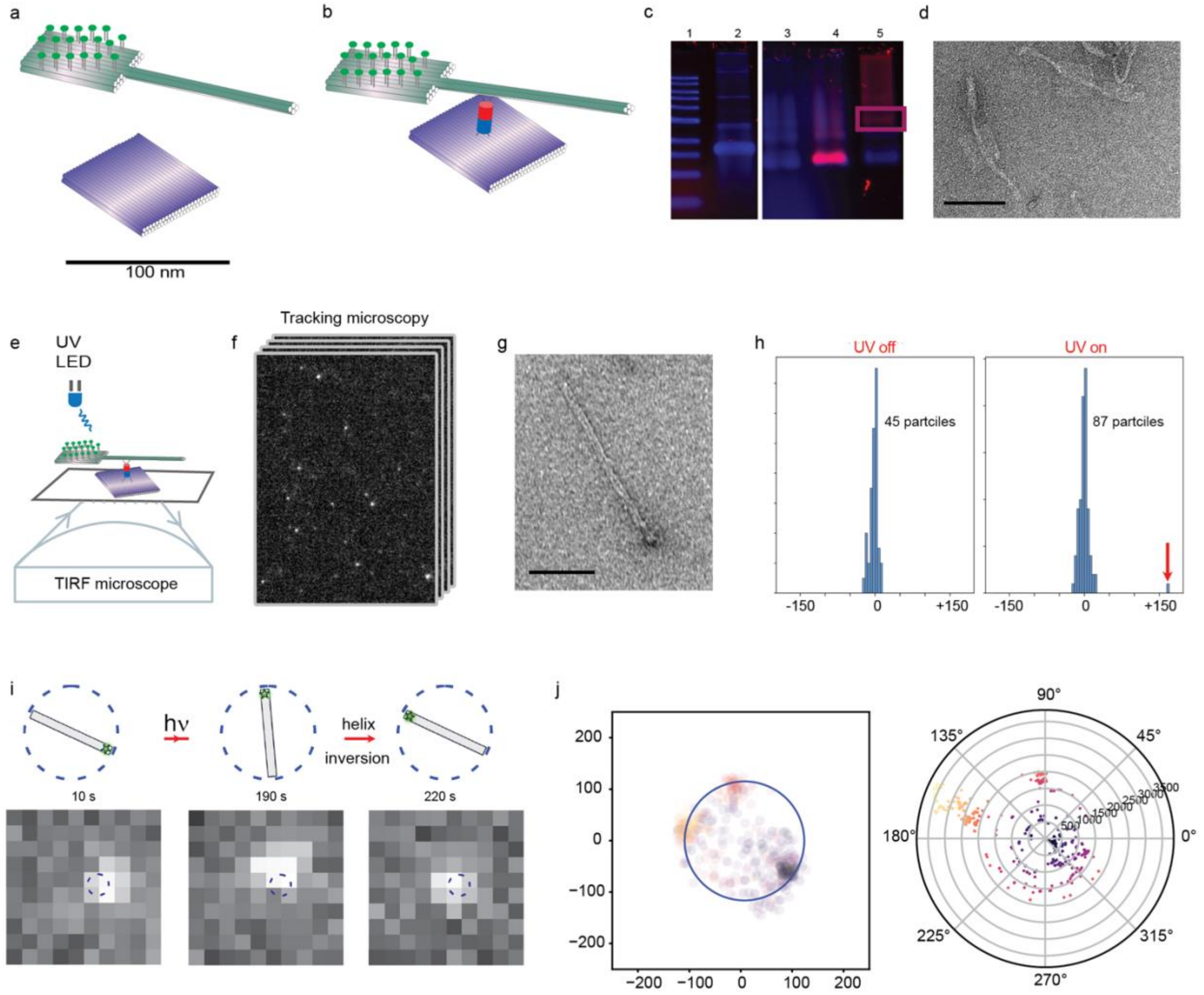

*__Fig 3. Assembly of molecular motors into rotary DNA devices.__ (a) Schematic of device components: a 217 nm 4HB rotor arm bearing Cy3-labelled oligonucleotides at one end, and a square baseplate. (b) The motor-ABCD conjugate is designed to link the two components via hybridization of adapters A, B and C, D to the scaffolds of the baseplate and rotor arm, respectively. (c) 2% agarose gel showing: 1 kb ladder (lane 1), scaffold p8064 (lane 2), assembled motor base (lane 3), assembled Cy3-labelled rotor arm with motor-ABCD annealed during-assembly (lane 4), and assembled rotary devices formed by combining the motor–arm complex with the baseplate (lane 5); the dimer band corresponding to fully assembled devices is highlighted. (d) Transmission electron microscopy (TEM) confirms the correct assembly of the rotary structures. (e) Schematic of the motion detection apparatus: rotary devices were immobilized on streptavidin-coated coverslips and imaged using a TIRF microscope under simultaneous UV illumination. (f) Devices were tracked at 100 Hz by localizing the fluorescence from the labelled rotor arms in each frame. (g) TEM showing the new 10-HB rotor arm. (h) Histograms of rotation during up to 60 secs observation (the difference between the mean angles of the rotor arm in the first and last 100 frames) in structures without (N=45) and with UV illumination (N=87); a single device (arrow) exhibits significant rotation under UV. (i) Fluorescence images recording the motion of this device are consistent with a rotation of 120° triggered by photoisomerization of the motor followed by an additional 60° rotation on thermal helix inversion, indicated by the schematics above. Circular trajectories fitted are indicated by dashed circle. (j) Left: localization map of the 10HB rotor arm with a fitted circle indicating the tracked trajectory; right: polar plot representing rotor orientation over*

*time, with time as the radial coordinate from the start of tracking. Colour map indicated temporal order of localization from dark to light.*

The 10-HB rotor (Fig. 3g, Figure S10) is significantly stiffer, with a persistence length of well over 1 μm[51], and is positioned further from the baseplate. With the 10HB rotor, and software-based filtering to remove low-precision localizations, devices showed little or no motion in the absence of UV illumination. We imaged and tracked selected devices for up to 60 seconds at 100 Hz. Strong localization clusters corresponding to DNA origami devices were identified automatically and extracted. We then fitted a circle with a diameter matching the rotor arm to each cluster, allowing us to parameterize the motion of each device over time. (For motors that display only two states, it is not possible to determine the direction of rotation.)

Most 10HB devices did not move during tracking, with or without UV illumination (Fig. 3h). However, in initial measurements under continuous UV illumination, we observed a single device that reoriented significantly, exhibiting two abrupt changes in orientation, first by approximately 120° then by 60° (Fig. 3i, j). This motion is consistent with a 120° rotation triggered by photoisomerization of the motor, followed by a 60° rotation corresponding to thermal helix inversion. The small number of active devices can be attributed to the mechanical constraints imposed by the tight coupling of the motor to the baseplate and rotor coupled to frictional drag on the rotor and also potential interference between baseplate and rotor.

We redesigned the system to eliminate the baseplate, increase the distance of the rotor from the surface and increase the compliance of the coupling between motor and rotor. On either side of the motor, the adapters form 16-nt duplexes with two oligonucleotides that then hybridize to each other. The motor thus bridges two double-stranded DNA segments (Fig. 4a). On the rotor side of the motor, the first 90-bp segment is formed between two extended staple strands at the mid-section of the 10HB rotor arm. Each strand then includes a 2-nt spacer followed by a 16-nt region complementary to adapter C or D, enabling hybridization to the rotor part of the motor-ABCD conjugate. On the stator side, adapters A and B form 16-nt duplexes with two oligonucleotides which, after 2-nt spacers, hybridize to each other to form the second 30-bp double-stranded segment. These oligonucleotides continue via 10-nt single-stranded spacers to terminal biotin modifications which anchor it to the streptavidin-coated cover slip. The adapter-conjugated motor was mixed at a 0.5:1 ratio with the pre-hybridized biotinylated double-stranded DNA segment and allowed to hybridize, forming a motor-biotin complex. This complex was then combined with the 10HB rotor arm and incubated to complete assembly of the full rotor arm-motor-biotin construct.

## Motion Analysis – Simplified System:

In subsequent measurements, we either introduced UV illumination after an initial period of tracking or applied it in pulses during imaging, allowing us to compare the behaviour of individual devices with and without UV exposure. Devices were tracked for 30 seconds at 100 Hz, and their angular trajectories were inferred by fitting circles to the localization data. Based on their total angular displacement during the observation period, we classified each device as stationary (displacement <25°), moderately rotating (25-75°), or highly rotating (>75°). This classification enabled comparative analysis of rotational behaviour across different assembly configurations and illumination regimes. Under no-UV conditions, the majority of devices (83%) remained stationary; only about 2% exhibited angular displacements exceeding 75° (Figure S16).

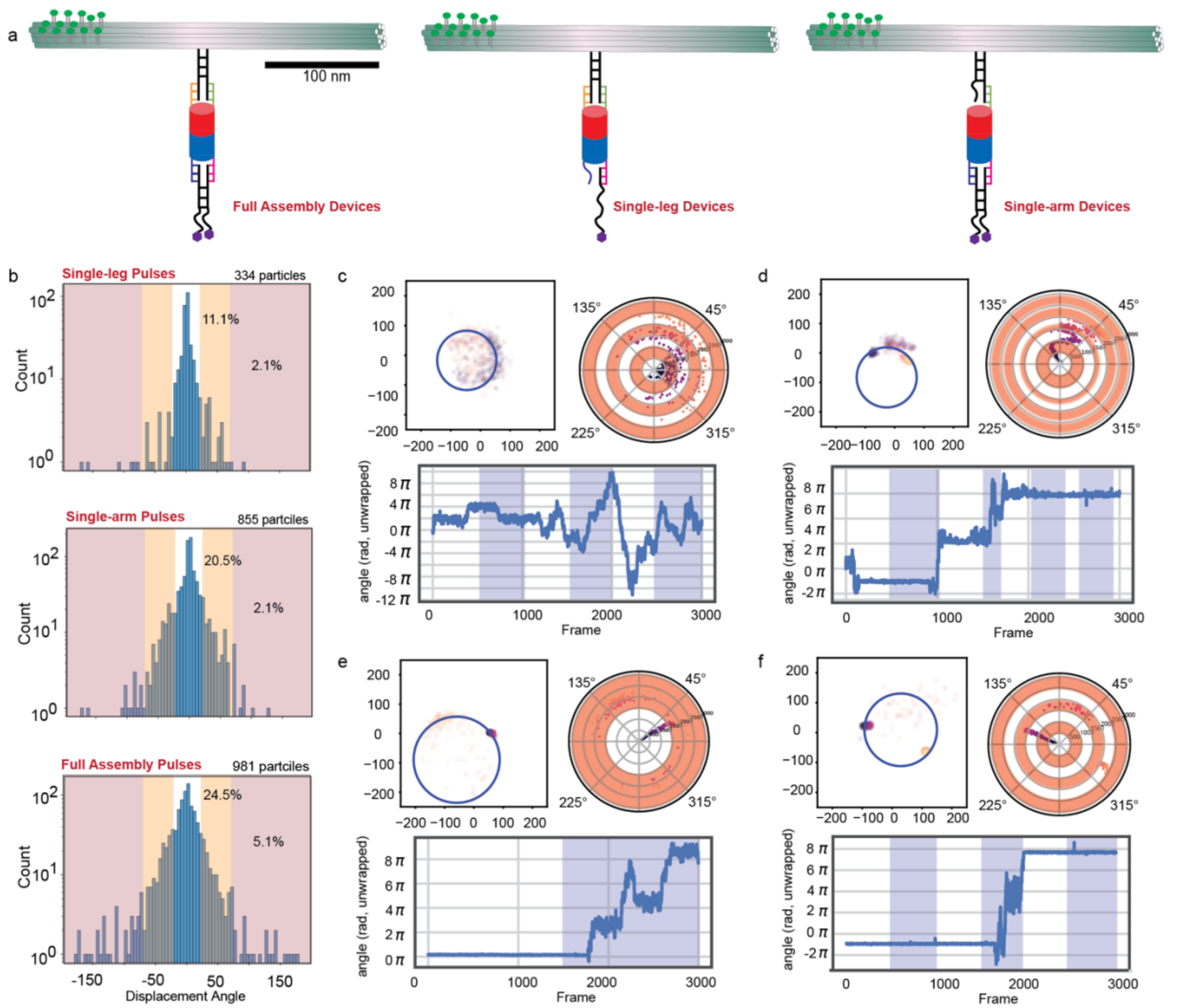

*Fig. 4. Motion tracking of DNA origami rotary devices during UV exposure. (a) Schematic device designs. Controls: single-leg devices lacking one biotinylated strand for surface anchoring, and single-arm devices assembled with a motor missing one of the rotor-binding adapters. (b) Histograms of angular displacement after three UV pulses of 5 secs each. Fully assembled devices displayed a higher fraction of large-angle displacements (>75°), than the control configurations. Representative trajectories for the control devices are shown in (c) and (d), where devices either displayed random circular motion with small radius (c) or low-amplitude displacements below 90° (d), independent of UV exposure. In contrast, examples from the fully assembled group (e, f) show directional rotation initiating during UV illumination, with angular displacement trajectories aligned to the UV-on periods. For each example, the left panel shows the localization map of the rotor arm end with a fitted circle indicating the tracked trajectory, while the right panel is a polar plot representing rotor orientation over time, with time as the radial coordinate from the start of tracking (0 to 3000 frames over 30 secs, 100Hz), and the bottom panel shows the angular displacement in "π" representing the device's rotation over time from the start of tracking. UV exposure intervals are shaded in blue.*

Negative controls shown in Fig. 4a have one broken motor connection, allowing rotation about the remaining connection on the same side of the motor and thus impairing transduction of torque from the motor to the rotor. In the "single-leg" control one of the biotin-functionalized anchor oligonucleotides was omitted so the motor-rotor complex was tethered to the surface only through adapter B. This configuration resulted in a noticeably lower number of surface-immobilized particles, likely due to reduced attachment efficiency. In the "single-arm" control, the motor-ABC conjugate was used in place of the fully conjugated motor-ABCD. When exposed to UV (Fig. 4b), only a small fraction of control devices of either type (consistently in the range 2-2.5%) exhibited net angular displacements greater than 75°. These results were consistent across both the single-pulse and multiple-pulse illumination conditions (Figures S17 and S18). These controls establish a robust baseline for the identification of motor-driven devices. Interestingly, the recorded devices showed small average change in orientation (less than 3°), possibly due to the experimental difficulty in tracking direction of motion with few states visited during tracking. Upon testing the fully assembled devices, we observed a clear increase in the proportion of devices exhibiting rotation following UV exposure. In contrast to all control configurations, a single 10-second UV pulse triggered a higher fraction of devices undergoing moderate to large angular displacements (Figure S19); this response was further enhanced by repeated activation through multiple UV pulses (Fig. 4b).

To probe the nature of the observed motion, we examined individual trajectories from devices – fully assembled and controls - that rotated more than 60°. In the control groups, most devices either displayed confined, random circular motion with a smaller-than-expected radius (~100 nm) (Fig. 4c), or exhibited angular displacements below 90° even under UV illumination (Fig. 4d). These results can be attributed to the mechanical decoupling between rotor and anchoring surface corresponding to unconstrained rotation about the unpaired connection on one side of the motor. - It should be noted that, in traces showing large discontinuities in the angular trajectory, it is difficult to determine unambiguously the direction and number of full rotations, as the analysis assumes stepwise progression below 180° per frame interval; an example of this can be seen in Fig. 4d.

In contrast, among the fully assembled devices, we observed trajectories that exhibited UV-triggered directional rotation in response to both single and multiple illumination pulses (Fig. 4e, f). These devices displayed sustained reorientation that was temporally aligned with UV illumination, indicating effective photoactivation of the motor. Quantitative analysis revealed an average rotation rate of 0.213 rotations/sec and an average angular speed of 76.7°/sec during UV illumination, compared to 0.087 rotations/sec and 31.4°/sec in the absence of UV, suggesting successful motor engagement and light-driven actuation.

We further analysed the angular trajectories of the fully assembled devices. The largest group (Fig. 5a) exhibited robust, unidirectional rotation, initiated during or shortly after UV illumination. These trajectories featured either continuous angular progression or large discrete steps, with motion consistently biased in one direction. While occasional reversals or fluctuations were observed, they did not disrupt the overall directional trend. On average, this class exhibited a UV-on rotation rate of 0.21 rotations/sec and an angular speed of 75.8°/sec, compared to 0.09 rotations/sec and ~35°/sec when UV was off-supporting a strong coupling between UV activation and motor function.

A smaller group of devices were initially static or exhibited only minimal fluctuations in the absence of UV illumination but displayed clear stepwise angular transitions synchronized with UV exposure (Fig. 5b). This suggests sudden light-induced conformational changes competing with rotor-surface interactions. (Fig. 5b).

A third group of high-displacement trajectories from fully assembled devices did not exhibit a consistent directional bias or link to UV exposure (Figure S20). These traces resembled those observed in control experiments and may represent false positives retained through histogram-based filtering.

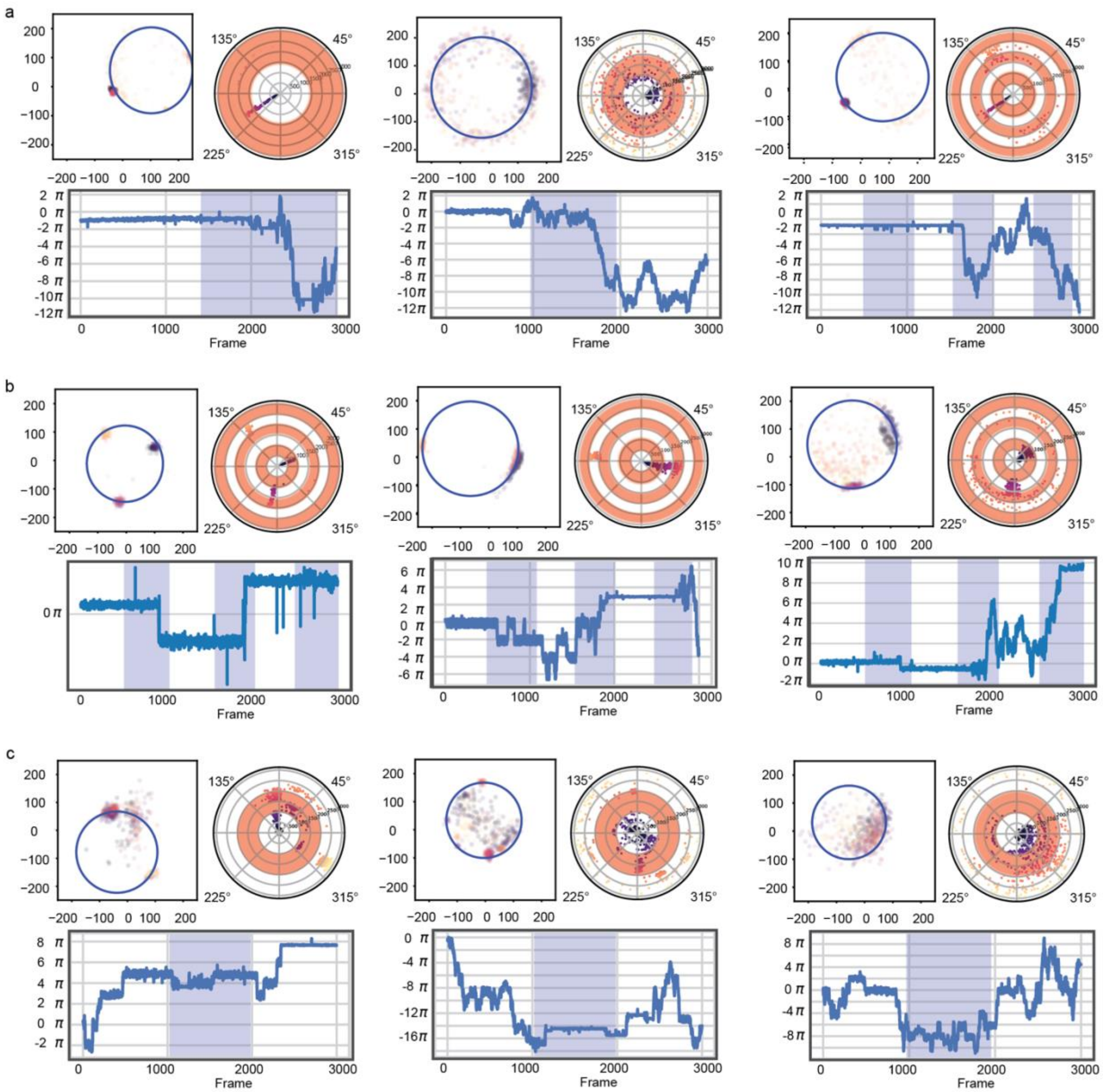

*Fig. 5. Device Motion classes. a, shows examples of individual devices that undergo directional rotation following UV activation. b, shows examples of stepping motion synchronized with each UV exposure. c, shows devices that exhibit directional motion before UV or transition into step-like behaviour during UV, indicating more complex or pre-activated dynamics.*
*Similar to fig.4, in each case, the left panel shows the localization map with a fitted circle, the right panel is a polar plot and the bottom panel shows the angular displacement in "π". UV-on intervals are shaded in blue.*

## Summary and Discussion

Our observations of directional motion and step changes in rotor orientation synchronised with UV illumination support the interpretation that motion of the rotor is driven by light-activated rotation of the mechanically coupled molecular motor. The sizes of discrete angular steps and the direction of UV-induced rotation are consistent with prior studies of related second-generation molecular motors using both single-molecule imaging[17] and mechanically constrained conditions[7–9]. Control experiments, confirm that directional motion arises specifically from motor activity.

Some of the variability observed between individual device behaviours can be attributed to imperfect device assembly, interactions between rotor and surface and photochemical degradation of motors. Another contributing factor - for the simplified system - is likely to be our deliberate introduction of flexibility in the coupling between motor and rotor. Single-stranded sections of DNA and the short polyethylene glycol linkers on one side of the motor contribute to the compliance of the connections between surface, motor and rotor, relaxing the temporal coupling between conformation changes of the motor and the response of the rotor. Optimization of the mechanical properties of the connections between molecular motors and larger, stiffer DNA-based superstructures will the focus of future work.

The modular DNA-motor interface presented here provides a generalisable framework for the integration of synthetic rotary motors into DNA nanostructures, enabling exploration of motor photophysics and the development of new generations of nanoscale robotic devices.

## References

1. Koumura, N., Zijistra, R. W. J., Van Delden, R. A., Harada, N. & Feringa, B. L. Light-driven monodirectional molecular rotor. *Nature 1999 401:6749* **401**, 152–155 (1999).
2. Koumura, N., Geertsema, E. M., Van Gelder, M. B., Meetsma, A. & Feringa, B. L. Second generation light-driven molecular motors. Unidirectional rotation controlled by a single stereogenic center with near-perfect photoequilibria and acceleration of the speed of rotation by structural modification. *J Am Chem Soc* **124**, 5037–5051 (2002).
3. Vicario, J., Walko, M., Meetsma, A. & Feringa, B. L. Fine tuning of the rotary motion by structural modification in light-driven unidirectional molecular motors. *J Am Chem Soc* **128**, 5127–5135 (2006).
4. Foy, J. T. *et al.* Dual-light control of nanomachines that integrate motor and modulator subunits. *Nature Nanotechnology 2017 12:6* **12**, 540–545 (2017).
5. Pfeifer, L. *et al.* Photoefficient 2nd generation molecular motors responsive to visible light. *Chem Sci* **10**, 8768–8773 (2019).
6. Deng, Y. *et al.* Photo-responsive functional materials based on light-driven molecular motors. *Light: Science & Applications 2024 13:1* **13**, 1–14 (2024).
7. Gao, C., Jentzsch, A. V., Moulin, E. & Giuseppone, N. Light-Driven Molecular Whirligig. *J Am Chem Soc* **144**, 9845–9852 (2022).
8. Li, Q. *et al.* Macroscopic contraction of a gel induced by the integrated motion of light-driven molecular motors. *Nature Nanotechnology 2015 10:2* **10**, 161–165 (2015).
9. Schiel, P. *et al.* Supramolecular polymerization through rotation of light-driven molecular motors. doi:10.26434/CHEMRXIV-2024-QBD2Q-V2. & *Nature Nanotech*. **2025**, *in press*, DOI: 10.1038/s41565-025-01933-0
10. Orlova, T. *et al.* Revolving supramolecular chiral structures powered by light in nanomotor-doped liquid crystals. *Nature Nanotechnology 2018 13:4* **13**, 304–308 (2018).
11. Morimoto, M. & Irie, M. A diarylethene cocrystal that converts light into mechanical work. *J Am Chem Soc* **132**, 14172–14178 (2010).
12. Block, S. M., Goldstein, L. S. B. & Schnapp, B. J. Bead movement by single kinesin molecules studied with optical tweezers. *Nature 1990 348:6299* **348**, 348–352 (1990).
13. Mehta, A. D. *et al.* Myosin-V is a processive actin-based motor. *Nature 1999 400:6744* **400**, 590–593 (1999).
14. Yildiz, A. *et al.* Myosin V walks hand-over-hand: Single fluorophore imaging with 1.5-nm localization. *Science (1979)* **300**, 2061–2065 (2003).
15. Yildiz, A., Tomishige, M., Vale, R. D. & Selvin, P. R. Kinesin Walks Hand-Over-Hand. *Science (1979)* **303**, 676–678 (2004).
16. Reck-Peterson, S. L. *et al.* Single-Molecule Analysis of Dynein Processivity and Stepping Behavior. *Cell* **126**, 335–348 (2006).
17. Krajnik, B. *et al.* Defocused Imaging of UV-Driven Surface-Bound Molecular Motors. *J Am Chem Soc* **139**, 7156–7159 (2017).
18. Van Delden, R. A. *et al.* Unidirectional molecular motor on a gold surface. *Nature 2005 437:7063* **437**, 1337–1340 (2005).
19. van Beek, C. L. F. & Feringa, B. L. Coupled Rotary Motion in Molecular Motors. *J Am Chem Soc* **146**, 5634–5642 (2024).
20. Wang, P.-L. *et al.* Transducing chemical energy through catalysis by an artificial molecular motor. *Nature 2025 637:8046* **637**, 594–600 (2025).
21. Seeman, N. C. Nanomaterials based on DNA. *Annu Rev Biochem* **79**, 65–87 (2010).

22. Seeman, N. C. & Sleiman, H. F. DNA nanotechnology. *Nature Reviews Materials 2017 3:1* **3**, 1–23 (2017).
23. Rothemund, P. W. K. Folding DNA to create nanoscale shapes and patterns. *Nature 2006 440:7082* **440**, 297–302 (2006).
24. Douglas, S. M. *et al.* Self-assembly of DNA into nanoscale three-dimensional shapes. *Nature 2009 459:7245* **459**, 414–418 (2009).
25. Aghebat Rafat, A., Sagredo, S., Thalhammer, M. & Simmel, F. C. Barcoded DNA origami structures for multiplexed optimization and enrichment of DNA-based protein-binding cavities. *Nature Chemistry 2020 12:9* **12**, 852–859 (2020).
26. Wagenbauer, K. F. *et al.* Programmable multispecific DNA-origami-based T-cell engagers. *Nature Nanotechnology 2023 18:11* **18**, 1319–1326 (2023).
27. Huang, J. *et al.* A modular DNA origami nanocompartment for engineering a cell-free, protein unfolding and degradation pathway. *Nature Nanotechnology 2024 19:10* **19**, 1521–1531 (2024).
28. Helmi, S. *et al.* A scalable, reproducible platform for molecular electronic technologies. (2025).
29. Marras, A. E., Zhou, L., Su, H. J. & Castro, C. E. Programmable motion of DNA origami mechanisms. *Proc Natl Acad Sci U S A* **112**, 713–718 (2015).
30. Ketterer, P., Willner, E. M. & Dietz, H. Nanoscale rotary apparatus formed from tight-fitting 3D DNA components. *Sci Adv* **2**, (2016).
31. List, J., Falgenhauer, E., Kopperger, E., Pardatscher, G. & Simmel, F. C. Long-range movement of large mechanically interlocked DNA nanostructures. *Nature Communications 2016 7:1* **7**, 1–7 (2016).
32. Ramezani, H. & Dietz, H. Building machines with DNA molecules. *Nature Reviews Genetics 2019 21:1* **21**, 5–26 (2019).
33. Kosuri, P., Altheimer, B. D., Dai, M., Yin, P. & Zhuang, X. Rotation tracking of genome-processing enzymes using DNA origami rotors. *Nature 2019 572:7767* **572**, 136–140 (2019).
34. Benson, E., Marzo, R. C., Bath, J. & Turberfield, A. J. A DNA molecular printer capable of programmable positioning and patterning in two dimensions. *Sci Robot* **7**, 5459 (2022).
35. Pumm, A. K. *et al.* A DNA origami rotary ratchet motor. *Nature 2022 607:7919* **607**, 492–498 (2022).
36. Majikes, J. M., Ferraz, L. C. C. & LaBean, T. H. PH-Driven Actuation of DNA Origami via Parallel I-Motif Sequences in Solution and on Surfaces. *Bioconjug Chem* **28**, 1821–1825 (2017).
37. Ijäs, H., Hakaste, I., Shen, B., Kostiainen, M. A. & Linko, V. Reconfigurable DNA Origami Nanocapsule for pH-Controlled Encapsulation and Display of Cargo. *ACS Nano* **13**, 5959–5967 (2019).
38. Gerling, T., Wagenbauer, K. F., Neuner, A. M. & Dietz, H. Dynamic DNA devices and assemblies formed by shape-complementary, non-base pairing 3D components. *Science (1979)* **347**, 1446–1452 (2015).
39. Turek, V. A. *et al.* Thermo-Responsive Actuation of a DNA Origami Flexor. *Adv Funct Mater* **28**, 1706410 (2018).
40. Yurke, B., Turberfield, A. J., Mills, A. P., Simmel, F. C. & Neumann, J. L. A DNA-fuelled molecular machine made of DNA. *Nature 2000 406:6796* **406**, 605–608 (2000).

41. Liu, M. *et al.* A DNA tweezer-actuated enzyme nanoreactor. *Nature Communications 2013 4:1* **4**, 1–5 (2013).
42. Walbrun, A. *et al.* Single-molecule force spectroscopy of toehold-mediated strand displacement. *Nature Communications 2024 15:1* **15**, 1–15 (2024).
43. Kopperger, E. *et al.* A self-assembled nanoscale robotic arm controlled by electric fields. *Science (1979)* **359**, 296–301 (2018).
44. Shi, X. *et al.* A DNA turbine powered by a transmembrane potential across a nanopore. *Nature Nanotechnology 2023 19:3* **19**, 338–344 (2023).
45. Nickels, P. C. *et al.* Molecular force spectroscopy with a DNA origami-based nanoscopic force clamp. *Science (1979)* **354**, 305–307 (2016).
46. Derr, N. D. *et al.* Tug-of-war in motor protein ensembles revealed with a programmable DNA origami scaffold. *Science (1979)* **338**, 662–665 (2012).
47. Nickels, P. C. *et al.* Molecular force spectroscopy with a DNA origami-based nanoscopic force clamp. *Science (1979)* **354**, 305–307 (2016).
48. Hu, Y. *et al.* Quantifying T cell receptor mechanics at membrane junctions using DNA origami tension sensors. *Nature Nanotechnology 2024* 1–12 (2024) doi:10.1038/s41565-024-01723-0.
49. Helmi, S. & Turberfield, A. J. Template-directed conjugation of heterogeneous oligonucleotides to a homobifunctional molecule for programmable supramolecular assembly. *Nanoscale* **14**, 4463–4468 (2022).
50. Schnitzbauer, J., Strauss, M. T., Schlichthaerle, T., Schueder, F. & Jungmann, R. Super-resolution microscopy with DNA-PAINT. *Nature Protocols 2017 12:6* **12**, 1198–1228 (2017).
51. Pfitzner, E. *et al.* Rigid DNA Beams for High-Resolution Single-Molecule Mechanics. *Angewandte Chemie International Edition* **52**, 7766–7771 (2013).

# Supporting Information

## Integration of a Synthetic Molecular Motor Into a Rotary DNA Nanostructure: A Framework for Single-Molecule Actuation

Seham Helmi[1,2]*, Erik Benson[3], Jan Christoph Thiele[1,2], Emilie Moulin[4], Nicolas Giuseppone[4] and Andrew J Turberfield[2,3*]

[1] University of Oxford, Department of Chemistry, Mansfield Road, Oxford OX1 3TA United Kingdom
[2] Kavli Institute for Nanoscience Discovery, University of Oxford, Department of Biochemistry, Parks Road, Oxford OX1 3QU United Kingdom
[3] Science for Life Laboratory (SciLifeLab), Department of Microbiology, Tumor and Cell Biology, Karolinska Institutet, 171 77 Solna, Sweden
[4] SAMS Research Group, Université de Strasbourg, CNRS, Institut Charles Sadron UPR22, ,67000 Strasbourg, France, and Institut Universitaire de France (IUF), Paris, France

**Materials**

Unless otherwise specified, all reagents and solvents were obtained from commercial suppliers and used without further purification. The light-driven rotary molecular motor was obtained from the lab of Nicolas Giuseppone and was synthesized according to previously published procedures[1]. DMT-MM was purchased from Fluka, and all other chemicals, including copper sulfate, sodium ascorbate, THPTA ligand, DMF, MOPS buffer, and TEAA salts, were obtained from Sigma-Aldrich. DNA oligonucleotides were ordered from Integrated DNA Technologies (IDT). The M13-derived single-stranded DNA scaffold p8064 was purchased from Tilibit nanosystems (Germany).

# I. Methods

## 1. Conjugation of the Molecular Motor to DNA Adapters

To enable programmable integration into DNA nanostructures, the light-driven molecular motor was conjugated to four DNA adapters via orthogonal chemistries. The motor scaffold used features two terminal alkyne groups (on the stationary part) and two carboxylic acid groups (on the rotary part). Two distinct types of DNA adapters were synthesized for this purpose: azide-modified strands for CuAAC ("click") reactions and amine-modified strands for amide bond formation.

We first conjugated the motor to DNA adapters using two separate DNA-templated reactions, following our previously reported protocol[2]. Each conjugation step was facilitated by hybridizing two DNA adapters to a short complementary DNA template strand, thereby bringing the reactive groups into close proximity to increase local concentration and promote selective conjugation. Fluorophores (3'-Cy3 or 5'-Cy5) were incorporated at the distal ends of the adapters to allow visualization by PAGE and later optical tracking.

### 1.1 Templated CuAAC Reaction (Stationary Part)

The stationary part of the motor was functionalized with azide-modified DNA adapters A and B using a templated copper-catalyzed azide–alkyne cycloaddition (CuAAC) reaction. The DNA adapters (A and B) and the complementary template (T-AB) were annealed in 100 mM MOPS buffer (pH 7.0) containing 500 mM NaCl, each at 10 μM final concentration. The motor was added at a 10× molar excess relative to the DNA duplex, in a reaction mixture containing 15% DMF.

To initiate the click reaction, sodium ascorbate (600 equivalents), THPTA ligand (420 equivalents), and $CuSO_4$ (60 equivalents) were added. The reaction mixture was degassed and incubated overnight at room temperature on a shaker. The templated design ensured that once the first click occurred (adapter A or B), the proximity of the second adapter promoted selective conjugation, reducing the formation of off-target products (e.g., A-A or B-B).

### 1.2 Templated Amide Coupling (Rotary Part)

The rotary part of the motor was conjugated to DNA adapters C and D using amine-acylation chemistry. Amine-modified DNA adapters (C and D) were annealed to a DNA template (T-CD) strand (10 μM) in 100 mM MOPS buffer (pH 7.0) with 500 mM NaCl and 15% DMF. After annealing, the motor was added in 10× molar excess, followed by the addition of DMT-MM to a final concentration of 50 mM. The reaction was incubated for 15 hours at 25 °C.

## 2. Failed Attempts of Second conjugation

We tested different condition in which the HPC purified motor-AB or motor-CD conjugates were mixed to the missing adapters (C and D in case of motor-AB, and A and B in case of motor-CD) to facilitate their conjugation and obtained the intended motor-ABCD.

For motor-AB to form motor-ABCD, the following conditions were tested (Fig S3):

### 2.1 Templated to Templated

HPC purified motor-AB conjugate was initially annealed in a 1 to 1 molar ratio to the template strand (T-AB) in 100 mM MOPS buffer (pH 7.0) containing 500 mM NaCl. The annealed mixture was then added to the templated C and D adapters (using T-CD strand), in a 2.5 to 1 molar ratio of templated adapters to templated motor-AB conjugate. This was followed by the addition of DMT-MM to a final concentration of 50 mM, and reaction incubation for 15 hours at 25 °C.

### 2.2 Templated to Non-Templated

HPC purified motor-AB conjugate was initially annealed in a 1 to 1 molar ratio to the template strand (T-AB) in 100 mM MOPS buffer (pH 7.0) containing 500 mM NaCl. The annealed mixture was then added to the C and D adapters (non-templated), in a 2.5 to 1 molar ratio of adapters to templated motor-AB conjugate. This was followed by the addition of DMT-MM to a final concentration of 50 mM, and reaction incubation for 15 hours at 25 °C.

### 2.3 Non-Templated to Templated

HPC purified motor-AB conjugate was added to the templated C and D adapters (using T-CD strand), in a 2.5 to 1 molar ratio of templated adapters to motor-AB conjugate. This was followed by the addition of DMT-MM to a final concentration of 50 mM, and reaction incubation for 15 hours at 25 °C.

### 2.4 Non-Templated to Non-Templated

HPC purified motor-AB conjugate was added to the C and D adapters, in a 2.5 to 1 molar ratio of adapters to motor-AB conjugate. This was followed by the addition of DMT-MM to a final concentration of 50 mM, and reaction incubation for 15 hours at 25 °C.

For motor-CD to form motor-ABCD, the following conditions were tested (Fig. S4):

### 2.5 Templated to Non-Templated

HPC purified motor-CD conjugate was initially annealed in a 1 to 1 molar ratio to the template strand (T-CD) in 100 mM MOPS buffer (pH 7.0) containing 500 mM NaCl. The annealed mixture was then added to the A and B adapters (non-templated), in a 2.5 to 1 molar ratio of adapters to templated motor-AB conjugate. This was followed by the addition of sodium ascorbate (600 equivalents), THPTA ligand (420 equivalents), and $CuSO_4$ (60 equivalents) were added. The equivalents in the click reaction were to the motor-CD conjugates. The reaction mixture was degassed and incubated overnight at room temperature on a shaker.

### 2.6 Non-Templated to Templated

HPC purified motor-CD conjugate was added to the templated A and B adapters (using T-AB strand), in a 2.5 to 1 molar ratio of templated adapters to motor-CD conjugate. This was followed by the addition of sodium ascorbate (600 equivalents), THPTA ligand (420 equivalents), and $CuSO_4$ (60 equivalents) were added. The equivalents in the click reaction were to the motor-CD conjugates. The reaction mixture was degassed and incubated overnight at room temperature on a shaker.

## 3. Two-Template Conjugation Strategy

Attempts to sequentially perform the two conjugation reactions failed maybe due to steric hindrance imposed by previously attached DNA, which seems to prevent the second reaction to proceed efficiently. To overcome this limitation, we implemented a two-template conjugation strategy (see Fig. 2 in the main text).

In this method, HPC purified motor-AB conjugate was initially annealed in a 1 to 1 molar ratio to the two template strands (T-AD and T-BD) in 100 mM MOPS buffer (pH 7.0) containing 500 mM NaCl in a 1 to 1 molar ratio of templates to motor-AB conjugate. C and D adapters were then added to the annealed mixture in a 1 to 1 molar ratio of adapters to templated motor-AB conjugate. This was followed by the addition of DMT-MM to a final concentration of 50 mM, and reaction incubation for 15 hours at 25 °C.

The reaction products were then gel-purified using 20% denaturing PAGE followed by Ethanol precipitation.

4. **Motor-ABC and motor-ABD to motor-ABCD, Fig. S7:**

Gel-purified motor-ABC and motor-ABD conjugates, we annealed in separate reactions to template strand (T-CD) in a 1.5 to 1 molar ratio of template T-CD to motor-conjugate in 100 mM MOPS buffer (pH 7.0) containing 500 mM NaCl. The missing adapter (C in case of motor-ABD conjugate and D in case of motor-ABC conjugate) was the added to the annealed mixture was in a 2.5 to 1 molar ratio of adapters to templated motor- conjugate. This was followed by the addition of DMT-MM to a final concentration of 50 mM, and reaction incubation for 15 hours at 25 °C.

5. **Short annealing programs 1 and 2:**

The short annealing program **1** began at 96 °C , while short annealing program **2** began at 40 °C, each program decreased to 15 °C at a rate of 1 °C every 3 seconds. Once the target temperature was reached, the sample was held at 15 °C.

6. **Denaturing PAGE gel:**

Denaturing polyacrylamide gel electrophoresis (PAGE) was performed using 20% gels (29:1 acrylamide : bis-acrylamide) containing 25% formamide and 30% w/v urea in TBE buffer. Gels were pre-run at 300 V for 30 minutes before sample loading. Samples were denatured by heating to 95 °C for 10 minutes, then immediately placed on ice. After loading, electrophoresis was continued at 300 V for 50 minutes. The loading buffer contained 80% formamide, 10% Tris-glycine, and 10% urea.

7. **Gel purification and ethanol precipitation**

The products bands were cut from the gel, and incubated- separately- in 1x Tris-EDTA buffer containing 500 nM NaCl. The sample was subjected to 5 cycles of freeze and sawing, then left overnight.

Up to 220 μL of sample was transferred to a clean tube, followed by the addition of 1/10 volume of 3 M sodium acetate and 1/100 volume of glycogen. The mixture was vortexed briefly, and 5× volume of ice-cold (-20 °C) 100% ethanol was added, followed by a second vortexing. The samples were incubated for 2 hours at -20 °C or for 45 minutes at -80 °C. Precipitated DNA was pelleted by centrifugation at >20,000 rcf for 30 minutes at 4 °C. The supernatant was carefully removed, and the pellet was washed with 100 μL of cold (-20 °C) 70% ethanol, followed by a second centrifugation for 5 minutes at >20,000 rcf and 4 °C. The pellet was left to air dry for 10 min, followed by resuspension in MQ.

## 8. Native PAGE Gel:

Native polyacrylamide gel electrophoresis (PAGE) was performed using 10% gels (19:1 acrylamide : bis-acrylamide) in TAE buffer. Electrophoresis was carried out at 300 V for 15 minutes. To assess the nature of the reaction products, we tested them against the complementary adapter strands. Gel-purified motor–DNA conjugates (motor-ABC, motor-ABD, and motor-ABCD) were individually hybridized with the complementary adapter strands in a stepwise fashion. After each addition, an aliquot of the mixture was collected and loaded onto the gel for analysis, enabling visualization of sequential hybridization and confirming the nature of reaction products.

## 9. Gel Scanning and Analysis:

All gels were scanned using the Typhoon FLA 9500 gel scanner in each relevant fluorophore fluorescence channel prior to staining. Gels were then stained with SYBR® Gold and rescanned in the SYBR® Gold channel. Gel images were analysed using ImageJ software.

## 10. HPLC (Agilent 1200):

HPLC purification was performed using a Waters XBridge® Oligonucleotide BEH C18 column (130 Å, 2.5 μm, 4.6 × 50 mm) at 60 °C. Buffer A consisted of 5% acetonitrile (MeCN) in 0.1 M triethylammonium acetate (TEAA), while Buffer B contained 70% MeCN in 0.1 M TEAA.

| Time / min | %B | Flow |
|---|---|---|
| 0.0 | 0.0 | 1 mL/min |
| 1.0 | 0.0 | 1 mL/min |
| 20.0 | 100.0 | 1 mL/min |
| 23.0 | 100.0 | 1 mL/min |
| 23.10 | 0.0 | 1 mL/min |

## 11. DNA Origami Design

In this study, three DNA nanostructures were designed to assemble functional rotary devices: a baseplate for motor anchoring, an initial rotor arm design (v1), and a rigidified 10-helix bundle (10HB) rotor arm used for optical tracking of rotary motion.

The motor baseplate is a two-layer DNA origami tile measuring approximately 60 nm × 64 nm × 5.4 nm. It is composed of two orthogonally stacked layers of parallel double helices. Ten biotinylated staples extend from the bottom layer to enable immobilization on streptavidin-coated glass coverslips, while the top layer contains two central binding sites for the A and B adapters of the motor conjugate. These binding sites are positioned to geometrically align with the rotary component of the device, ensuring correct spatial configuration for torque transduction.

The rotor arm (v1) is a long duplex DNA structure approximately 220 nm in length. It is built from a scaffold extended with 66 identical staple extensions at one end. Each extension is complementary to a short fluorophore-labelled oligonucleotide, enabling optical tracking via fluorescence microscopy. The C and D adapter binding sites are located at a central region along the scaffold, 110 nm from either end, allowing the motor to connect the rotor arm to the baseplate with minimal slack.

To improve mechanical stiffness and reduce bias in rotational trajectories, the 10HB rotor arm was designed as a rigid rod-like structure based on a ten-helix DNA bundle with an overall length of approximately 280 nm. The bundle was assembled from parallel double helices connected by periodic crossovers, providing enhanced persistence length and directional stability. Two central staples from the mid-section were extended to hybridize with the C and D adapters of the motor, while a short fluorophore-labelled duplex was attached to the distal end for motion tracking.
Each origami component was prepared in a separate folding reaction, assembled using p8064 scaffold DNA and analysed by agarose gel electrophoresis and TEM. The integration of the motor conjugate with the rotor and baseplate structures was achieved via site-specific hybridization of the four motor adapters (A-D) to their respective positions on the DNA scaffold.

## 12. DNA origami assembly and purification:

The assembly of both origami designs involve combining 10 nM of the p8064 scaffold with each staple oligonucleotide at a concentration of 200 nM in a folding buffer solution of 10 mM Tris, 1 mM EDTA (pH of 8), and 16 mM $MgCl_2$. This mixture was then heated to 65 °C for 20 minutes to denature all DNA strands, followed by a gradual reduction in temperature to 20 °C over 16 hours, cooling at a rate of 1 °C every 45 minutes.
For assembling the devices to include the motor molecule, HPLC-purified motor-ABCD conjugate was added to the assembly mix of the rotor arm at a concentration of 300 nM.
All the different origami structures (baseplate, rotor-arm V1, 10HB rotor-arm and control versions) were prepared in separate batch reactions.

### 12.1 Dimerization of Baseplate and Rotor Arm V1 or 10HB rotor- arm

For the dimerization step, the rotor arm-preassembled with the motor-ABCD conjugate was mixed with the baseplate structure at a 0.5:1 molar ratio of rotor arm to baseplate. The integration was carried out using a post-assembly annealing program starting at 40 °C and decreasing by 1 °C every 15 minutes until reaching 15 °C. The assembled structures were analysed by 2.0% agarose gel electrophoresis.
In the case of the 10HB rotor-arm, the motor-ABCD conjugate was first annealed to a pre-assembled, Cy3-labelled 10HB at o.5 to 1 ratio motor-10HB, then combined with a pre-assembled baseplate using the post-assembly annealing program. The mixture was then mixed with the preassembled baseplate and left for another post-assembly annealing.

### 12.2 Assembly of Simplified Devices: 10HB Rotor Arm Without Baseplate

The 10HB rotor arm was first purified using three rounds of PEG precipitation[3] to remove unbound staple strands. For each cycle, an equal volume of PEG precipitation buffer (12.5 mM $MgCl_2$, 0.5 M NaCl, 15% PEG [w/v], and 1× TAE) was added to the sample, followed by centrifugation at 20,000 rcf for 30 minutes at 20 °C.
The gel-purified motor-ABCD conjugate was hybridized to the first biotinylated DNA strand at a 1:5 molar ratio of motor conjugate to biotinylated strand. The mixture was annealed in folding buffer (10 mM Tris, 1 mM EDTA, pH 8.0, 16 mM $MgCl_2$) using the short annealing program 1 (see Methods §5). Following this step, the second biotinylated strand was added to the mixture at a 5:1 molar ratio relative to the initially added biotinylated strand, and the sample was subjected to a second annealing step using short annealing program 2.
The resulting motor-biotin-DNA hybrid was then mixed with the PEG-purified rotor arm at a 0.5:1 molar ratio (motor to rotor arm). A Cy3-labelled oligonucleotide was also added at this stage to serve as a fluorescence reporter. The final assembly was subjected to a post-assembly annealing program starting at 40 °C and decreasing by 1 °C every 15 minutes until reaching 15 °C. Assembled structures were analysed by 2.0% agarose gel electrophoresis.

**13. Agarose gel:**

DNA origami structures were analysed using 2% agarose gels prepared in 1× TAE buffer (40 mM Tris, 40 mM acetic acid, 1 mM EDTA, pH 8) supplemented with 11 mM $MgCl_2$. Electrophoresis was performed at 60 V for 2 hours on ice to maintain structural integrity during separation.

**14. Transmission electron microscopy (TEM) analysis:**

For TEM analysis, a 5 μL of the diluted origami sample solution (1 nM) was applied to glow-discharged, carbon-coated grids. These samples were then stained with a 2% solution of filtered uranyl format in 5 mM NaOH, for a duration of 2 minutes. TEM imaging was conducted using a (to be checked) transmission electron microscope, operating at an acceleration voltage of 100 kV.

**15. Flow-Chamber Preparation and Sample Loading:**

Biotinylated BSA was prepared by diluting one 20 μL aliquot (10 mg/mL) in 180 μL of buffer A+. Streptavidin was prepared by diluting one 10 μL aliquot (10 mg/mL) in 190 μL of buffer A+. The imaging buffer was prepared fresh by mixing 5 μL of 100× Trolox, 5 μL of 100× PCD, and 12.5 μL of 40× PCA with 477.5 μL of buffer B+ to a final volume of 500 μL. Flow chambers were assembled using double-sided tape between a microscope slide and a coverslip. For the sample loading, surfaces were treated as follows: 20 μL of diluted biotin–BSA was introduced and incubated for 2 minutes, followed by a wash with 50 μL of buffer A+. Next, 20 μL of diluted streptavidin was added and incubated for 2 minutes, then washed with 50 μL of buffer A+. The surface was then washed with 50 μL of buffer B+, followed by the introduction of 20 μL of DNA nanostructures and a 2-minute incubation. A final wash with 50 μL of buffer B+ was performed before adding 20 μL of imaging buffer. The chamber was sealed with epoxy prior to imaging.

## II. Gels

### 1. Templated Conjugation CuAAc Reaction- Preparing motor_AB conjugate (main text Fig. 1d)

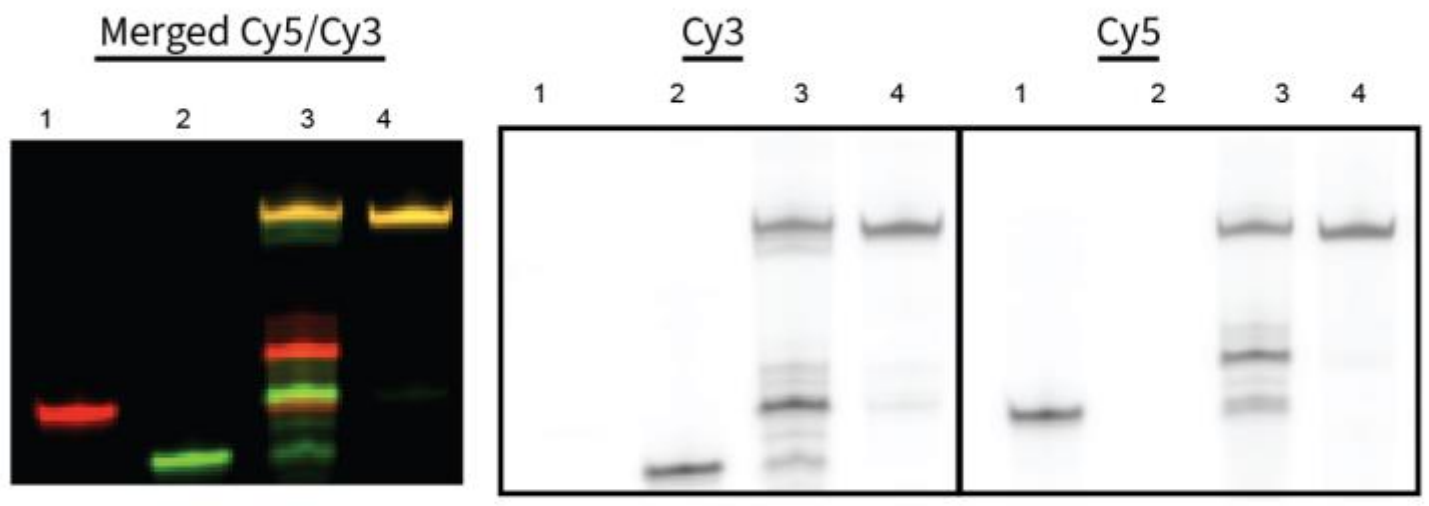

*Figure S1: CuAAc Reaction. Split-colour image of the gel shown in the main text, Fig. 1d.*

### 2. Templated Conjugation Amide Reaction- Preparing motor_CD conjugate (main text Fig. 1e)

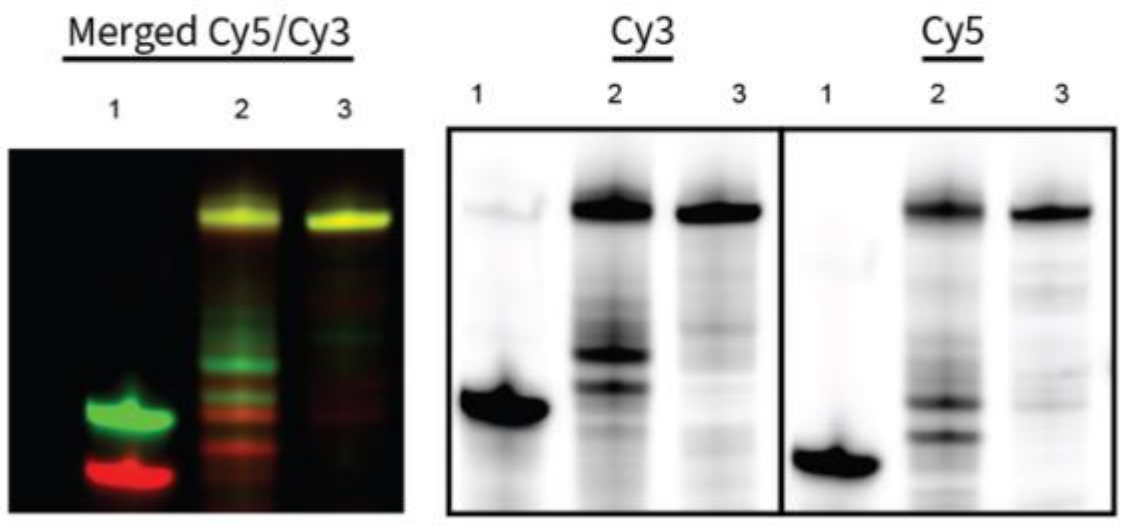

*Figure S2: Amide Reaction. Split-colour image of the gel shown in the main text, Fig. 1e.*

### 3. Initial Attempts to complete motor-AB conjugation to adapters C and D

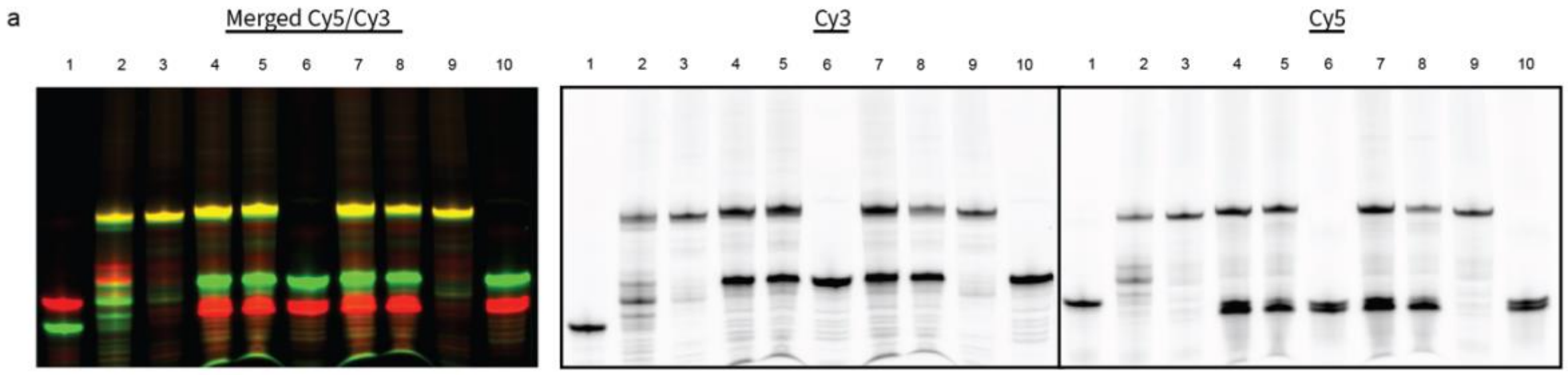

*Figure S3. Attempts to complete conjugation of motor-AB to form motor-ABCD.*
*Merged and Split-color image of the denaturing PAGE gel corresponding to Methods section §2. Lanes: (1) adapters A and B; (2) CuAAC reaction of A and B; (3) HPLC-purified motor-AB conjugate; (4) templated-to-templated reaction (§2.1); (5) templated-to-non-templated (§2.2); (6) adapters C and D; (7) non-templated-to-templated (§2.3); (8) non-templated-to-non-templated (§2.4); (9) motor-AB; (10) adapters C and D.*

## 4. Initial Attempts to complete motor-CD conjugation to adapters A and B

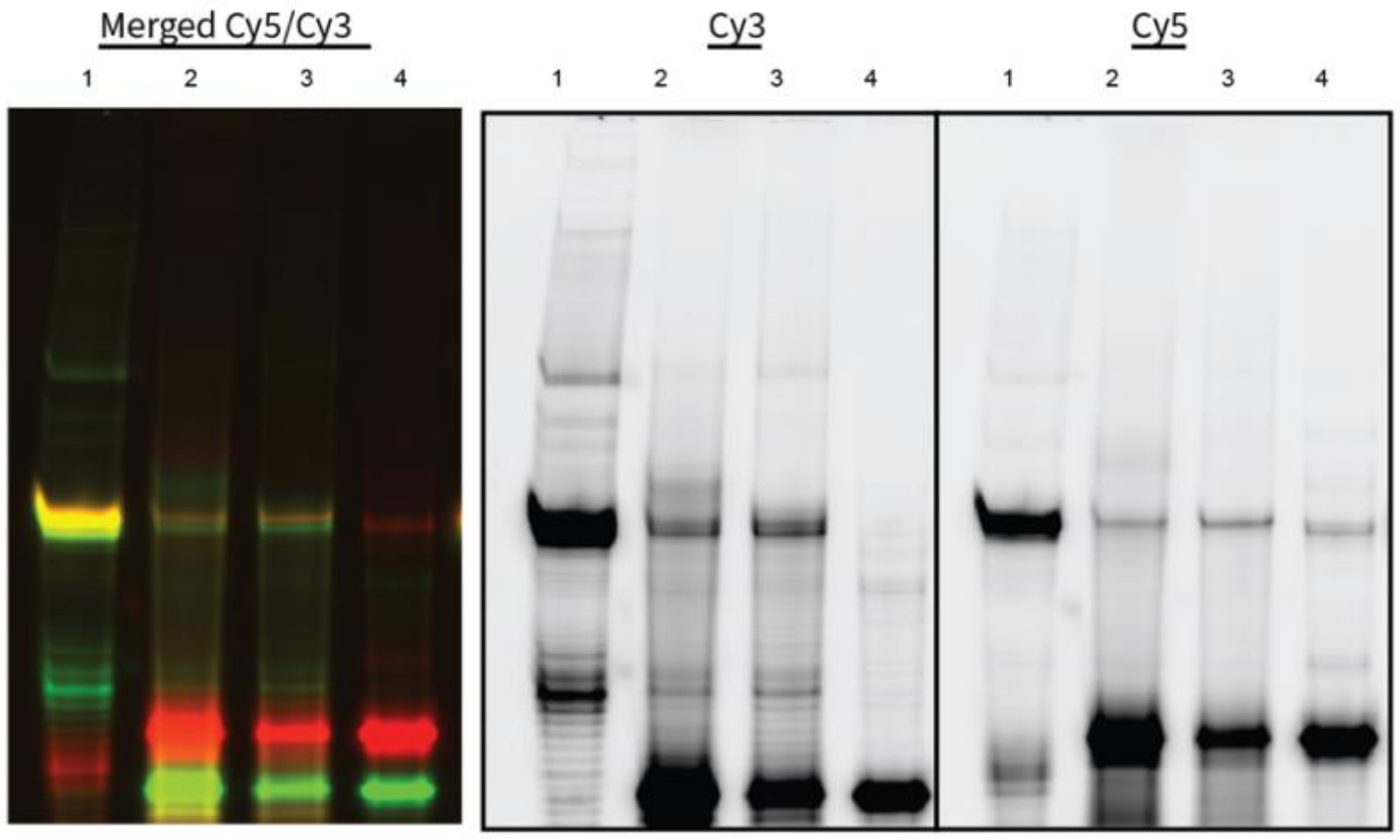

*Figure S4. Attempts to complete conjugation of motor-CD to form motor-ABCD. Merged and Split-colour image of the denaturing PAGE gel corresponding to Methods Section §2.Lanes: (1) HPLC-purified motor-CD conjugate; (2) templated-to-non-templated reaction (§2.5); (3) non-templated-to-templated reaction (§2.6); (4) adapters A and B.*

## 5. Two template conjugation scheme (main text Fig. 2b, and c)

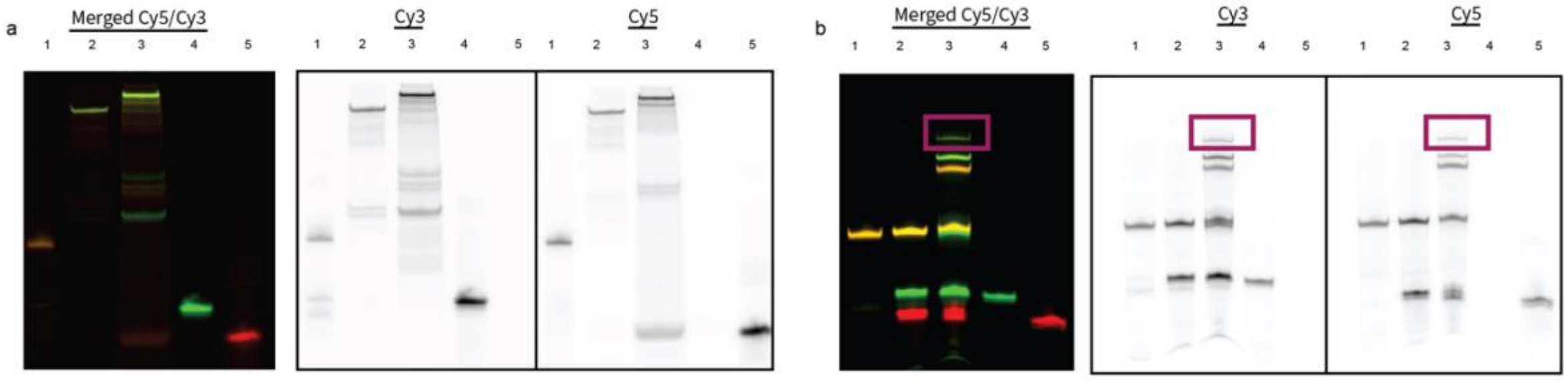

*Figure S5: (a) Split-colour image of the gel shown in the main text, Fig. 2b, and (b) Split-colour image of the gel shown in the main text, Fig. 2c.*

## 6. Motor-adapter conjugates against complements (main text Fig. 2d, e, and f)

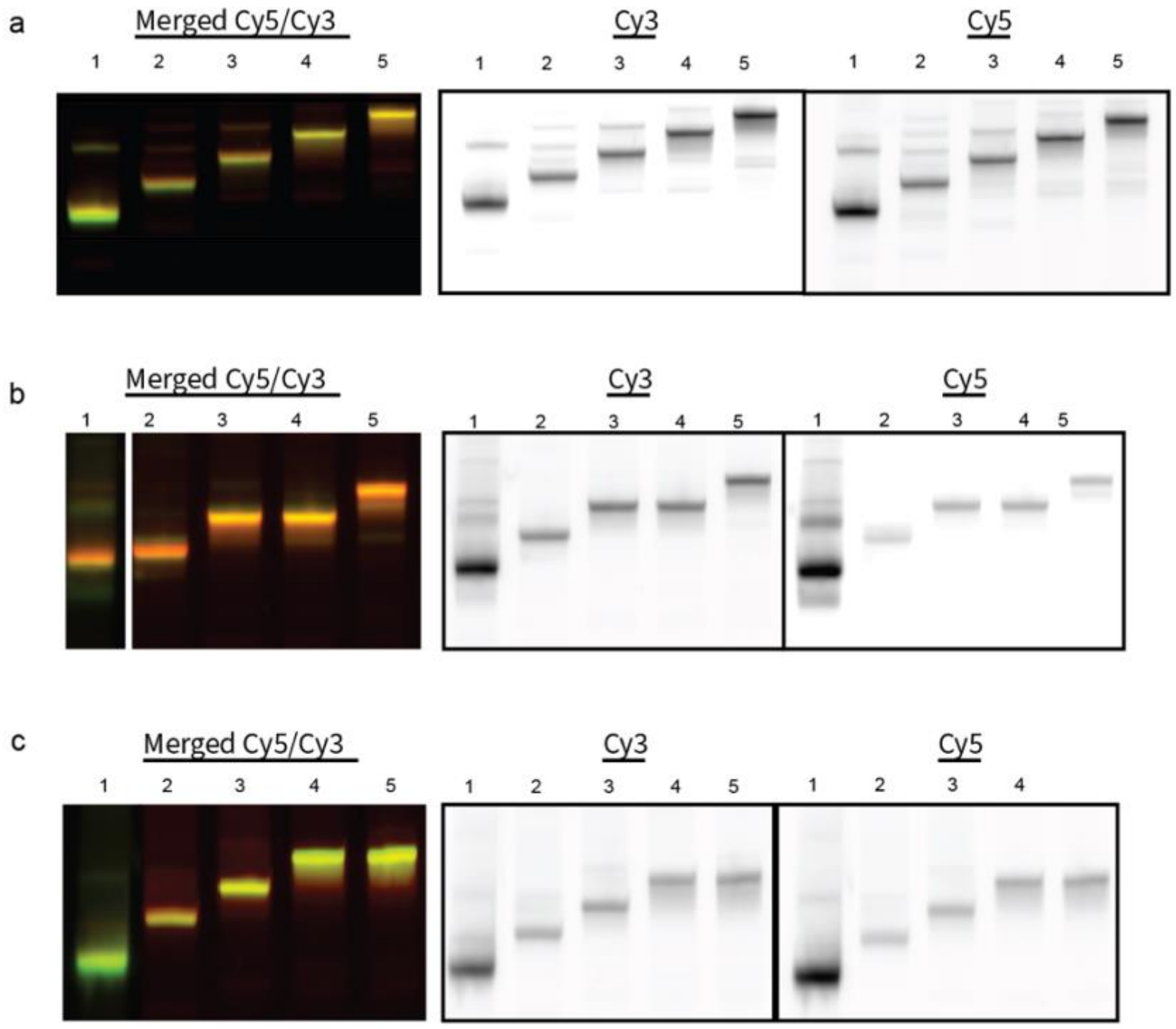

*Figure S6: (a) Split-colour image of the gel shown in the main text, Fig. 2d. (b) Split-colour image of the gel shown in the main text, Fig. 2e. (c) Split-colour image of the gel shown in the main text, Fig. 2f.*

## 7. Motor-ABC and motor-ABD to motor-ABCD

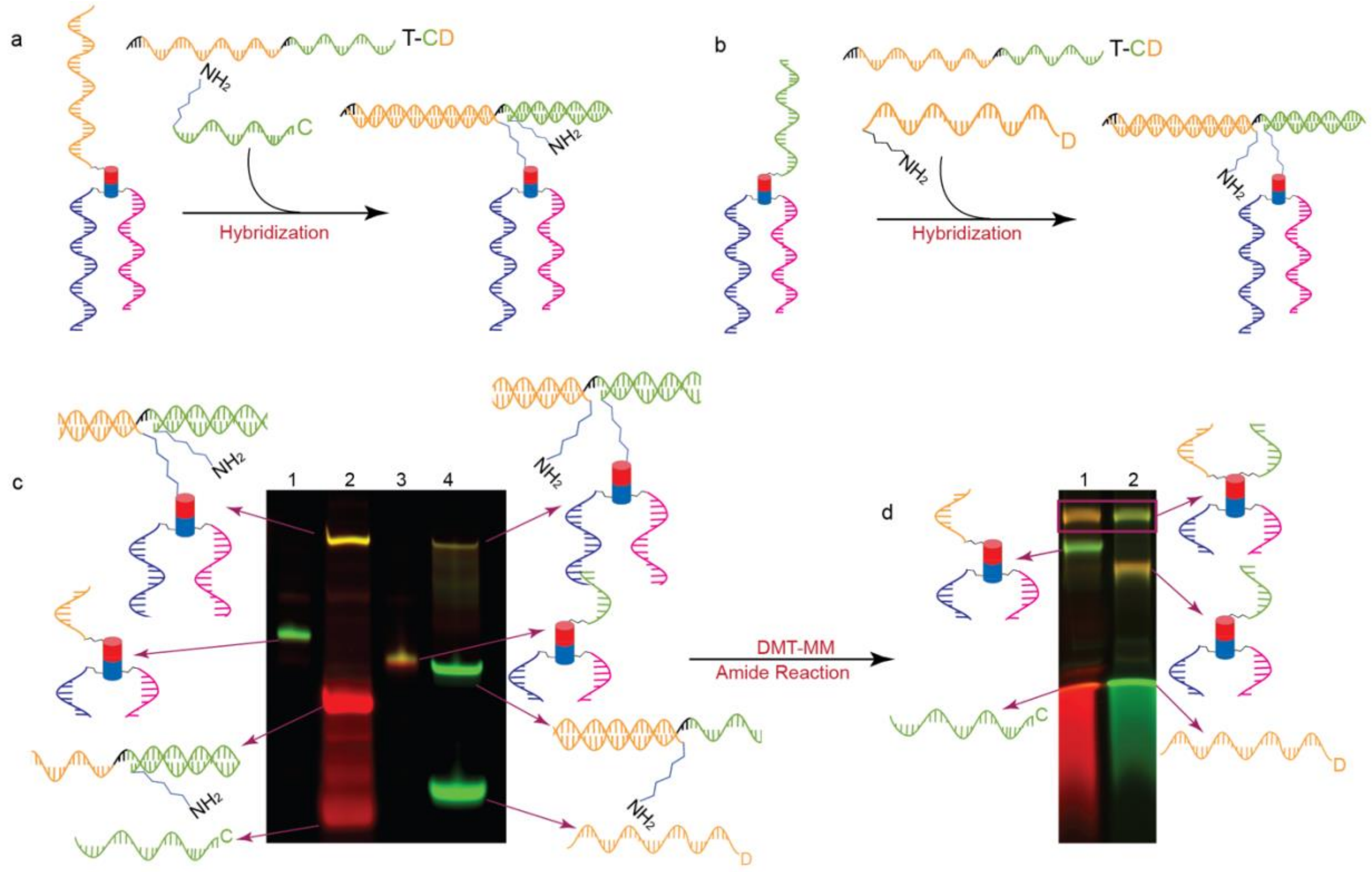

*Figure S7: Adding a fourth adapter to a triply conjugated motor. Panels a and b show the templated reactions designed to generate motor-ABCD from motor-ABC and motor-ABD, respectively. (c) A 10% native PAGE shows purified motor-ABC (lane 1) and motor-ABD (lane 3), annealed to template T-CD and adapters D and C in lanes 2 and 4, respectively, prior to amide bond formation. (d) A 20% denaturing PAGE shows the products of the templated amide bond formation reactions. Motor-A,B,C (lane 1) and motor-A,B,D (lane 2) were each reacted with the missing adapter to yield the fully conjugated motor-ABCD, visible as the lowest-mobility band (indicated by a box). All gels were scanned using Cy3 and Cy5 fluorescence channels.*

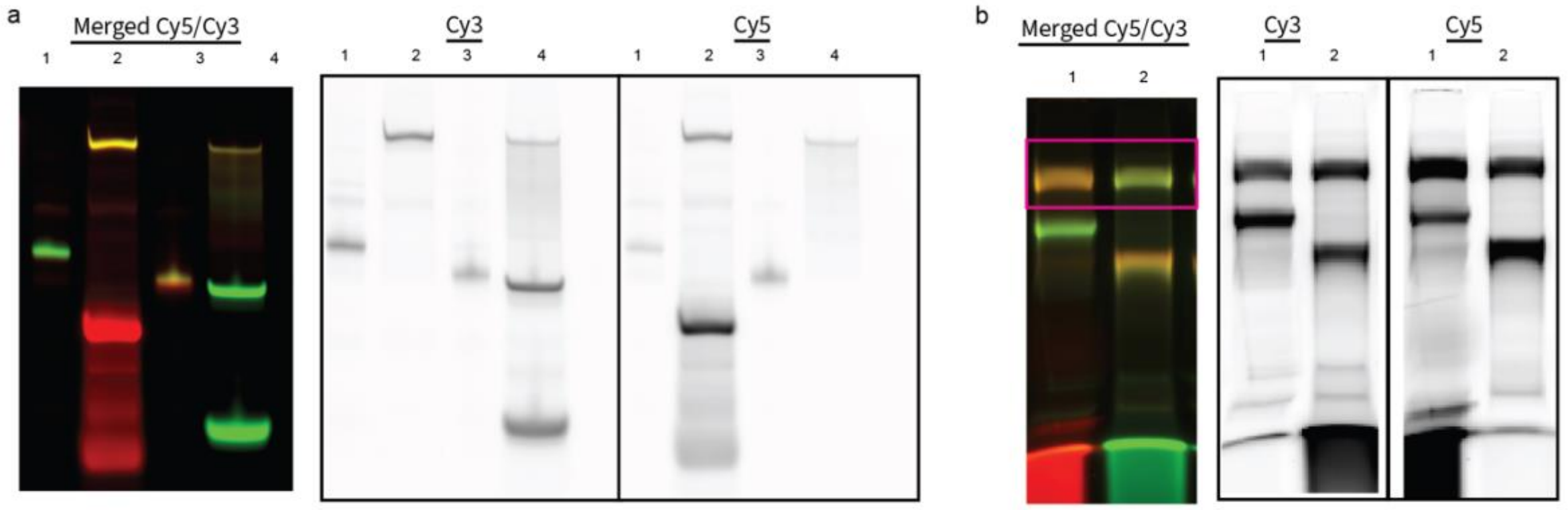

*Figure S8. (a) Split-color image of the gel shown in Fig. S3c. (b) Split-color image of the gel shown in Fig. S3d.*

## 8. Devices Assembly (main text Fig. 3c)

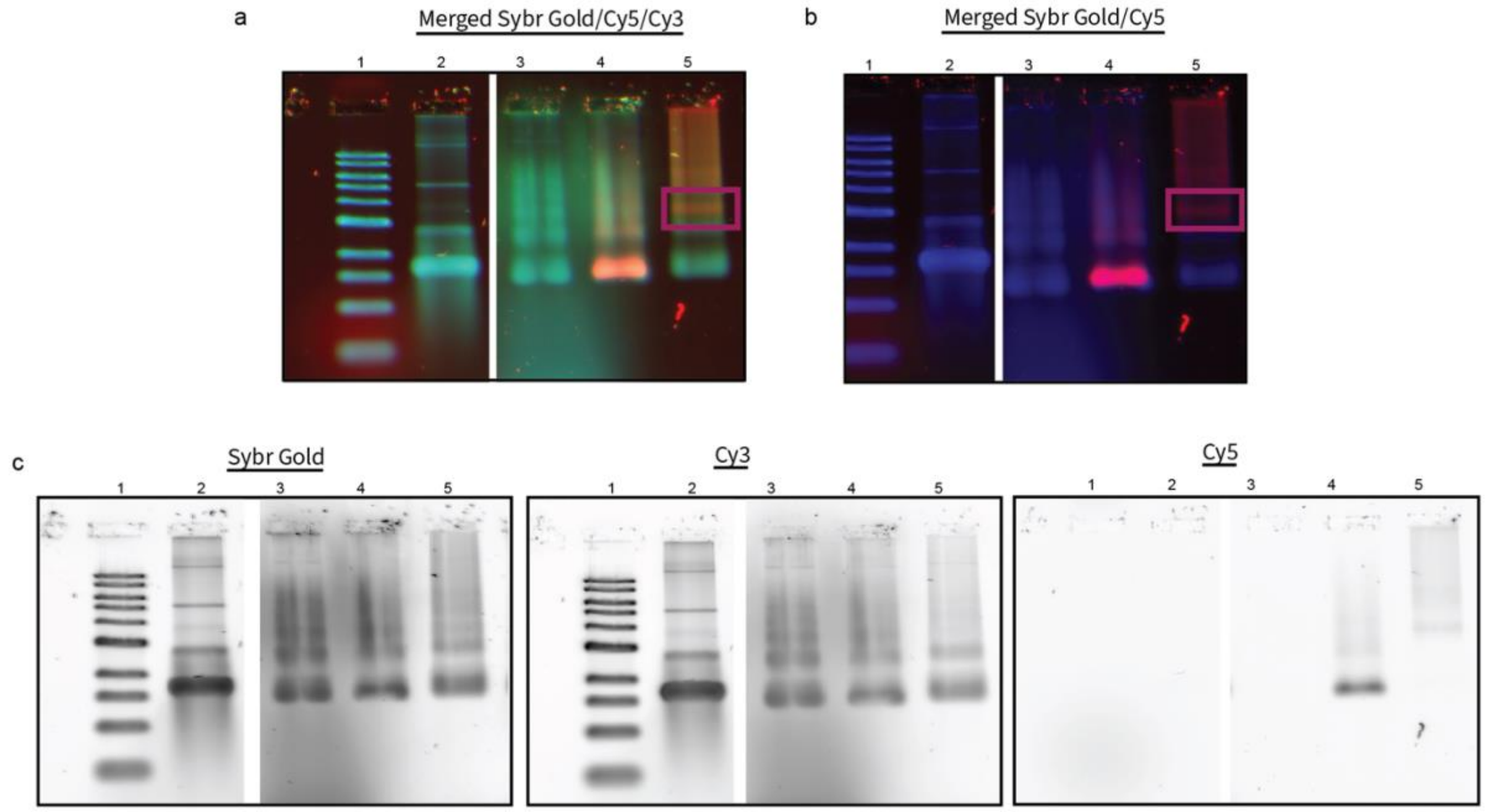

*Figure S8. (a), (b) merged colour of the gel shown in Fig. 3c. (c) Split-color image of the gel shown in Fig. 3c.*

## III. HPLC chromatograms of adapters A and B and motor-AB conjugate:

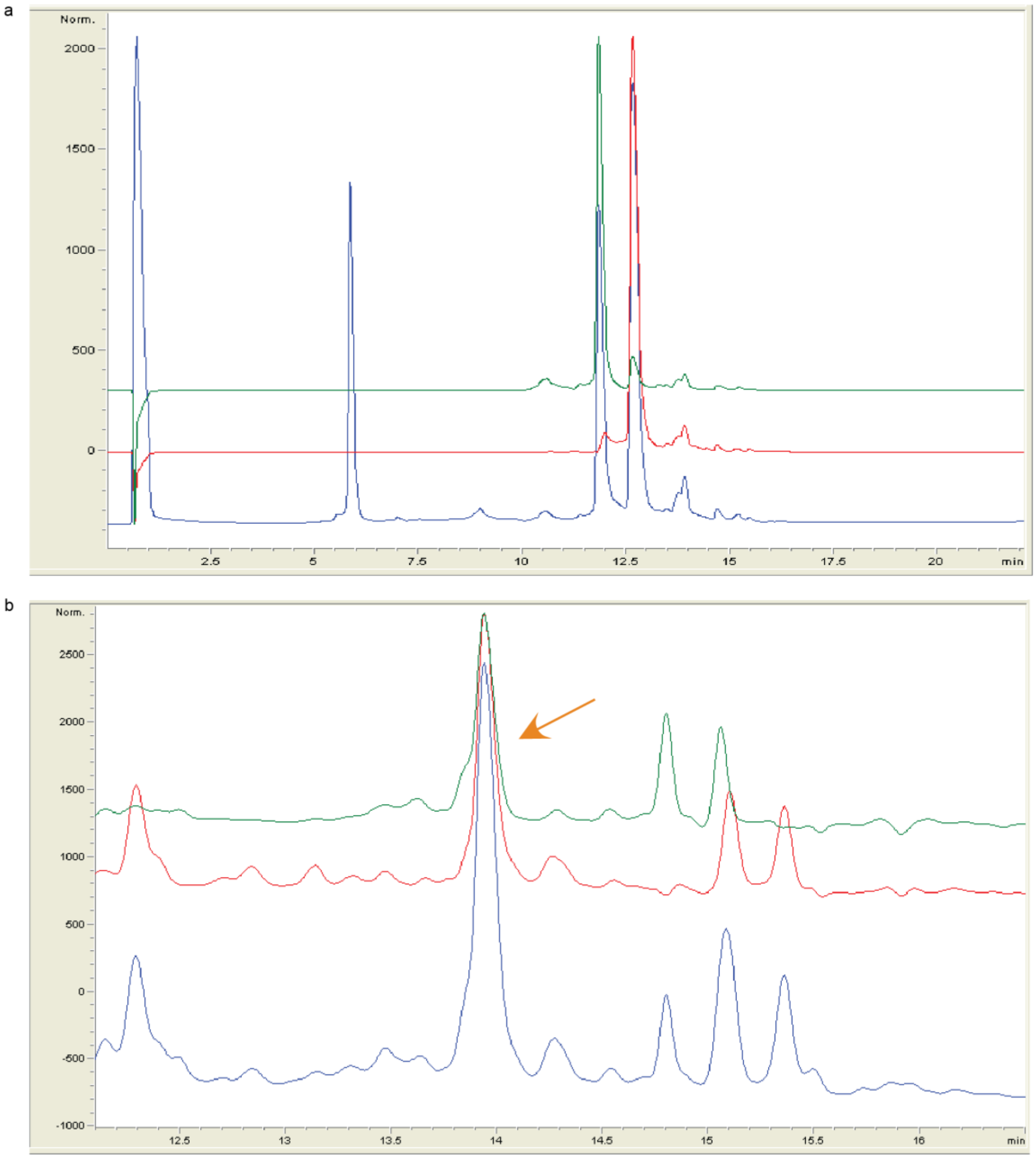

***Figure S9. HPLC trace** of (a) templated adapters A and B, and (b) templated CuAAC reaction of the motor with adapters A and B. Chromatograms show the absorbance signals at 260 nm (DNA, blue), 648 nm (Cy5, red), and 554 nm (Cy3, green). The peak where all three signals coincide (indicated with an orange arrow) corresponds to the motor–AB conjugate product in panel (b) after the .*

# IV. TEM

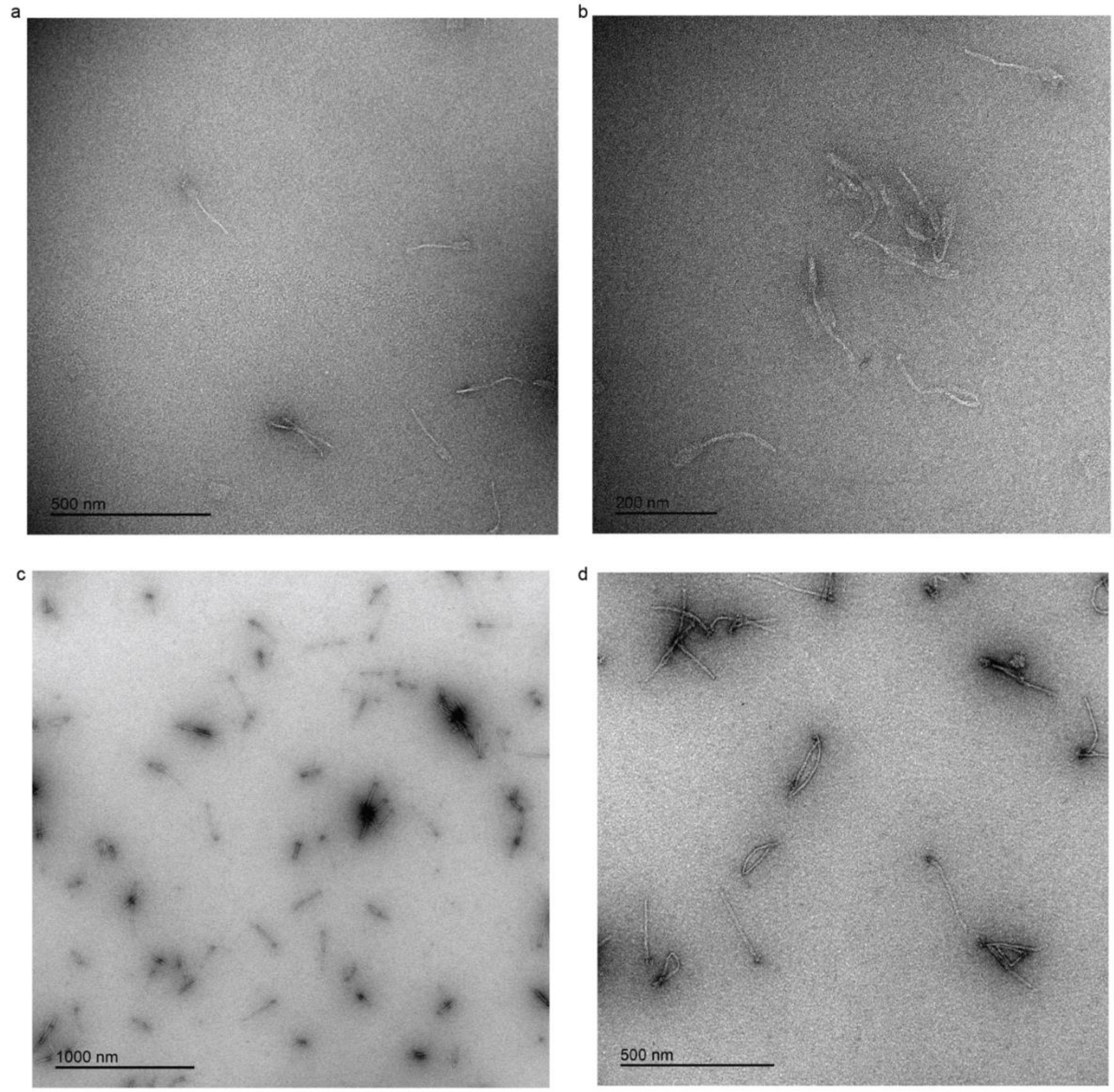

*Figure S10. (a) TEM image of rotor-arm V1 , (b) the baseplate- rotor-arm dimmer. (c) and (d) TEM images of the rotor-arm V2 (10HB).*

# V. Origami Designs:

## a. Baseplate

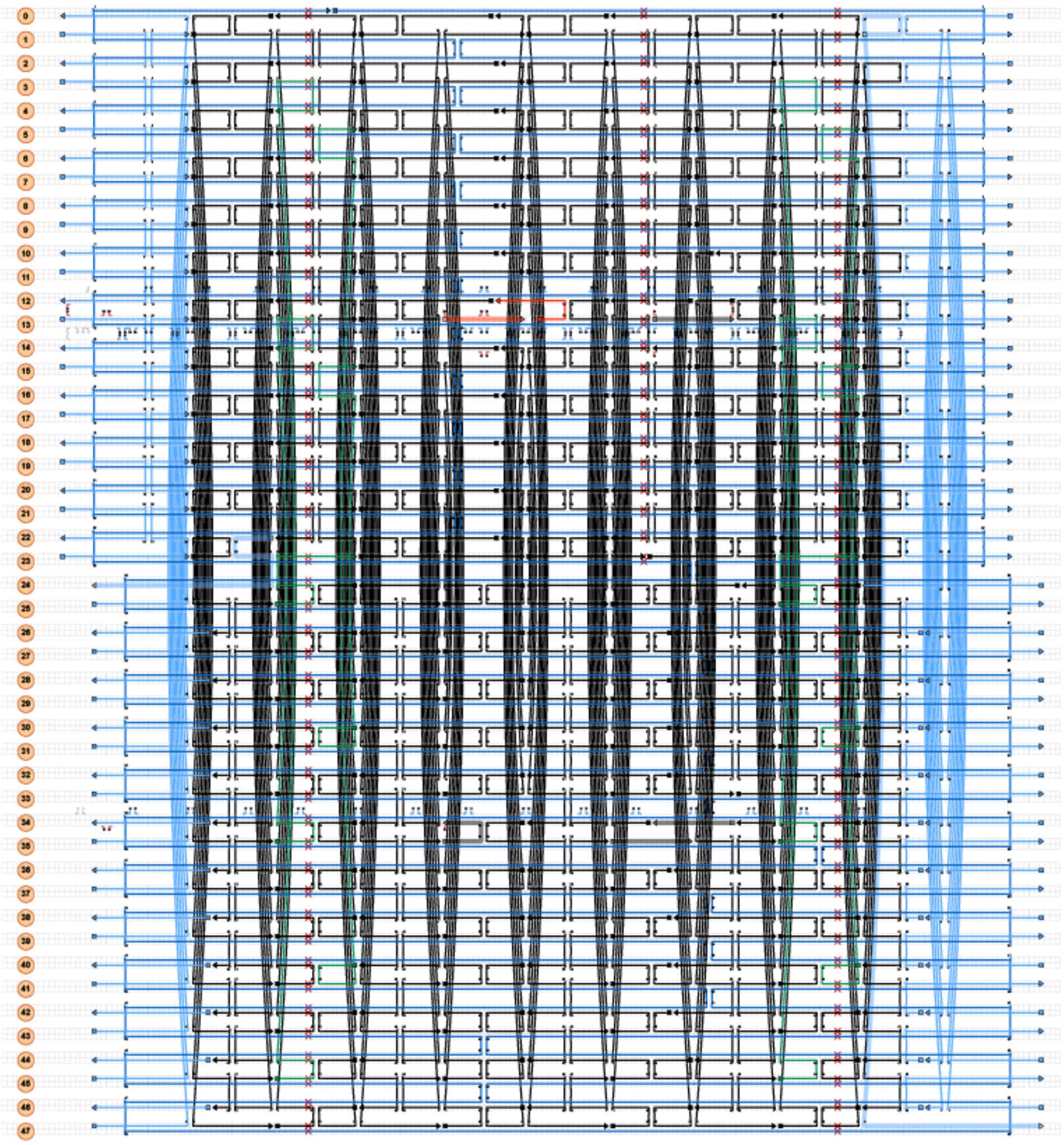

***Figure S11. DNA origami baseplate design***. *CaDNAno schematic showing the strand diagram (top) and cross-sectional view (bottom) of the DNA origami baseplate. The scaffold routing is shown in dark blue, with end staples in light blue and core staples in black. Attachment adapters A and B for motor integration are highlighted in red. Staple strands modified with biotin for surface anchoring are shown in dark green.*

## b. Rotor-Arm V1

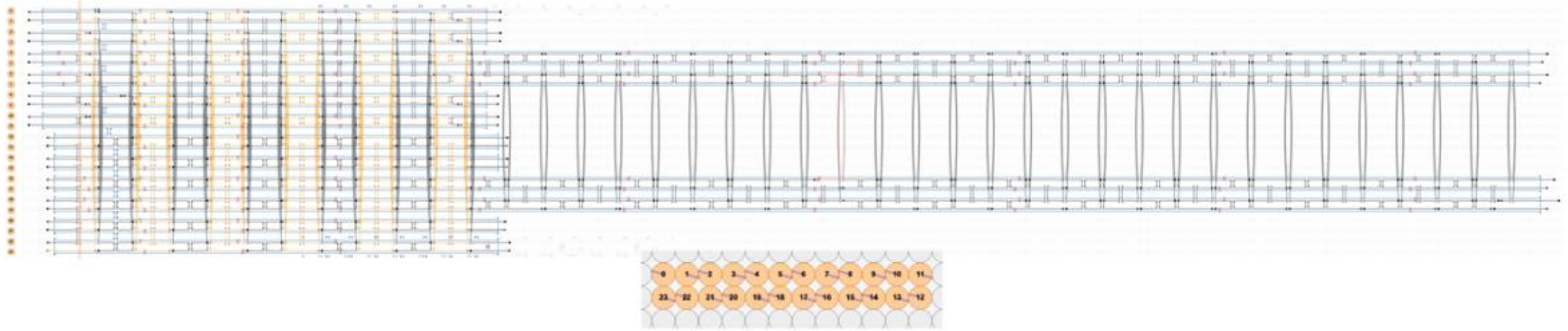

***Figure S12. DNA origami rotor-arm V1 design.*** *CaDNAno schematic showing the strand diagram (top) and cross-sectional view (bottom) of the rotor-arm V1 design. The scaffold path is shown in dark blue, with core staples in black. Staples containing overhangs for Cy3-labeled oligo hybridization are highlighted in yellow. Attachment adapters C and D for motor integration are shown in red.*

## c. 10HB Rotor-Arm

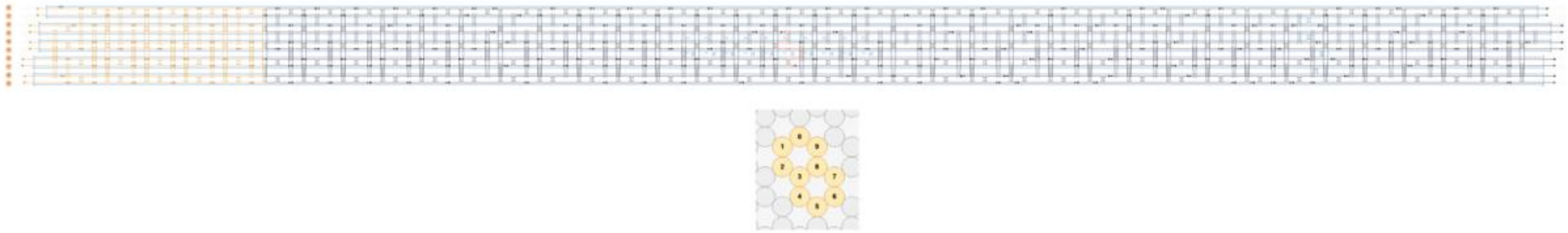

***Figure S13. DNA origami 10 HB rotor-arm design****. CaDNAno schematic showing the strand diagram (top) and cross-sectional view (bottom) of the 10-helix bundle (10HB) rotor-arm design. The scaffold path is shown in dark blue, with core staples in black. Attachment adapters C and D for motor integration are highlighted in red.*

## VI. Scripts and Single Particles Analysis

The data was localized using Picasso with the 'lq' (low quality) fit method and an initial Gaussian width guess of 100 nm (-g 100). This step produced localization coordinates from the raw TIRF image stack of the recorded measurements for subsequent reconstruction and analysis. All subsequent analyses were performed using custom Python scripts developed for this study.

### 1. Dimer Colocalization Analysis:

For the analysis of the baseplate-rotor-arm dimers, to analyse the motion dynamics, dimers of interest were first manually selected from time-lapse fluorescence images of individual rotor devices captured by TIRF microscopy. K-means clustering was then applied to localize the two distinct fluorescent particles within each selected region. The centre coordinates of these clusters were used to define a rotation axis and reorient the positional data accordingly. The x-coordinate of the rotated trajectory was then plotted against time to generate a switching trace, allowing visualization of dynamic reorientation events. For each region, the x-coordinate of the rotated trajectory was then plotted against time to generate a switching trace, allowing visualization of dynamic reorientation events.

### 2. Single-Particle Tracking and Rotation Analysis:

A difference-of-Gaussian filter was applied to detect particles, followed by Gaussian fitting for sub-pixel localization. Coordinates were recorded in pixel units (117 nm/pixel) and converted to nanometres for downstream rotational analysis. Particles are then linked across frames to form trajectories based on spatial continuity. Only trajectories present in at least 80% of the total frames are retained for further analysis. For each particle, angular displacement is calculated relative to the centre of mass of its trajectory, and traces are filtered to exclude those with significant radial fluctuations. From the angular displacement data, cumulative angle, angular velocity, and stepwise transitions are computed. A threshold-based step detection algorithm is used to identify discrete angular changes. In addition to per-particle plots of rotation angle over time, angular velocity, and step histograms, the script outputs a .csv file containing the net angular change for each particle, calculated as the difference in mean angle between the first and last 100 frames (compound rotation). A .pickle file containing all computed data is also generated to facilitate downstream analysis. A summary Excel file is compiled with all particle-level measurements to enable batch comparison and classification.

### 3. Histogram Analysis of Net Angular Changes:

A histogram of net angular changes was generated from the .pickle file containing unwrapped angular trajectories (produced in Scripts and Single-Particle Analysis §1). For each particle, the mean angle over the first 100 frames was subtracted from the mean angle over the last 100 frames to calculate the total angular change. These values were compiled and plotted as a histogram to assess the distribution of angular displacements across the dataset.

### 4. Angular Trace Extraction of Pre-Selected Rotating Particles:

Rotational dynamics were quantified for particles pre-selected from prior localization data generated using Picasso. For each selected group, localizations were re-centred and reconstructed into a 2D histogram, which was then cross-correlated with a library of annular templates of varying radii to refine the particle centre. The localization coordinates were transformed into polar form relative to the best-fit centre, and the angular coordinate was unwrapped over time to generate rotation traces. This unwrapping procedure assumes that angular changes between consecutive frames do not exceed 180°, which can introduce ambiguity in traces with large angular jumps. In traces with large angular jumps, this can make the determination of rotation direction and number more challenging.
Angular dynamics were plotted against frame number, and UV illumination periods were overlaid for temporal reference. Additional plots - including radial reconstructions, correlation maps, and angular histograms - were generated for each particle. Results were saved as annotated images and exported to .h5 files for further analysis.

5. **Quantification of Angular Displacement and Velocity During UV On and Off:**

Angular displacement data were analysed from an Excel workbook (.xlsx), where each sheet corresponds to a single particle and contains its unwrapped angular trajectory. Each sheet contains a frame index, an unwrapped angular coordinate (theta_unwrapped), and a binary indicator (UV_on) marking UV illumination frames. For each particle, the script calculates frame-by-frame angular changes ($\Delta\theta$) and segments them into UV-on and UV-off intervals based on the UV_on mask. It computes both signed and absolute angular displacements and velocities, separately for UV-on and UV-off frames. All UV-on and UV-off segments within a trace are pooled together to compute a single average per condition. For each particle, the following metrics are extracted and saved: mean signed velocity during UV-on, mean signed velocity during UV-off, mean absolute velocity during UV-on, mean absolute velocity during UV-off, total angular displacement ($\Delta\theta$) during UV-on, and the total number of UV-on frames. These results are compiled across all particles. The script does not compute per-segment values but provides direct quantitative readouts of rotation behaviour under photoactivation versus baseline conditions.

6. **Extraction of Particle-centred Image Crops from Time-Lapse Movies:**

Cropped time-lapse image sequences were extracted from raw TIRF microscopy data, centred on selected particle coordinates. One movie per particle is generated to enable direct visual comparison of rotational or dynamic behaviour in the raw fluorescence data.

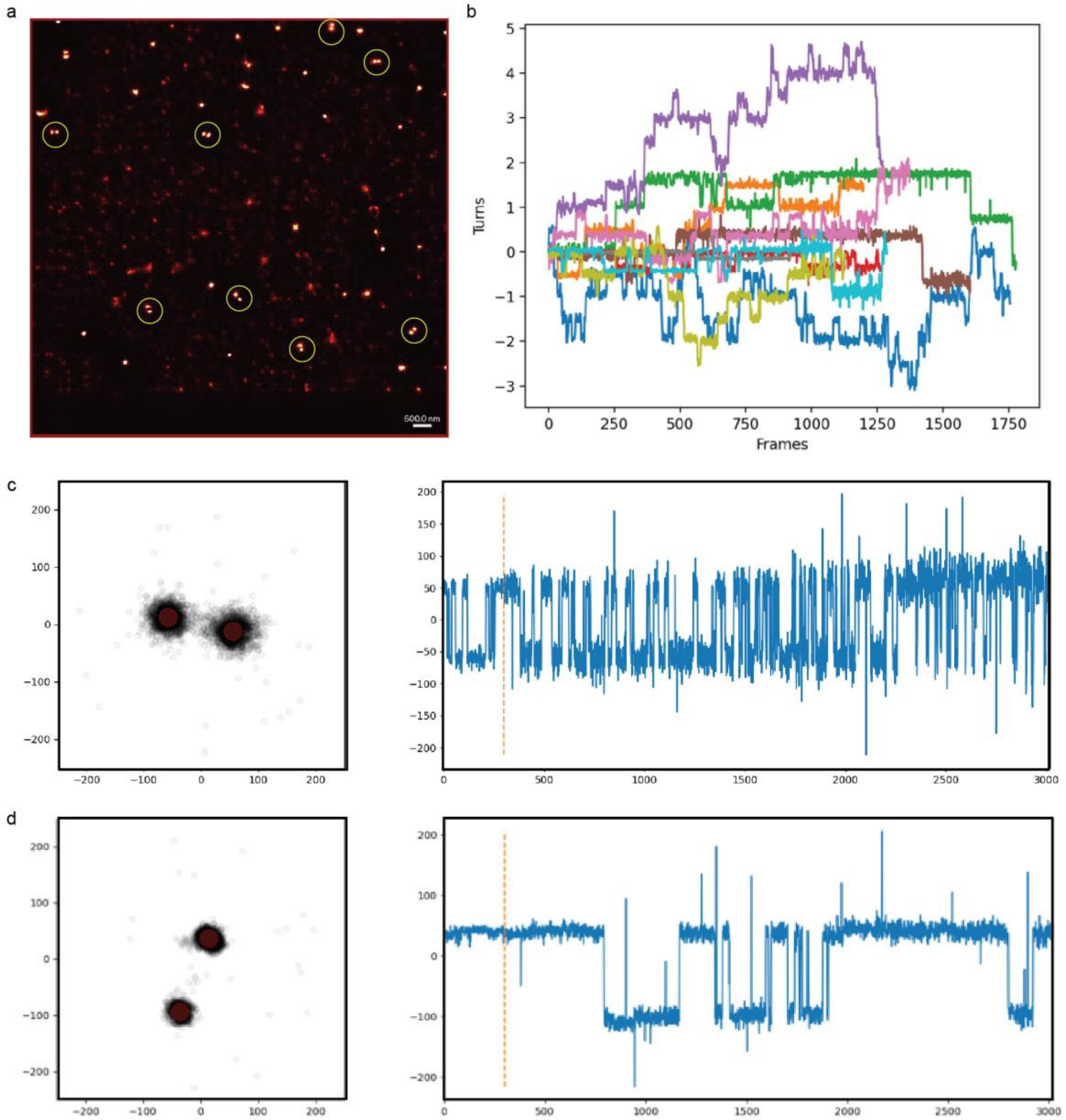

***Figure S14. Tracking of initial motor design without UV illumination.*** *Example localization images of first-generation devices (baseplate–rotor-arm V1) recorded without UV illumination. Devices were tracked at 100 Hz during imaging for 3000 frames (30 seconds), and several particles (highlighted by yellow circles) display switching behaviour between two states. (b) shows the corresponding cumulative rotation traces from tracked particles. (c) presents an example trace showing rapid switching both before and after UV exposure (orange line indicates UV-on period), while (d) shows a trace with slow switching occurring only after UV exposure.*

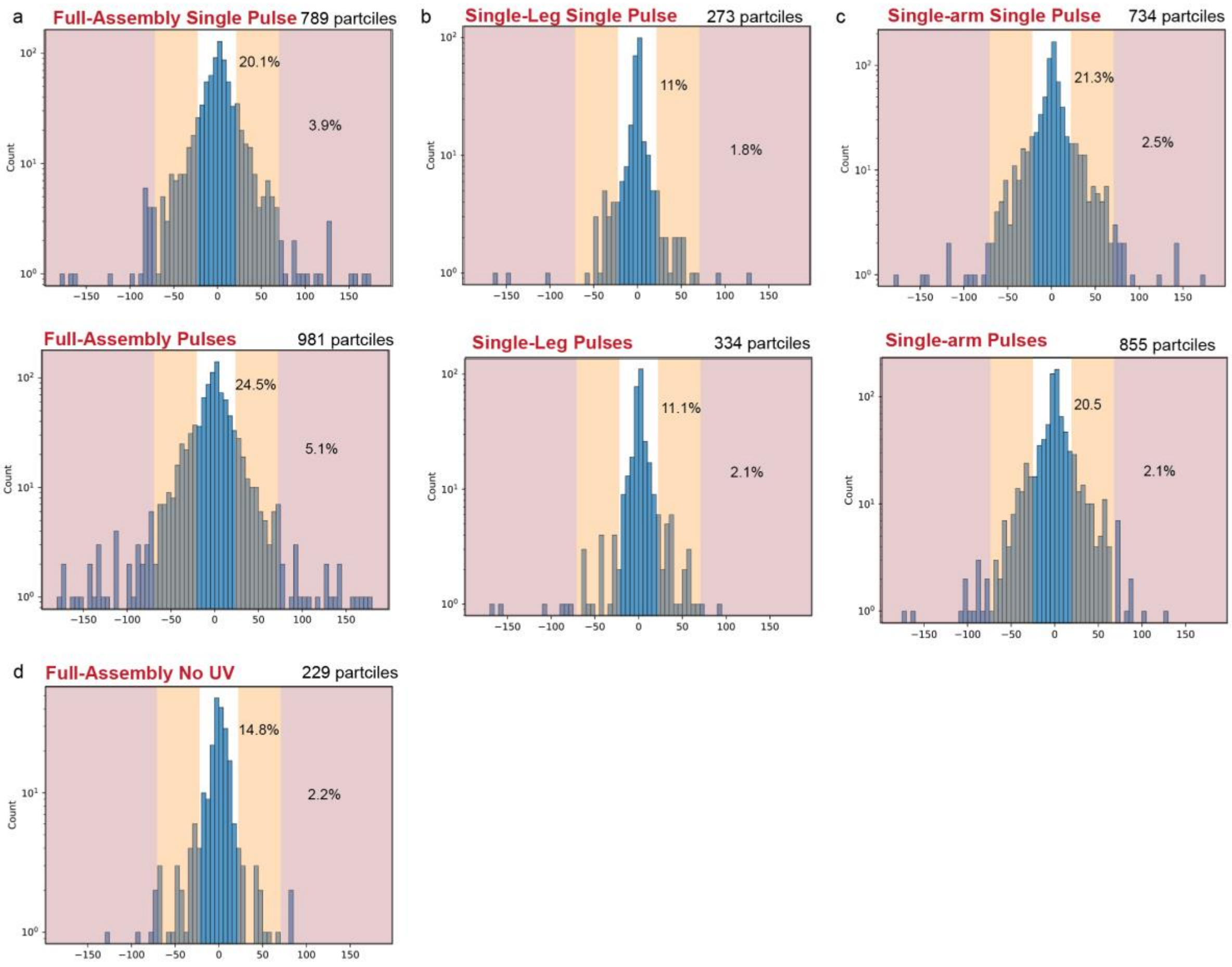

***Figure S15. Angular Displacement Analysis.*** *Histograms of angular displacements for (a) full-assembly, (b) single-leg, and (c) single-arm devices following single (top) or multiple (bottom) UV pulses. Panel (d) shows the displacement distribution for full-assembly devices under no UV illumination.*

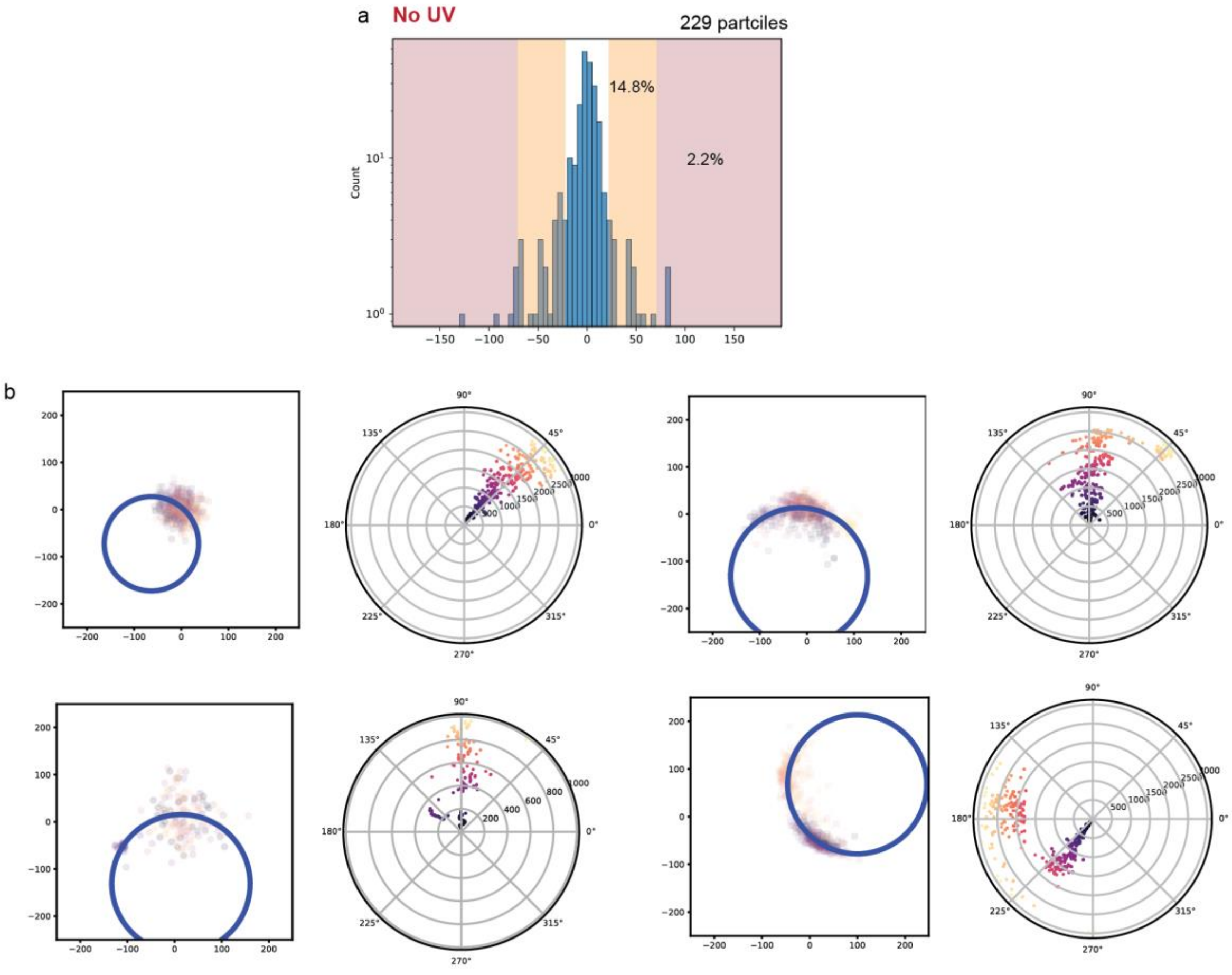

***Figure S16. Device examples – Full-Assembly case – No UV.*** *(a) Histogram of net angular displacements for 229 tracked full-assembly devices recorded without UV illumination. Shaded regions indicate thresholds used to define significant displacement, with 14.8% of particles exceeding ±25°, and 2.2% exceeding ±75°.*

*(b) Examples of individual rotor-arm trajectories showing large angular displacements. For each device, the left panel shows the localization map of the rotor arm end, with a fitted circle indicating the tracked trajectory. The right panel displays the corresponding polar plot of rotor orientation over time, where angle is plotted on the circumference and time increases radially from the centre outward.*

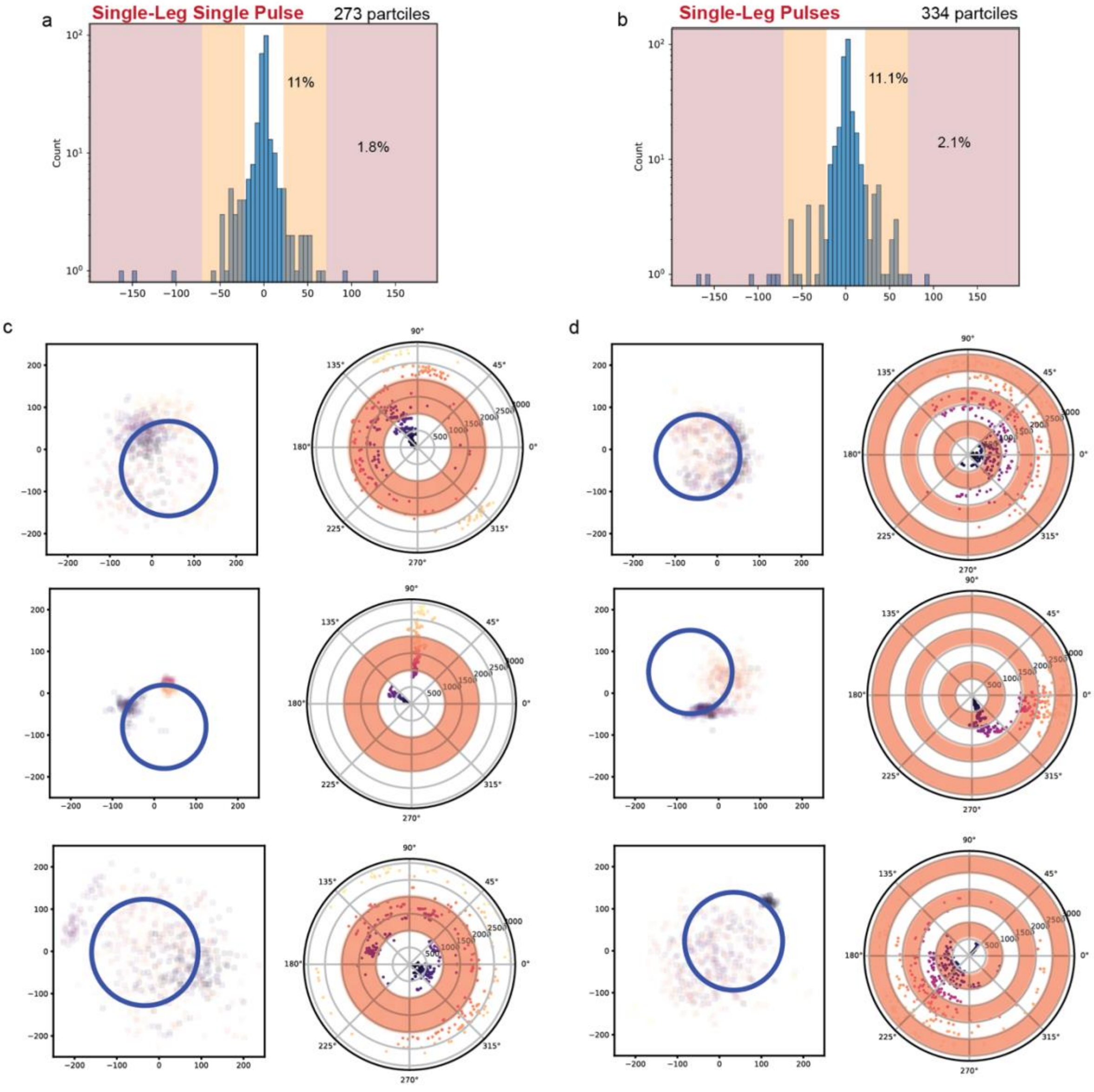

***Figure S17. Device examples – Single-Leg Control.*** *(a, b) Histograms of net angular displacements for single-leg devices under (a) a single UV pulse (273 particles) and (b) multiple UV pulses (334 particles). Shaded pink regions indicate thresholds for significant displacement; 11% (a) and 11.1% (b) of particles exceeded ±25°, and 1.8% (a) and 2.1% (b) exceeded ±75°.*
*(c, d) Example traces from individual single-leg devices showing large angular displacements under a single UV pulse (c) and multiple UV pulses (d). For each device, the left panel shows the localization map of the rotor arm end with a fitted trajectory circle, and the right panel presents a polar plot of rotor orientation over time. Time increases radially from the centre outward. Red-shaded rings in the polar plots indicate UV illumination periods.*

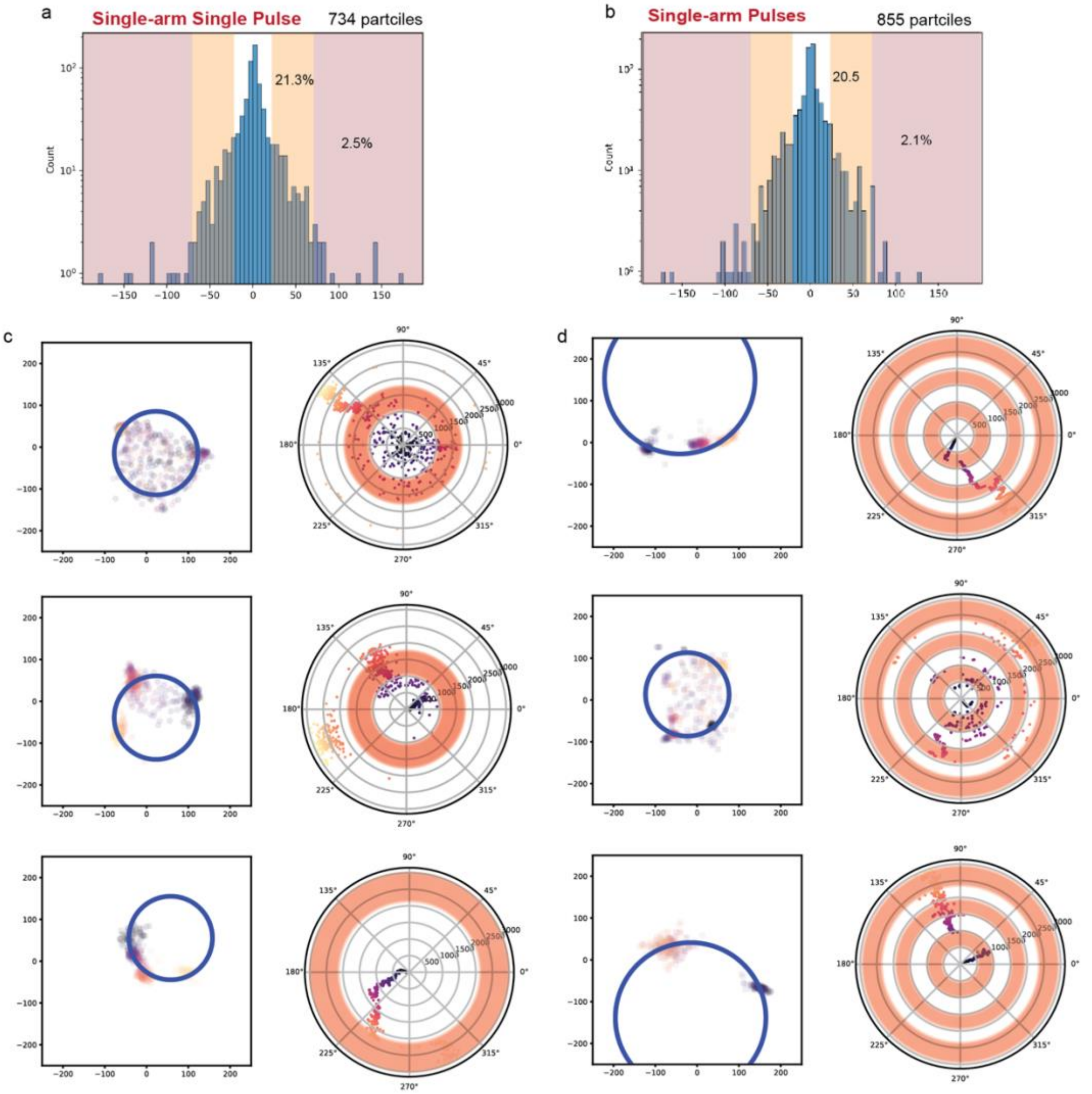

***Figure S18. Device examples – Single-Arm Control.*** *(a, b) Histograms of net angular displacements for single-arm devices under (a) a single UV pulse (734 particles) and (b) multiple UV pulses (855 particles). Shaded pink regions indicate thresholds for significant displacement; 21.3% (a) and 20.5% (b) of particles exceeded ±25°, and 2.5% (a) and 2.1% (b) exceeded ±75°.*
*(c, d) Example traces from individual single-arm devices showing large angular displacements under a single UV pulse (c) and multiple UV pulses (d). For each device, the left panel shows the localization map of the rotor arm end with a fitted trajectory circle, and the right panel presents a polar plot of rotor orientation over time. Time increases radially from the centre outward. Red-shaded rings in the polar plots indicate UV illumination periods.*

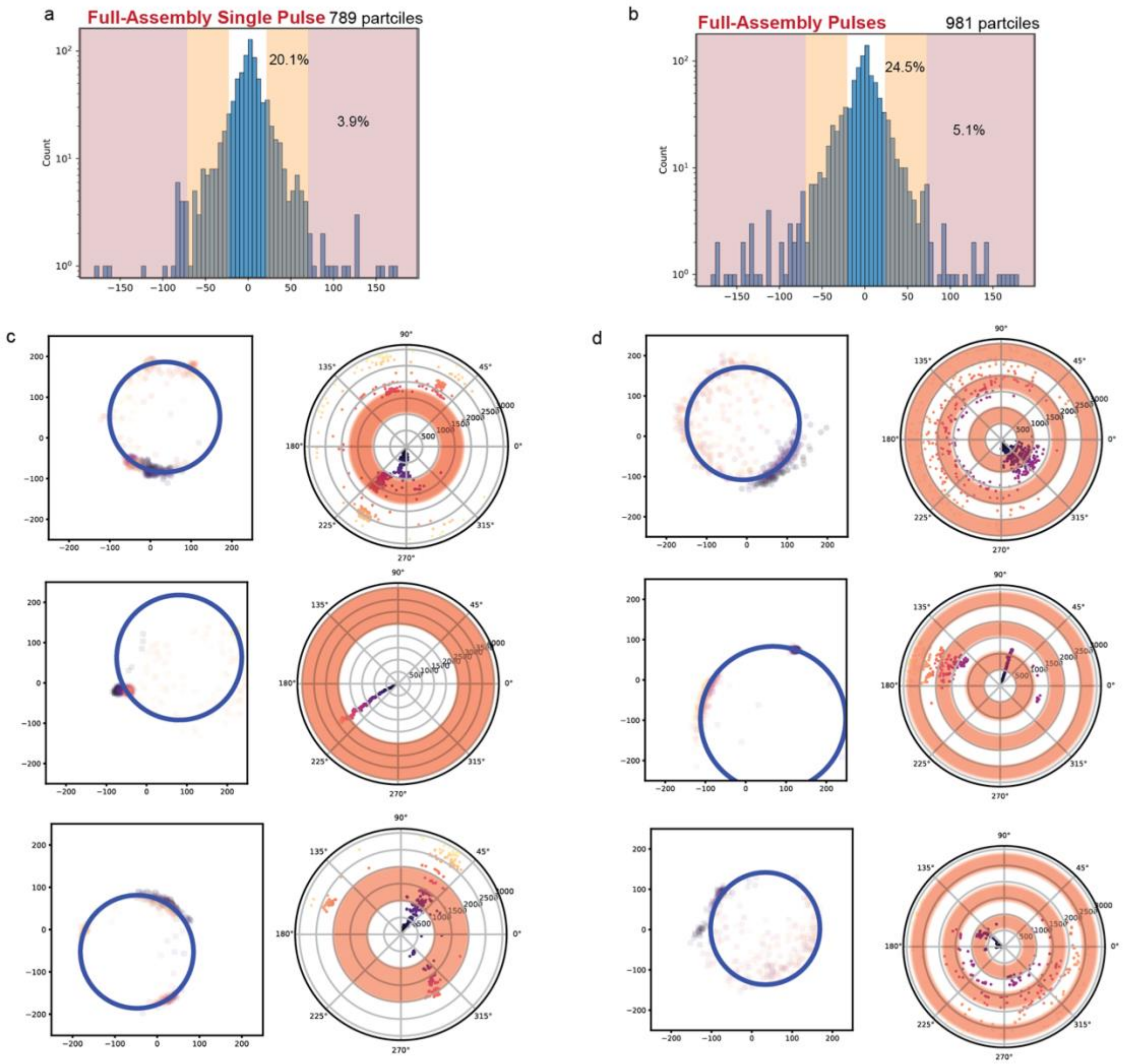

***Figure S19. Device examples – Full-Assembly Devices.*** *(a, b) Histograms of net angular displacements for full-assembly devices under (a) a single UV pulse (789 particles) and (b) multiple UV pulses (981 particles). Shaded pink regions indicate thresholds for significant displacement; 20.1% (a) and 24.5% (b) of particles exceeded ±25°, and 3.9% (a) and 5.1% (b) exceeded ±75°.*
*(c, d) Example traces from individual full-assembly devices showing large angular displacements under a single UV pulse (c) and multiple UV pulses (d). For each device, the left panel shows the localization map of the rotor arm end with a fitted trajectory circle, and the right panel presents a polar plot of rotor orientation over time. Time increases radially from the centre outward. Red-shaded rings in the polar plots indicate UV illumination periods.*

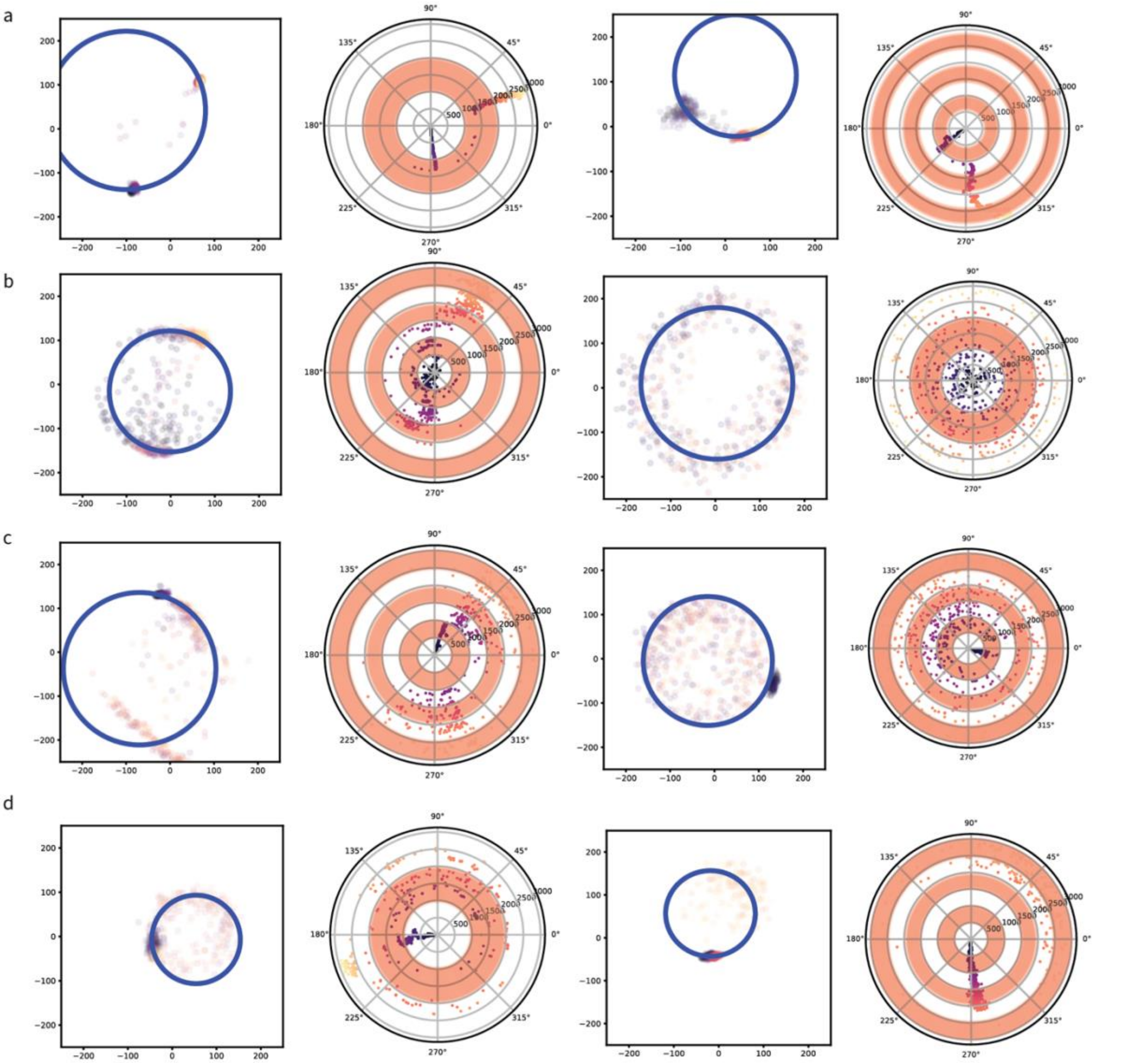

***Figure S20. Full-assembly devices showing ambiguous or control-like behaviour.***

*Representative examples of full-assembly devices with large angular displacements exhibiting motion characteristics resembling those of control configurations. (a) Devices showing small angular displacements upon UV activation (up to 120°). (b) Devices displaying random motion with no clear correlation to UV exposure. (c) Devices that initiate motion during UV pulses but exhibit inconsistent directionality or lack sustained rotation. (d) Devices with localized movement patterns confined to smaller-than-expected radii.*

*For each device, the left panel shows the localization map of the rotor arm end with a fitted trajectory circle, and the right panel shows the corresponding polar plot of rotor orientation over time. Time increases radially from the centre outward, and red-shaded rings indicate UV illumination periods.*

VII. **Oligo Sequences:**

## 1) Baseplate

| Name | Sequence |
| --- | --- |
| Core_Oligo1 | CCAACGTCAAAGGGCGAAGCTTGACGCTTTCCAGCGTAACGA |
| Core_Oligo2 | CCACTATTGAGAGAGTCTTTTGCGCGGAACGA |
| Core_Oligo3 | CCACTACGTCGGAACCATGGGATTGTAGCATT |
| Core_Oligo4 | GTTGAGTGTTGCCCCAGGTCGCTGTGAGGACT |
| Core_Oligo5 | CTGGCCCTAAAGAACGAAAAAAGGCTCCAAAA |
| Core_Oligo6 | GCACTAAATGAACCATAGAATAGAAAGGAACA |
| Core_Oligo7 | ACGCTGGTTTGTTCCATCGGTTTATCAGCTTG |
| Core_Oligo8 | CAGCTGATAAAGCCGGCGAGGCGCGTATCATTTTCCCGCCACC |
| Core_Oligo9 | AAAGCGAACACACCCGCCCATGTAACCGCCAC |
| Core_Oligo10 | TTTCACCATGCCTAATAGCGATTATGTATCAT |
| Core_Oligo11 | TTTGCGTAGCGTTGCGAGAATACACGCGACCT |
| Core_Oligo12 | CATTAATTTTGGGCGCAGAGGCTTAGGCTTGC |
| Core_Oligo13 | GCCTGGGGGTGAGACGGACAGCATGGATCGTC |
| Core_Oligo14 | CGCTACAGACGTGGCGATAGTTAGACGTTAGT |
| Core_Oligo15 | CCAGTCGGGGCCAACGAGGAAGTTGCCCACGC |
| Core_Oligo16 | GTAACCACAGGAGCGGCAACGCCTTTGC |
| Core_Oligo17 | GCCGGAAGTGGTTGCTTTGATCAGTGAACCAGGCGACAGTTAA |
| Core_Oligo18 | ACAATTCCTTTTCACGGAAGGATTATTATTCT |
| Core_Oligo19 | CTCGTTAGGTACGCCAATATAAGTGTCAGTGC |
| Core_Oligo20 | GCTGTTTCCACGCGTGTGTACAGAAGGCTGGC |
| Core_Oligo21 | CGTGAGCCTCGAATTCGCCGGAACCAACCTAA |
| Core_Oligo22 | GCGGGCCGACACAACACGGAGATTTACCAAGC |
| Core_Oligo23 | CAGGAACGAATCAGAGGGTTTAGTCCGT |
| Core_Oligo24 | TCTGCCAGCTGTGTGATCGAAATCCTAAAACA |
| Core_Oligo25 | TTTTTATAACGAGCACCCTCAGAAAGGGATAG |
| Core_Oligo26 | CCCCCTGCCGTCATAATGCCCTGACAGAACGA |
| Core_Oligo27 | TGGTGCTGCACCGAGTAAAAATATTACGTCTCTGACCACCACC |
| Core_Oligo28 | ACTGCGCGGCGGCTGGAAATCCTCAGGTTGAG |
| Core_Oligo29 | AACCGTTGAACTATCGTACATGGCAGAGCCAC |
| Core_Oligo30 | GTTGCCCTCCTGTGCACGGAACCTAGGATTAG |
| Core_Oligo31 | AGAACTCATAGCAATATTAACGGGATAG |
| Core_Oligo32 | TCTTTGCTATCAGACGAAGCGCATCCAGGAGGCTGA |
| Core_Oligo33 | TCCAGAACAGAGTCTGCCGTATAAGATAAGTG |
| Core_Oligo34 | GTGTGTTCACCTGCAGGTAATCTTTTTGAAAG |
| Core_Oligo35 | GCATCAGCCTGGTCAGGTCAGGACAAAATCTA |
| Core_Oligo36 | TGCGGTATCAGCCATTGCAAACAGAGATTCATAATGCAGCACC |
| Core_Oligo37 | ACCTACATCACGACCACCACCGGAGCCGGAAA |
| Core_Oligo38 | ACTCAATCCGGACTTGGTCATAGCTTTGCCTT |
| Core_Oligo39 | GTGCTGGTGGGGTCATGCCAACACCGAGATTGATAT |
| Core_Oligo40 | ACCAGTCATTTGACGCCCACCCTCTTTT |
| Core_Oligo41 | TTCTGGCCACAGGAAACACCAGAAATTTACCG |
| Core_Oligo42 | GAACGTGCCGCCGGGCTTGACAGGATTAAAGC |
| Core_Oligo43 | TGCCAACGCAGCTTACTGAGATGGCTGCTCAT |
| Core_Oligo44 | CCAGAGCAAACCCTTCTGAAGAAGATATGAATTATTATAAAAG |
| Core_Oligo45 | CACAGACACGCCATTATGGGAATTCGTAGAAA |
| Core_Oligo46 | ACGCCGCGAAGAGTTGGGTTTTCA |
| Core_Oligo47 | CCGTTCCGGTGTACATAAGGGCGAATGGTTTA |
| Core_Oligo48 | CCGTAAAAGTAAAGTTCTAACGGATTAAGAAC |
| Core_Oligo49 | CCACCAGCCCTGAAAGCATCGATACAAAATCA |
| Core_Oligo50 | TTAAATTTTGGCAGCCGAATCAAGCCCCTTAT |
| Core_Oligo51 | TAAAACATATATTTTTTTAGCAAGACCG |
| Core_Oligo52 | TCCGTGGTCAGCAGTTGGGCGGTTGCAA |
| Core_Oligo53 | TAGCTCTCTTGTAAAAAAGCAGATATGAAATA |
| Core_Oligo54 | AACAGTGCCAAATATCCAAAAGAACGCTAATA |
| Core_Oligo55 | ATCTGGTCCGGTCAGTTGGCAACACACCGTCA |
| Core_Oligo56 | GATAACCTATGTTTACCAGATACAGGAATACC |
| Core_Oligo57 | CTGAACCTCACGCTGATTAGCAAAAGAG |
| Core_Oligo58 | GCTGCAAGAAGATCGCAACGTCAAAGCCATAT |
| Core_Oligo59 | GCTATTACGCTTCTGGCATTGAATTGCTTTAA |
| Core_Oligo60 | AATATCTTATTTAGAAAGACGGGACTGAATCT |
| Core_Oligo61 | GTAACGCCCAACAGTTGAAGACAACTCAGAGAATACCAGAGCC |
| Core_Oligo62 | AACAATTCAGGAATTGACAAGAATAAACGCAA |
| Core_Oligo63 | CCGGCACCGCCAGCTGCGAAAATA |
| Core_Oligo64 | GCCTCAGGGCGATTAAAAGAAACAAGCCGAAC |
| Core_Oligo65 | ATTTGAGGTAGGAGCATAATTGAGCTGG |
| Core_Oligo66 | AAAGCGCCGGGCGATCTTTTGCCATCATAACC |
| Core_Oligo67 | AGTTTGAGCATATTCCAGCCTTAAACGGGTAT |
| Core_Oligo68 | TAACCGTGATCAAAAATTATCCGGGGAATCAT |
| Core_Oligo69 | ATAGGTCAAGCCAGCTAAGATTAAGAAAGACT |
| Core_Oligo70 | GAGGGGACTTAAATCCTTTATCAATATAGCGAACCAGAACAAG |
| Core_Oligo71 | GGAACGCCCATCTGCCAAATAAACAAATGAAA |
| Core_Oligo72 | GGCAATTCGCCCGAACGCGTCTTTACATAAAA |
| Core_Oligo73 | CTTCCTGTCGTTGGTGAATTCAAACCCCCAAGAAAC |
| Core_Oligo74 | TGTGAGCGAACGGCGGAGAATGACACTGCGGA |
| Core_Oligo75 | ATTATCATTAACATTAATTTTATCGAAT |
| Core_Oligo76 | CGCATTAATGGAGCAAGCCTGTTTGTCCAGAC |
| Core_Oligo77 | AGGGTTAGGATTTTCAGCATGTAGCAGAGGCA |
| Core_Oligo78 | CAGCTCATCCTGATTGTTTAGTAACAGAAGAAAAATATAAAGT |
| Core_Oligo79 | AAATTGTATCGTAAAACTCCTTTTCATTTTTG |
| Core_Oligo80 | ATTGCGTAAACCTACCTTCCAAGAATCA |
| Core_Oligo81 | TGAGAGTCATTTTTGTTCATCGTATATTCTAA |
| Core_Oligo82 | AACGGTAAAACGTTAAAAGAGCCCGAGGACAAATCA |
| Core_Oligo83 | TATGTACCGAAGATTGTAATTCGAGAAGCAAA |
| Core_Oligo84 | GAATATACGGATTATAACTCATCGTCCCGACT |
| Core_Oligo85 | TTCTAGCTAATGCCTGATTCCATATTCCCAAT |
| Core_Oligo86 | AGCTATTTAATTTTTAGCCTGTTTAGGCGTTA |
| Core_Oligo87 | TGATTGCTGAGCAAAAGGGCTTAATTTAGTTA |
| Core_Oligo88 | AAAGGCTACTTTTACATCGGAAAACAACTTACCAGAATGGTTT |
| Core_Oligo89 | TTAAATGCGATAAATTATGGAGGTGATAATTCAGCT |
| Core_Oligo90 | AATTACCTTTGAATACAATTTAGGAAAC |
| Core_Oligo91 | AACATCAAGGAGAAACTAAGAGAATAATATCC |
| Core_Oligo92 | AAGATTCAACCATCAATTAATTGCGGATTAGA |
| Core_Oligo93 | AGGATAAATTGAGAGAGTAATTCTATCAACAA |
| Core_Oligo94 | AAGCTAAAACTAATAGAGCTGAA |
| Core_Oligo95 | AGAAGCCTTAATTACATTTTGAAAACATTGGGTTATATAACTA |
| Core_Oligo96 | ACCCTGTATGAGAAGATGAGAGACTACCTTTT |
| Core_Oligo97 | AGAATCCTAACAATTTCTAAATTTTATAAAGC |
| Core_Oligo98 | TAAGACGCATACTTTTGATAAATAAGTATCAT |
| Core_Oligo99 | CCAATAAAAAGCAATAATTTAGTTAAAGTACG |
| Core_Oligo100 | GTAAATCGGAAACAGTAAATATATTTGA |
| Core_Oligo101 | AGTGAATTTTCAATTCTTCGGTTGTGGAGTTGATAACAATCATAA |
| Core_Oligo102 | AAGGTGGCAATCAAAATTAAACACCACCAAAAACATATATT |
| Core_Oligo103 | TATGTAAACTTCTGACCATTTGAAATGAAACA |
| Core_Oligo104 | TTCATTTCGAGTAGAAGCCTCGTAGGTA |
| Core_Oligo105 | TAACCTCCGACCGTGTGCGGG |
| Core_Oligo106 | GAAATACCGGCTTAGGTAGCGATAGCTTAGAT |
| Core_Oligo107 | AATAAGAACATAGGTCGTCAAT |
| Core_Oligo108 | TTACTAAACAACAATGCCGGAATCGATG |
| Core_Oligo109 | CAACGCTCGCCAGTAAAATAACGGCGTCAGAT |
| Core_Oligo110 | ATGCGTTAAAGGTAAATCTAC |
| Core_Oligo111 | GTGTCTGTTAGAGCTATGATATCAATCA |
| Core_Oligo112 | ACCGACAATACAAATTAATTTATTTCAACGCA |
| Core_Oligo113 | TTTTCGAAACAGTAGAAGATGTTACCTT |
| Core_Oligo114 | GACGACAATAGAAAAAGAACCCTCATTATG |
| Core_Oligo115 | CGGATGGCGAAGTTTCAGTAATGTAGAGCATA |
| Core_Oligo116 | GAGTACCCGCGTTTTATAAGCCATTAAA |
| Core_Oligo117 | AATGCCAATAGCTATTTTGTCGTCTGGC |
| Core_Oligo118 | CATCCTAAAGTACCGCCTTCTGAACAGATGAT |
| Core_Oligo119 | TAGATAAGTTTTATTTTAAAT |
| Core_Oligo120 | TACCGCGCCAGAACGCACAAGAGAGAGGGT |
| Core_Oligo121 | CAAGCCGTTCCTGAACTACTCAGGTCATTGCC |
| Core_Oligo122 | TCAAATATTTTAATTGCTAGCATGTTCAACCG |
| Core_Oligo123 | TAAACCATTTACGAGGTTTAAATTCGCC |
| Core_Oligo124 | GATACCAATCCATAGATGGGCCAGCTTT |
| Core_Oligo125 | GAACGCGACAGTTACAAGTTT |
| Core_Oligo126 | TAATTTGCGGCGTTTTAATTTTTTAACCAATA |
| Core_Oligo127 | TATTTATCTAGAAGGCTAATTCGTAAAATT |
| Core_Oligo128 | ACAGTTCAGCATCAAATTCATCAAAAATATTT |
| Core_Oligo129 | TACCAACGTTTTGATGATTATTAATGGA |
| Core_Oligo130 | GATTTATCTTACGCGAAAGG |
| Core_Oligo131 | ATCGTCAAAAGAAGGGTGCGGGGGAACG |
| Core_Oligo132 | ATAGCAGCATAAGAGCGTTGG |
| Core_Oligo133 | ACAGGGAAGATAACCCAGGAAGGTCAATCAAT |
| Core_Oligo134 | GCCCAATACTTTACAGGTAGACGACAGTATCG |
| Core_Oligo135 | GCGAGAGGCTTTTGCATAAATATTTGCCGGAAGTAATGGG |
| Core_Oligo136 | TCAGAGAGCGCATTGTATTAGTTTTAAA |
| Core_Oligo137 | GCAATAGCTTTTGTTTACTCCAGCGCATCG |
| Core_Oligo138 | AAAGTTACAGACACCACGGAATAATTGACGGA |
| Core_Oligo139 | CTCGTTTACTAATGCAGTCCCAAAAATC |
| Core_Oligo140 | TAATAACGCATAAAGGATTAACACCGAACGAA |
| Core_Oligo141 | AATTATTCTAGCGACATCCGG |
| Core_Oligo142 | CCGACTTGAATGAAACCGTAAGAAAAGGGACA |
| Core_Oligo143 | ACATTCAAAACGAAAAACGATGCAAGAA |
| Core_Oligo144 | TGAGGACTGTAGGGTCCGTTTCAGCGTG |
| Core_Oligo145 | GTAATCAGATTAAAGGAAACAGCGGATCAAAC |
| Core_Oligo146 | CGTCACCAGCCATTAAAATACCGCCTGC |
| Core_Oligo147 | CGTTAATAAAAGACAACGACATAAGGAATTTG |
| Core_Oligo148 | TCGGGACGATTGTGCAGGCGTAAAGGTT |
| Core_Oligo149 | CCGGAACCAGAGCCGCAACGCTCATGGTAATA |
| Core_Oligo150 | TGGCTCATTGGGCTGGCTGGATACACTG |
| Core_Oligo151 | TAGCGTTTCGCCAGCAGCGGT |
| Core_Oligo152 | GCAGGTCACATTTTCGTAGAACGTTTTCGT |
| Core_Oligo153 | CACCCTCAGAGCCAGTAATAATACGTGG |
| Core_Oligo154 | GTAGTAAATTATACCACAGCAACCGCTGATTG |
| Core_Oligo155 | AGAGCCGCGCCATCTTTAGCATCCTCATAACG |
| Core_Oligo156 | CAGAATGGGCCTATTTCTCTG |
| Core_Oligo157 | TCACAAAGTATTATCCAGCGCGGGGGTT |
| Core_Oligo158 | CTTGAGTAGCGTCAGCCTTGCTGGAAAT |
| Core_Oligo159 | GAAACATGAAACAAATTAATGGGCTTTCGC |
| Core_Oligo160 | TGCCCCCTAAAGCGCACGCGAGCCGGGTCACT |
| Core_Oligo161 | TGACCTTCATAAGGCTACATCCCTGGTGTCCA |
| Core_Oligo162 | CGGGGTTTAAGTACAATACGA |
| Core_Oligo163 | CCGTCGAGACCGCCACGTATAACGTAATGCGC |
| Core_Oligo164 | GACTAAATTGTGAATTGTTACTAACTCA |
| Core_Oligo165 | AGGACAGTTACTTAGTAATCACCGCTTT |
| Core_Oligo166 | CTCAGAGCTGCTCAGTGGCCGGCCAGAATGCG |
| Core_Oligo167 | CCTCAGAAGGGTTGGAATCCTGCAAATT |
| Core_Oligo168 | GCTCCATGATGAACGGCCTGTTCTTGCCGGTG |
| Core_Oligo169 | CGCCTGATCCTCAAGAGTCATACCAGTGTC |
| Core_Oligo170 | AACGAAATTTCATGCGCGGGGAAATCCT |
| Core_Oligo171 | CAAGCCCACAGCCCTCAGAAAGGAGAGCCCCCGATTTAGAAAACCGTCAATAATAATTTTTTCA |
| Core_Oligo172 | GCGAAACACAGCGAAAGGCAA |
| Core_Oligo173 | CTCATCTTACGGCTACCAGGGTGGAGCGGTCC |
| Core_Oligo174 | GGGTAGCATGACCCCCGAGTGAGTCCGCTC |
| Core_Oligo175 | TCTAAAGTCACCACCCCTACATAAAGTGTAAA |
| Core_Oligo176 | ACCCTCAGATCTCCAATGGACT |
| Core_Oligo177 | AGGGAGTTTAATTGTAGTTTGGAACAAGAGT |
| Core_Oligo178 | AAATGAATGAATTGCGTATCAGGGCGATGGC |
| Core_Oligo179 | ATAACCGGTGAATTATAGCCCGAGATAGG |
| Core_Oligo180 | CGTTGAAATTTGTCGTGGGTGCCCTTCACCGC |
| Core_Oligo181 | CTTTCGAGATATATTCGCAGGCGAAGAGGCGG |
| Core_Oligo182 | GGAGCCTTAAAGGCCGTGCAGCATTTTTCT |
| Core_Oligo183 | ACTAAAGTTTCTGTCTAAAGGAGGGAAG |
| Core_Oligo184 | CTCGTCGCCTGCTCATGGTAAATAGTTTATTT |
| Core_Oligo185 | TGAGAGATGAGGTGGAACGATAAA |
| Core_Oligo186 | AGACTTTCAAATTCATCATTCAACCGAT |
| Core_Oligo187 | CAGTGCCAAGCTTTCA |
| Core_Oligo188 | AACCAAGCCCTTTTTA |
| Core_Oligo189 | AGAAAAGTCGACGGCGGGATGT |
| Core_Oligo190 | TGTCACAATCAATAGA |
| Core_Oligo191 | TAGCGTCAGGAGGGAATTGCCGCGAAGGGA |
| Core_Oligo192 | GATTTAGAAAACCGTCAATAATAATTTTTTCA |
| Core_Oligo193 | CAAGCCCACAGCCCTCAGAAAGGAGAGCCCCC |
| Core_Oligo194 | AGACGCAGAAACAG A |
| Core_Oligo195 | GGTGAGGAGTTGGCAAACCGAGGTGAGTTAA |
| Core_Oligo196 | AAACGCAACAGAAGGA |
| Ends_Oligo1 | TACACTCAAAATCCCTTATAAAGTGGTTCCCAAC |
| Ends_Oligo2 | TACACTGGGGTCGAGGTGCGCTGGCAAGTACACT |
| Ends_Oligo3 | GTTTGATGTCAAAAGATCTTAAACAGCTTGATACCGATAGTTGCGTACACT |
| Ends_Oligo4 | TACACTGCTGCATTAATGAATCGAAACCTGGAAG |
| Ends_Oligo5 | TACACTTGTAGCGGTCAAAACAGGAGGCTACACT |
| Ends_Oligo6 | TACACTTCCCCGGGTACCGAGCTCCTCACATGAC |
| Ends_Oligo7 | TACACTCGATTAAAGGGATTAGTAATAATACACT |
| Ends_Oligo8 | TACACTCATCAGATGCCGGGTTAGCAAATCTAAC |
| Ends_Oligo9 | TACACTCATCACTTGCCCTGAAATGGATTACACT |
| Ends_Oligo10 | TACACTTGCCATCCCACGCAACGCAGCACCATGC |
| Ends_Oligo11 | TACACTTATTTACATTGATTAGTCTTTATACACT |
| Ends_Oligo12 | TACACTCCTTTAGTGATGAAGGAAAGCCGCGAGA |
| Ends_Oligo13 | TACACTATGCGCGAACTAGCAAATGAAATACACT |
| Ends_Oligo14 | TACACTCGAAACGTACAGCGCCCACCGGAACAAC |
| Ends_Oligo15 | TACACTAATCTAAAGCATAATAGATTAGTACACT |
| Ends_Oligo16 | TACACTGCGCAACTGTTGGGAAATTCGCCACGTC |
| Ends_Oligo17 | TACACTAGCCGTCAATACGGAACAAAGATACACT |
| Ends_Oligo18 | TACACTATTCTCCGTGGGAACAAGTAACAATATA |
| Ends_Oligo19 | TACACTAACCACCAGAAATTATTTGCACTACACT |
| Ends_Oligo20 | TACACTAAAGCCCCAAAAACAGCCGGTTGAACAG |
| Ends_Oligo21 | TACACTGTAAAACAGAAAAAATCGCGCATACACT |
| Ends_Oligo22 | TACACTCGGAGACAGTCAAATCAAAGGGTGATGC |
| Ends_Oligo23 | TACACTGAGGCGAATTACAATATATGTGTACACT |
| Ends_Oligo24 | TTTTAATGTCGCTATTGCAAATCCAATCGCAAGACAAAGAACGCGTACACT |
| Ends_Oligo25 | TACACTAAGAATTAGCAAAATTTCATACAGCCTGTTTAGCTATATT |
| Ends_Oligo26 | TACACTAGTGAATAACCTTGCTTCT |
| Ends_Oligo27 | TACACTAAATGGTCAATAAGCAAGGCATACACT |
| Ends_Oligo28 | TACACTAGAAAACTTTTTCACATAAATTTCATTTC |
| Ends_Oligo29 | GAATCGCCATATTTAACTACACT |
| Ends_Oligo30 | TACACTACATGTTTTAAATAGAAAGGCTACACT |
| Ends_Oligo31 | AACTTGACCATTAGATACATTTCGCTACACT |
| Ends_Oligo32 | TACACTAACGCCAACATGTCAAGTTACATAAAGAA |
| Ends_Oligo33 | GTCATGAATATAATGCTGTAGCTCATACACT |
| Ends_Oligo34 | TACACTGAAGCAAACTCCATAATCAGATACACT |
| Ends_Oligo35 | CAATCAATAATCGGCTGTACACT |
| Ends_Oligo36 | TACACTTCTTTCCTTATCAATATCAAAGGAGCGGA |
| Ends_Oligo37 | TACACTTTACCCTGACTATCCCGTCGGTACACT |
| Ends_Oligo38 | AGATTAGTTGCTATTTTTACACT |
| Ends_Oligo39 | GTCAGCTTCAAAGCGAACCAGACCGTACACT |
| Ends_Oligo40 | TACACTGCACCCAGCTACATCATTTTGGATAATAC |
| Ends_Oligo41 | TACACTTTAGACTGGATAGTTCAGGCTTACACT |
| Ends_Oligo42 | TAACTGAACACCCTGAATACACT |
| Ends_Oligo43 | CAATCATAAATCAAAAATCAGGTCTTACACT |
| Ends_Oligo44 | TACACTCAAAGTCAGAGGGCTAACAACTCACCTTG |
| Ends_Oligo45 | CATGATTAAGACTCCTTTACACT |
| Ends_Oligo46 | TACACTGGCATAGTAAGAGACAATCGGTACACT |
| Ends_Oligo47 | ACTAGAGGGGGTAATAGTAAAATGTTACACT |
| Ends_Oligo48 | TACACTATTACGCAGTATGGAGCCAGCGATAGCCC |
| Ends_Oligo49 | TACACTAGATTCATCAGTTACAGGCGGTACACT |
| Ends_Oligo50 | TTTATAACGCCAAAAGGAATTACGATACACT |
| Ends_Oligo51 | CCAGCAAAATCACCAGTTACACT |
| Ends_Oligo52 | TACACTAGCACCATTACCAGAATGGCTGCAGATTC |
| Ends_Oligo53 | GATTACAACATTATTACAGGTAGAATACACT |
| Ends_Oligo54 | TACACTTGTGAATTACCTTGTCGGTGGTACACT |
| Ends_Oligo55 | CCTCCCTCAGAGCCGCCTACACT |
| Ends_Oligo56 | TACACTACCCTCAGAACCGTCAATCGTTGAGTAGA |
| Ends_Oligo57 | TACACTACCCAAATCAACGGTTAACGGTACACT |
| Ends_Oligo58 | AAAGTTTAATTTCAACTTTAATCATTACACT |
| Ends_Oligo59 | GATGATACAGGAGTGTATACACT |
| Ends_Oligo60 | TACACTCTGGTAATAAGTTCTTCTTTGATTTTAGA |
| Ends_Oligo61 | CAACGACAAGAACCGGATATTCATTTACACT |
| Ends_Oligo62 | TACACTAAGGGAACCGAACGTTGAGGATACACT |
| Ends_Oligo63 | CCCGGAATAGGTGTATCTACACT |
| Ends_Oligo64 | TACACTACCGTACTCAGGACGGGAGCTCGCTGCGC |
| Ends_Oligo65 | GCACGAGGCGCAGACGGTCAATCATTACACT |
| Ends_Oligo66 | AACACTGAGTTTCGTCATACACT |
| Ends_Oligo67 | TACACTGTAATGCCACTACTCGTGCCATACACT |
| Ends_Oligo68 | TACACTCCAGTACAAACTAGCGCTAGGGCCGTAAA |
| Ends_Oligo69 | TACACTCCGACAATGACAAGAAATCGGTACACT |
| Ends_Oligo70 | TAAACAACTTTCAACAGTACACT |
| Ends_Oligo71 | CATCTCCATTAAACGGGTAAAATACTACACT |
| Ends_Oligo72 | TACACTTTTCAGCGGAGTGCACCCAAATCAAGTTTTTTTACACT |
| 5'Biotin_Oligo1 | CCCATTTCATTGCTGATAATTAATTTTCCCTT |
| 5'Biotin_Oligo2 | CCCTCTGCGAAGGGGCGCGTAGTAGCATTAACAT |
| 5'Biotin_Oligo3 | CCCGCGGATTGAAAACGATTGACCACCAGGC |
| 5'Biotin_Oligo4 | CCCTGCGGGAGGCTAACGAGTTATTAAACTTTACA |
| 5'Biotin_Oligo5 | CCCATACATAGAATACCAAACCCTTATCTAA |
| 5'Biotin_Oligo6 | CCCCCAGCGCCACCAGACGGCCGCCACGCCTCTTC |
| 5'Biotin_Oligo7 | CCCTTCCAGTAAACAGTGCTCCATCACGAGAAGTG |
| 5'Biotin_Oligo8 | CCCTCAGTGAATCAAGACCAGCGGTCGCGTC |
| 5'Biotin_Oligo9 | CCCAAAGACTTGAGGCAAACTCACTGCTGGTCATA |
| 5'Biotin_Oligo10 | CCCCCACAGAATAGGAACCGCGCTTGCTTTC |

## 2) Rotor-Arm_V1

| Name | Sequence |
|---|---|
| Core_Oligo1 | TACACTCCAGCAGGCGAAAATCTCCA |
| Core_Oligo2 | TACACTATCACCCAAATCAAGTGCCCACTACGTGAACCTACACT |
| Core_Oligo3 | TACACTATGGTCATAGCTGTTTGCTC |
| Core_Oligo4 | TACACTCCCTAAAGGGAGCCCCAAGCACTAAATCGGAATACACT |
| Core_Oligo5 | GCCGGGCGTTGACCGTGGAATCATACCCTGAA |
| Core_Oligo6 | AAACAGCGACAGCGCCCGTTATACATTCAACC |
| Core_Oligo7 | CTGGTCAGGCTTCTGGTCCTAATTTTTAAGAA |
| Core_Oligo8 | CCGGACTTTTCGCCATGATAAGTCTTACCAGA |
| Core_Oligo9 | GCAGCACCAAGATCGCATAATCGGAAATAGCA |
| Core_Oligo10 | AAACGATGGATTAAGTATAAAGTATAGAAAAT |
| Core_Oligo11 | ACAGGCGGTCACGACGCAGAGGCAAAAAGAAA |
| Core_Oligo12 | CCGGCCAGGGCGATCGCTAATGCAAACGGAAT |
| Core_Oligo13 | GTCCGTTTCCAGCTGGTCCAGACGAGACTCCT |
| Core_Oligo14 | TACACTCCGTGGTGAAGGGATAGAGAGATAATTAC |
| Core_Oligo15 | TTTGCCGCATAACCTCACGCTCAAGGTTTACC |
| Core_Oligo16 | GGGGTCATGATGGGCGAGAACAAGTAATATCA |
| Core_Oligo17 | CAGCTTACTTTGAGGGACGGGTATGTTAAGCC |
| Core_Oligo18 | CGACATAAAGCTTTCACATATTTATTATTTTG |
| Core_Oligo19 | CAGTGCCAAAAAATCCATCGCAAGACAAAGAA |
| Core_Oligo20 | TTTCCCAGCCTTTAGTATAACTATATGTAAAT |
| Core_Oligo21 | CTATTACGTTTCGTCTTGAATTTATCAAAATC |
| Core_Oligo22 | GGGAACGGCAGCAGTTCAAATATATTTTAGTT |
| Core_Oligo23 | ACGGCGGACGGTTGCGACCTGAGCAAAAGAAG |
| Core_Oligo24 | GAAACGTGATCAAACTAAATTTAATGGTT |
| Core_Oligo25 | AAGCGCCAGTAGAACGTTCCCTTAGAATCCTT |
| Core_Oligo26 | TTGGGAAGAGCACATCCTTAGATTAAGACGCT |
| Core_Oligo27 | GAATTTGTGCTCTCACTGATAAAT |
| Core_Oligo28 | TGCAAGGCCTGATTGCACCTTTTTAACCTCCG |
| Core_Oligo29 | GCCTCAGGGTCGGTGGTACATAAATCAATATA |
| Core_Oligo30 | TCTGCCAGGGCTGGAGTTCATTTGAATTACCT |
| Core_Oligo31 | TTGGTGTATGCAGGCGCAAGAAAACAAAATTA |
| Core_Oligo32 | TACACTCGAGAAAGGAAGGGAAGCCGGCGAACGTGGTACACT |
| Core_Oligo33 | CGGCACCCAGCAACTGCTTCTGTAAATCGTCGCTATTATATCCCA |
| Core_Oligo34 | ATAAAGCTTCGAGCTTCTTATTAGCGTTTGCC |
| Core_Oligo35 | GATATTCAGTCCAATAGCCTTGATATTCACAA |
| Core_Oligo36 | AGGCCGGAATTGAATCTGACAGGAGGTTGAGG |
| Core_Oligo37 | TTATTTCAAAAGCGGAGCCTCCCTCAGAGCCG |
| Core_Oligo38 | GGCATCAACGGATGGCTCGATAGCAGCACCGT |
| Core_Oligo39 | ATGACCCTGCCCGAAAAAATCACCGGAACCAG |
| Core_Oligo40 | TACACTAATTCTGCGAACGAGTATAACAGTCGTCACCGACTTGAGC |
| Core_Oligo41 | ATTTCGCACTAAAGTACCAGCAAAATCACCA |
| Core_Oligo42 | ATTAGCAAACTCCAACGCGTTTTCATCGGCAT |
| Core_Oligo43 | CCGGAGAGGGGGTAATTAAAGCCAGAATGGAA |
| Core_Oligo44 | ACATCCAAAATTGCTCAATCAAGTTTGCCTTT |
| Core_Oligo45 | AATGTGTAAGAAAACGACCAGAACCACCACCA |
| Core_Oligo46 | ATATTTTCTGCTGTAGTAGCAAGGCCGGAAAC |
| Core_Oligo47 | TATGCAAAATGGTCGGGCGACAAATTCT |
| Core_Oligo48 | AAAAATCTATTTTAAGCCCTTTACGAGC |
| Core_Oligo49 | CATTCCATAGATTTAGGTAAATATAAAAAGCC |
| Core_Oligo50 | GGAAGCAAAATTAAGCAGCAAACGCCGACAAA |
| Core_Oligo51 | CATTTTTGTTCTACTAGAATAAGTACAACGCC |
| Core_Oligo52 | TCAGAAGCACGCAAGGGAACAAAGCTGAACAA |
| Core_Oligo53 | AAATATTCGACAGTCAAGAATTGATAAACCAA |
| Core_Oligo54 | TACACTTGTAGCGGTCACGCTTAGGGCGCTGGCAAGTACACT |
| Core_Oligo55 | AAGAGGAAGTAATACTGCAATAATGAACGCGC |
| Core_Oligo56 | GTACCTTTTAAATCATCAACATATTTTTCGAG |
| Core_Oligo57 | AACAGTTCGGTAAAGAAAACAATGCTGTCTTT |
| Core_Oligo58 | GAATATAAATTTGGGGATTCATATCAGTAGGG |
| Core_Oligo59 | TGCCAGAGGGTAGCTAAACTGAACTACCGCGC |
| Core_Oligo60 | GTTTTAATAAATCGGTCATGATTAACGACAAT |
| Core_Oligo61 | TGGATAGCACCGTTCTTTGAGCGCCAAGCCGT |
| Core_Oligo62 | TACACTAACACCAGAACGAGTAGCTT |
| Core_Oligo63 | CATTATTACCACACCCGCCTGGT |
| Core_Oligo64 | TACACTTACAGGGCGCGTACTAGCGCTTAATGCGCCGCTACACT |
| Core_Oligo65 | CGTAAGCTAAACAGGAGTCGTCACCCTC |
| Core_Oligo66 | TACACTATACACTAAAACACTCCGAA |
| Core_Oligo67 | TACACTTTAGAATCAGAGCGGGTAACGTGCTTTCCTCGTACACT |
| Core_Oligo68 | GTCGCTGATGAATTTTTACTTAGCGTACAGAC |
| Core_Oligo69 | GCTTGCTTTTTTTCACCCCCCAGCTCATTCAG |
| Core_Oligo70 | AACCATCGCTTTCAACATTGTGTCTCATCAAG |
| Core_Oligo71 | TGATACCGAGGAACATACAACGGTTCATTA |
| Core_Oligo72 | TACACTAAGGAGCCTTTAAGCAAAAGATACACT |
| Core_Oligo73 | CTTTTGCGAAGTTTTGTCATAAGGGAACCACGAGCA |
| Core_Oligo74 | GAATAGAAATAGTTGGTTTCCATTAAACGG |
| Core_Oligo75 | TTAGTAAAGGCTTGCACATCGGAACGAGGGTA |
| Core_Oligo76 | TACACTGAACGGTACGCAGGCCACCGAGTACACT |
| Core_Oligo77 | TGAGAAGTGGAGCCGATTAAAGGGATTTTAGACAGTACACT |
| Core_Oligo78 | ATAATAATTCGAGGTGCACTACGAAGGCACCA |
| Core_Oligo79 | CTAAACAACCCACGCAGCTTTGAGGACTAAAG |
| Core_Oligo80 | AAAAAGGCTCCAATACACT |
| Core_Oligo81 | TCGTCACCGTACCAGGGGCTCATTAAAAGGAA |
| Core_Oligo82 | TACACTCGCCACCCTCAGATGACGAGATACACT |
| Core_Oligo83 | AATAGGAAGATATAATTGTGAATCTATCAT |
| Core_Oligo84 | ACGATCTATGAAAGTAAACAA |
| Core_Oligo85 | CATTCCACAGAAGGATAAAAATCTCCACATTC |
| Core_Oligo86 | CCACCACCCCGTACTCGGCTTGAGAAAAACCAAAATAGCGATCAGGTC |
| Core_Oligo87 | AGAGGGTTCCCATGTACCGGATAAGATTTG |
| Core_Oligo88 | TCCTCAAGAGACAGCCTGAACGGTCGGAACGA |
| Core_Oligo89 | CCACCCTCAGAACTACACT |
| Core_Oligo90 | TACACTTAAAAGAGTCTACTTCTTTGATTACACT |
| Core_Oligo91 | TTTGCTCAAGTACAAAGCTGACCTGAAATCCG |
| Core_Oligo92 | CGCAAATTGTTTTTATTTGGAACTGACCAACTT |
| Core_Oligo93 | GTGTATCACTCATTTTAAAGCTGCGATTATAC |
| Core_Oligo94 | AGCGTCAGAAAGGTGGACAGGCAAGCCAGGGT |
| Core_Oligo95 | CTGAAACAGAGCGTCTATATTTAAATTGTGGAGCGG |
| Core_Oligo96 | TACACTGGTGAATTATCACTGATTCCCTACACT |
| Core_Oligo97 | GTACTGGTAAAAATGCGATGAACTCCTGTA |
| Core_Oligo98 | TTTCGGTCAGTATGTTAATAAAGCGGATGTGC |
| Core_Oligo99 | GCGTCATAATAACATAATTGCCTGTGAGCGAG |
| Core_Oligo100 | ACAAATAAGAGGGTAAAGCTGATAAGGTCACG |
| Core_Oligo101 | AGCCACCAGAGGAAACTTTGCGGGCGCAACTG |
| Core_Oligo102 | ATGCCCCCACAAAATAAGCCCCAATTGTTAAA |
| Core_Oligo103 | GCCACCACCCTCACTTACCGAAATGCAATCAGCTTTC |
| Core_Oligo104 | CATTTGGGGAGGGAAGTTTGACCACAGTCCCG |
| Core_Oligo105 | CCTTGAGTTCCAAATAAATCATATGAACGCCA |
| Core_Oligo106 | ATCTTTTGAACTGGTGTACCACTCTTCG |
| Core_Oligo107 | GTCACCAAAATAGAAACGCGAGCTGCCGCCAC |
| Core_Oligo108 | GAGCCGCCAGAGCAAGTTCAAAAGCAGTATCG |
| Core_Oligo109 | CAGGTCAGAACCCACAAATCACCAACCGTGCA |
| Core_Oligo110 | AGCGCAGTGGAGAATTTTTTTGAGGGGAACAA |
| Core_Oligo111 | CCACCCTCAGATAGCCATAAAAATACCAGGCA |
| Core_Oligo112 | AATCAGTAACACCACGATAGTAGTACGACGGC |
| Core_Oligo113 | GTAGCACCAGACAAAAAATAACCTCAATCGGC |
| Core_Oligo114 | TGCCAGTTTGCCTATTTAGGAATAACGTTAAT |
| Core_Oligo115 | AAGTAAGCAGAACCGCACCCTGACTATTATAG |
| Core_Oligo116 | CAATAATAGCCAGCATCCCCTCAAATGCTTTA |
| Core_Oligo117 | CCTGAGTAAACCGTTGCCACAGGTAGAAAGAT |
| Core_Oligo118 | CAAAGTCAATCCTCATAGTAAAATGTTTAGAC |
| Core_Oligo119 | AGCGCCAAATTACCATCTCAACATGTTTTAAA |
| Core_Oligo120 | ATAGCTATGAGCCGCCAGAATGACCATAAATC |
| Core_Oligo121 | TTAACGTCAATAAGTGAGCAACATACCTTA |
| Core_Oligo122 | AGGAAACCCCGGAACCTTGCATCAAAAAGATT |
| Core_Oligo123 | TCACAATCTGAAACCATTAGAGCTTAATTGCT |
| Core_Oligo124 | CGCAAAGGCGACAGCTTTTGATAAGAGGT |
| Core_Oligo125 | ACCCAAAACATAATCAGACTTCAAATATCGC |
| Core_Oligo126 | TATTACGCATAGCCCCCAAAGCGAACCAGACC |
| Core_Oligo127 | GATTGAGGAATTAGAGCGGTGTCTGGAAGTTT |
| Core_Oligo128 | TACACTTAGTAATAACATCGGCCTTGCTTACACT |
| Core_Oligo129 | CAGAGAGACATGGCTTACGACGATATGGTTTA |
| Core_Oligo130 | TTAGACGCTCTGAATTGCAAAAGAAGTTT |
| Core_Oligo131 | TATCCCAAAACAGTGCATAACGCCATACCAGT |
| Core_Oligo132 | ACATACATACTGTAGCAGGTCAGGATTAGAGA |
| Core_Oligo133 | GAGAGATACGATTGCTGCGGAATCGTCAT |
| Core_Oligo134 | ACGCTAACCACGTAAACGATCCAGCGCAGTTGACGG |
| Core_Oligo135 | AGGTAAAGGAGAGACTCGTTCCGGCAAACGCG |
| Core_Oligo136 | CCAATAGCTTTCAATTGTATGAGCCGGGTCA |
| Core_Oligo137 | TAGTGACGGAAATTATTCATTAAATACACT |
| Core_Oligo138 | AACATGTACAAATCCACGTAAAAAAAGCCGC |
| Core_Oligo139 | TACCAGTACTTCTGACCTTAAATTTCTGCTCA |
| Core_Oligo140 | AAACAACTCAATAGCGTCGCTGGCAGCCT |
| Core_Oligo141 | GAAAAATAAATTAATTTCAGCGTGGTGCTGGT |
| Core_Oligo142 | TTTTATTTACAAACATCTTTCGCACTCAATCC |
| Core_Oligo143 | GTACCGCATTAACAATGTGTCCAGCATCAGC |
| Core_Oligo144 | TACACTCACCGGAATCATAGACTTTCTTACACT |
| Core_Oligo145 | GCGAACCTAACGGATAACATCCCCTTCGCG |
| Core_Oligo146 | ATGTAGAAAATAACCTCGCAAGAATGCCAACG |
| Core_Oligo147 | CTGTTTATAGCGATAGCTCATAACGGAACGTG |
| Core_Oligo148 | CCCAGCTAATTTTCAGAGCCAGCGCCAGAATG |
| Core_Oligo149 | CTTAATTGAACTTTTTGGGCGGTTGTGTACAT |
| Core_Oligo150 | TGTTTAGTCGACCGTGGGAAAAAGAGACGCAG |
| Core_Oligo151 | CCAGTAATTGGGTTATGATGAAGGGTAAAGTT |
| Core_Oligo152 | CCTTATCAGGAAACAGTGCCATCCCACGCAAC |
| Core_Oligo153 | GCCTTAAAGTAACAGTCGTTAACGACCGGGGG |
| Core_Oligo154 | TATCCGGTGTTACAAAGGTAATGGGATCCCCG |
| Core_Oligo155 | ATTACATCTCATCGCATCGTATCAATAT |
| Core_Oligo156 | AATTTCATTAAAGCCAACCGGAAAGTTTAGCT |
| Core_Oligo157 | GCTTAGGTAAGAGAATTGGGTAACGGCAAAGA |
| Core_Oligo158 | TTATTCAAAGCAAATCTCCGTAGATCTA |
| Core_Oligo159 | ATAGGTCTTAATTCTGCGAAAGGGCTCAGAGC |
| Core_Oligo160 | AATATACATCAAGATTACCAATAGGTACCCCG |
| Core_Oligo161 | AATACCAAATTCTAAGATTAAATGAGAGTCTG |
| Core_Oligo162 | TGAAATACATCATATGATGTTTACTTAGATAC |
| Core_Oligo163 | TTTTTAATTTCCAAGAGACGACGAGGTGAGAA |
| Core_Oligo164 | CGCGAGAAAGAATCGCGAGGTGGAGAAAAGGT |
| Core_Oligo165 | GAGAAGAGATGTTCAGGTGCGGGCAAAACATT |
| Core_Oligo166 | TACACTGGTAATATCCAAAAAACGCTCATACACT |
| Core_Oligo167 | GAAACAATCCCGACTTCTGGCCTGGTAATC |
| Core_Oligo168 | TGTGAGTGACCAATCAACTCCAGCGCCTGAGT |
| Core_Oligo169 | ATGATGAATCATCGTAAATGGGATAATTAATG |
| Core_Oligo170 | ATTACCGCGAAGAACTAAAAAACGTTAATATT |
| Core_Oligo171 | GCTGATGATTTAGGTTGTAAAAGCATTA |
| Core_Oligo172 | TTGCGTAGCAATTTTATTAAATTTAAACAGGA |
| Core_Oligo173 | GAAAACATCAACAATATCAGGCTGAGAAGCCT |
| Core_Oligo174 | AAGGCGTTAAATAAGAATAAATACACT |
| Core_Oligo175 | TAATCCTGGATAATACGGGTGCCTGGTTTTTC |
| Core_Oligo176 | TACACTTAACATTATCATTTCGTAATCTACACT |
| Core_Oligo177 | CCACCAGATCCTTTGCAAATTGTTGAGTTGCA |
| Core_Oligo178 | ATTATTTGGAGGAAGGCGGGAAACCTGTCGAGGTGC |
| Core_Oligo179 | TAATGGAAACTAACAAATTGCGTTGAGAGGCG |
| Core_Oligo180 | CTGATTATCTTTACATACGAGCCGCTGATT |
| Core_Oligo181 | TACACTTGGAAATACCTATTTACATTGGTACACT |
| Core_Oligo182 | GTATTAGACAGATGATGCCTGTTTTACACT |
| Core_Oligo183 | ACGCTCAACAGCCATTAGCTGTCACTGCGCGC |
| Core_Oligo184 | GTATTAAAAGGAGCGGCAGTTGAGGTAAAGGT |
| Core_Oligo185 | CGTCAATAATTGTTTGACGGTCATGCATCAGA |
| Core_Oligo186 | TTAGGAGCGGGTTAGATGCTGCGGGTGCCGGT |
| Core_Oligo187 | TTAAAAGTTTGAGTACACT |
| Core_Oligo188 | TGCTGAACGGCACAGATGTTGTTCCAGTTTGG |
| Core_Oligo189 | ACACCGCCTCGCCATTTGGTGGTTCCGAAATC |
| Core_Oligo190 | AGCAGCAACTTTAATAATCAAAAGAATAGC |
| Core_Oligo191 | TACACTATAAAACAGAGGTGGTTTGCCTACACT |
| Core_Oligo192 | AAGGAATTTAAAAGGGGAAAAACCGTCTATCA |
| Core_Oligo193 | ATATCTGGACCCTTCTTTAAAGAACGTGGACTC |
| Core_Oligo194 | CTAAAACATGCAACAGCCCTGAGAATCCGCTC |
| Core_Oligo195 | CTATTAGTATGAAAAGGGCAACAGGAAGCA |
| Core_Oligo196 | CCACCAGCAGAAGTACACT |
| Core_Oligo197 | GAGATAGATCAGTTGGCGCGCGGGGCGCTCAC |
| Core_Oligo198 | ACCAGTAATCGTCTGAGTCGTGCCAGCTGCAT |
| Core_Oligo199 | GAATACGTCTCAAATAGCCAGGGTAATGAGTG |
| Core_Oligo200 | GGGCGATGTTTTTGGGAATGGATTACATTTTGAAAA |
| Core_Oligo201 | GCAAGCGGCTGTTTGAAAAAATACCGAACGAAAAAA |
| Core_Oligo202 | CGCTGAGGCGGTTATTAATTAAAA |
| Core_Oligo203 | CGTACGATTTAGGCAACAGGGAACAATAAAA |
| Core_Oligo204 | GGTACCGACCTGTGTGCCGAACGTCAGTATTAAAAA |
| Core_Oligo205 | GAATTTGCGGAAGAGGCGAAAAAA |
| Core_Oligo206 | CTGTTGCCCGTCGGATTCAGATATGAAGCGCAAAAA |
| Core_Oligo207 | GGAAAGAAAGCGCAAACTATCACTTGAAAA |
| Core_Oligo208 | TAACAACCCTGCGGCTATCGCGCACAAAGAAAAAAA |
| Core_Oligo209 | GCGCGCGCGTAATAGCAATGTCCATCAAAAA |
| Core_Oligo210 | CAAAGGCTAGAGGCTTTTTACCGTTAGTACCGAAAA |
| Core_Oligo211 | TGCTAATCAGTGCAGAATCCAAAA |
| Core_Oligo212 | TGAATAAGGTAAATTGAGGAGGTTTCCAGTAAAAAA |
| Core_Oligo213 | GCCCACCGCCACTCTCCAAAAAAA |
| Core_Oligo214 | ACCTAAAAATCTTTGAGTTGAAAACCTCAGAGAAAA |
| Core_Oligo215 | AGAGTTGTATCGGTTTATCAAAAA |
| Core_Oligo216 | AAAACAGGAGAAGGCTAAAA |
| Core_Oligo217 | TACACTCAGATTCACCAGTCACACGAAAA |
| Cy3_Oligo1 | CAACGTCTCGGCCAACAAATCAAAAATATCTAAAAAAAAAAAAAAA |
| Cy3_Oligo2 | AACAAGAATTGGGCTCAAACCTTAGAGCAAAAAAAAAAAAAAA |
| Cy3_Oligo3 | CCGAGATAAGTGAGACATCTAAAGATTTAGAAAAAAAAAAAAAAAAA |
| Cy3_Oligo4 | GGCAAAATCCGCCTGGTGCCACGCGACAACTCAAAAAAAAAAAAAAA |
| Cy3_Oligo5 | TAATGAAAAAGGGCACATTCTGGCCAACAAAAAAAAAAAAAAAA |
| Cy3_Oligo6 | GTTTGCGTGTCCACTAGACCTGAAAGCGTAAAAAAAAAAAAAAAAA |
| Cy3_Oligo7 | GCCCTTCACCCTTATAGCGCGAACTGATAGCCAAAAAAAAAAAAAAA |
| Cy3_Oligo8 | TTTTCACCGGGTTGAGCAATATTTTTGAATGGAAAAAAAAAAAAAAA |
| Cy3_Oligo9 | ACAATTCCCCTCCTCAAATTATCATTGCTTTGAAAAAAAAAAAAAAA |
| Cy3_Oligo10 | TGCCCGCTCTCTGTGGACCTACCATAAAGAAAAAAAAAAAAAAAAAA |
| Cy3_Oligo11 | TAAAGTGTAGCACGCGTGGCAATTCATCGGGAAAAAAAAAAAAAAAA |
| Cy3_Oligo12 | AGCTAACCGTTTTCGATTATATCAGATGAAAAAAAAAAAAAAA |
| Cy3_Oligo13 | TCCGTGAGACACAACAAACAATTCTGAGAGCCAAAAAAAAAAAAAAA |
| Cy3_Oligo14 | CTGTGCATTCCAGTTTATCTACAGTTGAAAAAAAAAAAAAAAA |
| Cy3_Oligo15 | CGGCGGGCTCACATTACTAATAGACTCAATCAAAAAAAAAAAAAAAA |
| Cy3_Oligo16 | TTTCTGCCAAAGCCTGATTTGAGGCATCACCTAAAAAAAAAAAAAAA |
| Cy3_Oligo17 | GCCCCCTGAATTCGCATCCTGAATGCCTAATTAAAAAAAAAAAAAAA |
| Cy3_Oligo18 | TTCTTTGCTCATCAACAACGCGAGAGCCTTTAAAAAAAAAAAAAAAA |
| Cy3_Oligo19 | GGTGTGTTAATTCGCGTGCGGGAGTTTTTTGTAAAAAAAAAAAAAAA |
| Cy3_Oligo20 | TGCCGGGTTTTTTAAGTTGCTTATTATTAAAAAAAAAAAAAAA |
| Cy3_Oligo21 | TCAAAAATCAGCAAATACCTTTTACATCAATAAAAAAAAAAAAAAAA |
| Cy3_Oligo22 | TTGTTAACATCAGAACAGAAATATCAAAAAAAAAAAAAAAAAA |
| Cy3_Oligo23 | GCCAGCTTTCGTCATATCGCCTGATCATATTCAAAAAAAAAAAAAAA |
| Cy3_Oligo24 | TCAGCTCATTACCTGCGTTTAACGCTTCTGAAAAAAAAAAAAAAAAA |
| Cy3_Oligo25 | GTAAAACTCATAGTAATTTAACGGTGCCGTCGAAAAAAAAAAAAAAA |
| Cy3_Oligo26 | GTTGATACAGATACCCGTATAGCGGGGTAAAAAAAAAAAAAAA |
| Cy3_Oligo27 | GAGCAAACTTTACCAGTTGATGATCGGAATAGAAAAAAAAAAAAAAA |
| Cy3_Oligo28 | AGATTGTATTGAGATTTCGGAACCGCTGAGACAAAAAAAAAAAAAAA |
| Cy3_Oligo29 | TTACGAGGAGCATGTCAGAAACGAGTTTTGAAAAAAAAAAAAAAAAA |
| Cy3_Oligo30 | TCATCAGTAAGCAATTCCAGACTTACCAAAAAAAAAAAAAAAA |
| Cy3_Oligo31 | AACCCTCGAAGAGAATAAAATAGCGCGTTTTAAAAAAAAAAAAAAAA |
| Cy3_Oligo32 | AACTAATGATCAGAAAAACAGCCAATTTTGCAAAAAAAAAAAAAAAA |
| Cy3_Oligo33 | CAGGACGTAGGCTGCTACAACGATTTTGAAAAAAAAAAAAAAA |
| Cy3_Oligo34 | TGCGATTTTGACAAGAACCGTAACCGGAGTGAAAAAAAAAAAAAAAA |
| Cy3_Oligo35 | ATTTCAACAACGTAACCAGGGATAAATTGCGAAAAAAAAAAAAAAAA |
| Cy3_Oligo36 | AAAACGAAAGGACAGACTCATAGTTCCAGACGAAAAAAAAAAAAAAA |
| Cy3_Oligo37 | TGAAAGCTAACGGTTAAGAGTATTATTAAAAAAAAAAAAAAA |
| Cy3_Oligo38 | CCCAAATCTTTAATCAGTATAGCCACAGGAGTAAAAAAAAAAAAAAA |
| Cy3_Oligo39 | CAGGCGCATTGGGAAGTAGGATTAAACAGTTAAAAAAAAAAAAAAAA |
| Cy3_Oligo40 | AGTAATCTTAAGAACTCGGATAAGGGTCAGTGAAAAAAAAAAAAAAA |
| Cy3_Oligo41 | CGACCTGTACAGAGTAACCGATATATTCGAAAAAAAAAAAAAAA |
| Cy3_Oligo42 | GGCGCAGAAAAGACAGGGGAGTTAAAGGCCGAAAAAAAAAAAAAAA |
| Cy3_Oligo43 | TATCATCGATGAGGAACGCCGACAATGACAACAAAAAAAAAAAAAAA |
| Cy3_Oligo44 | CAAGCGCGCGTAATGCAATTTCTTAAACAGCTAAAAAAAAAAAAAAA |
| Cy3_Oligo45 | GCAACGGCCTCCATGTCTGTATGGGCCTGTAGAAAAAAAAAAAAAAA |
| Cy3_Oligo46 | GTAAAATAAAACAAAGACTAAAGGGCAAGCCCAAAAAAAAAAAAAAA |
| Cy3_Oligo47 | AGCAGCGCGGTCAATCGTCTTTAGCGTAAAAAAAAAAAAAAAA |
| Cy3_Oligo48 | ACTTTTTCCCTGATAAAGTTTCAGACTGAGTTAAAAAAAAAAAAAAA |

### 3) Rotor-Arm- 10HB

| Name | Sequence |
|---|---|
| Core_Oligo1 | CGAGCCGTTAAATGGATTTAGACCCCCATATAAAC |
| Core_Oligo2 | AACAGCGTACTATGGTTGTTTAAAACCTTAGATTA |
| Core_Oligo3 | TACACTTGGTAATGGAACCAATAGGAATACACT |
| Core_Oligo4 | AAAGCCTTTTTTAGACTAATGACACTAAGCCTATT |
| Core_Oligo5 | CTCCTCAAAATTAAAGAACTGGAACGAGCTCTGAA |
| Core_Oligo6 | GCGAAAATTAGATAAAAATCAGCGCCGACAATAGG |
| Core_Oligo7 | GTCATACATTGGCAGATTCCCTAAAGCTCAACACA |
| Core_Oligo8 | CACTGGTTTAAAATTCAACATATCGGCCGGGAAGG |
| Core_Oligo9 | TGCCAGCAAAGCTATAAAAACTTCATGATTTGCTC |
| Core_Oligo10 | TTCCAGTAAAACATCCCTCGTAACGGGTAAGGATT |
| Core_Oligo11 | GCTCGTCTGTTAAACCTTCCTCTCCAGCCATTCAG |
| Core_Oligo12 | CCTGTGTGGGTGAGAACGGAAGATTTGTAATAA |
| Core_Oligo13 | GCCAACAGCAATAGCTTTTGGCTTTGAAAGTGCC |
| Core_Oligo14 | CACGCTGTCAATAAAACGAGAATCGCCCCATTTTC |
| Core_Oligo15 | GGTTTCTAAGGCTAAATTTCACAACTTTAAACAAA |
| Core_Oligo16 | AGAGAGTTTTTCATAATGCTTTATTCGGCACCCTC |
| Core_Oligo17 | GGTTCCGCGAGTAGCTATTATAACAGCTAACACTG |
| Core_Oligo18 | CGGTGCCAAAAGCCACCGTAACACGTTGCAA |
| Core_Oligo19 | GGTTTGCGGCAAAGCCAGAGGAGGGTAGTATAAGT |
| Core_Oligo20 | AATCATGTCAAATTCTACGTAATTGTGATGATAC |
| Core_Oligo21 | GTTACCTATTGTATGTGGGAAATCGTAATGGCGAA |
| Core_Oligo22 | GTGAGCTATTTCAACAAAAGGAAACGAATCTGAAA |
| Core_Oligo23 | GGTGGTTTCCAATATGTTTAGGCGAAAGGTGTATC |
| Core_Oligo24 | GCCCTTCGAAAAGGATATTCAAGGGAGTAACCGCC |
| Core_Oligo25 | TGCGGCCGAATCGAAGAAACACATAGGCCAGGAGG |
| Core_Oligo26 | TCAGCGGGGTCATTGCAGGGTGCTGCACAGGCGGA |
| Core_Oligo27 | CCTTATAGTTGATTCGGATTGGCTTTCGTACAAAC |
| Core_Oligo28 | GTATGAGATCCTCATGATTGCGTGAGAGGAACGGA |
| Core_Oligo29 | GTAATATGCCAGCAATAAATATCAATAATAATCAG |
| Core_Oligo30 | CCTGAGTTTATCCGCTCACAATGTGTAGGTAGAAA |
| Core_Oligo31 | AAATTAAATCAACAGAACGCGTTATTTTTTAGAGC |
| Core_Oligo32 | CCACACCACCACCAAATAACCCCTAATTACATACA |
| Core_Oligo33 | GAATCCTTAGATTACCGGCTTTCCGGTATATTGAC |
| Core_Oligo34 | CGCTATCATTTCGTCGCTCTAACGATATAAAA |
| Core_Oligo35 | ACGTGGCATACTTCGAAAACAGAAAATAGAAACGC |
| Core_Oligo36 | TTAGACATAATACAGAGAGACGGCGTTTGATTGAG |
| Core_Oligo37 | AAAACGCGCGGTCATACTAGAAAAATAATGTAGCG |
| Core_Oligo38 | CTCAAACTAAAGCAGTTTGAATTATCATAAACCAT |
| Core_Oligo39 | AATCCGCGGACTTGAAGGGTATGGTGAATCAGAGG |
| Core_Oligo40 | AGAAAGCATAATCCTAACAATTTTTTGTACCCAAA |
| Core_Oligo41 | TATAACGCTTTGCCAGCGATAATTTTGCTCACAAT |
| Core_Oligo42 | AGGAGGCAGTATTAATTTATCACTTGCGACAAAAG |
| Core_Oligo43 | AAAAGAGAGGAAGGCAAA |
| Core_Oligo44 | TGAAATGAATACCGTCTTACCGAACGCGTTAGCGT |
| Core_Oligo45 | TAATCAGTTTAGGAAACTATAAAATCAGAAAGGTG |
| Core_Oligo46 | GCAACCAAGCACCGGGTTGTGGCTCATTAGTTGGG |
| Core_Oligo47 | CAGTGGCGGGCCGTTTTCGAGAGTCAAATTGGGAA |
| Core_Oligo48 | GTGTCCAAACCGCAAAAATCCGCGGATCTCCCAGT |
| Core_Oligo49 | CTAGGGCAGATGATCTTTTTTCCAATCCTAAGACT |
| Core_Oligo50 | CTTTGCAAATCGTTAACGATTGTAATAACAACGTT |
| Core_Oligo51 | TGACTCACCAGTGAGACGCTAATAGAATACTGGCG |
| Core_Oligo52 | GAATCAGTTCGACAGCTGAGACTTAAATTATGGTT |
| Core_Oligo53 | CCAGTAACTTTAATGAGAATCCTGTCCAGAGCCAC |
| Core_Oligo54 | CTTCTTTCAATATCAATATATTCGAGAAGTAGCAC |
| Core_Oligo55 | CACCGCCCCATTGCGGATCCCCGGGTACTAT |
| Core_Oligo56 | TTGGATTGAGAAAGCAAAATC |
| Core_Oligo57 | CAACGGCGCTTACGGCGGTGC |
| Core_Oligo58 | GATTATCGCTGGCACCAGCAG |
| Core_Oligo59 | CAACTAAGAGAAGTGCCCGCT |
| Core_Oligo60 | GCTGAACCTCAAATTCTTCTGTTAAACCAGGCCGGAAACGTC |
| Core_Oligo61 | ATCAATAAATCTAAACAGATGTTAG |
| Core_Oligo62 | TTTTCGCCAACCCGACAACCCTCAG |
| Core_Oligo63 | TATTAGTTAAAAGGGTGGTGC |
| Core_Oligo64 | CATCAATGAAAGGATGATGGT |
| Core_Oligo65 | AAATATCTGAGGCCATTAATTGCGTTGCACT |
| Core_Oligo66 | ATTTAGACGATTAAATGAATCG |
| Core_Oligo67 | CAAACAAAGCGGGAGAGAGGC |
| Core_Oligo68 | CAAAGAACGCCGCGGGCCCTG |
| Core_Oligo69 | AATAGAGGCATGTAATAAGAACCGCCACCCTCCCC |
| Core_Oligo70 | CATTAAAGATTATTTACCGGG |
| Core_Oligo71 | AGATTATCATAATAGATATTTTCGGTCATAGCTCA |
| Core_Oligo72 | ACGTGCCCGGGCGCTCCCTTA |
| Core_Oligo73 | CGGAATTGCTGCGCAGCGGTC |
| Core_Oligo74 | TTAAATCTGCTTTCGCGCCAG |
| Core_Oligo75 | AACAAATAAACTTACGAGAGTTTGCCTTTAGCTTC |
| Core_Oligo76 | GTGGTGAGCGTCTGAACCTCAGACGATCCAGCTGT |
| Core_Oligo77 | TACACTCTCGTCGCTGGCAGCCCGGTCCGTGTT |
| Core_Oligo78 | AGGTGAGTCATGGACGTGAGC |
| Core_Oligo79 | GGAATTGTCTGTCCCTAATGA |
| Core_Oligo80 | CACCAGCGACGCTCCGCGTGCCTGTTCTGCT |
| Core_Oligo81 | CAATAGAGGAACGGCCTGTCG |
| Core_Oligo82 | CAGACAACAGAGATCACTGCGCGCCTGTAAA |
| Core_Oligo83 | TTTCGTTTGAGTAACATTACAGGGCGCTGATT |
| Core_Oligo84 | AACACATTCAAACCCTCATATATTGAAGCATAGCAATA |
| Core_Oligo85 | GGAAATACCAGCCGTTCTAGCTGATCAGTTGAAACAGGA |
| Core_Oligo86 | TTAAGCTTTCATCGCATTAAATTTTATAAACAGGTTGCG |
| Core_Oligo87 | CCGAACGAACTAAAGGCCGGAGACAGGTCATACTTGCTG |
| Core_Oligo88 | AAATAGGCATACGGGAGAAGCCTTTAACTCACACCGAGT |
| Core_Oligo89 | GCCGGTCTGGTCAGCAGCGCATCAGATGCCGG |
| Core_Oligo90 | ACATTAATCAAAATCACCAGGTCAGGTACAGA |
| Core_Oligo91 | TACACAATCGGCGAAACGTTACACT |
| Core_Oligo92 | CCTCCCGAAAATCAGCGAGAGAAGCCTC |
| Core_Oligo93 | AGAAACACGTAAAAATTCTCCAAGCAAATATTTAAGCA |
| Core_Oligo94 | TCCTCCACGGAATAAGTTTCAGGAGCGTCACC |
| Core_Oligo95 | GCCAGAGCCATTTCGCAAATCAACCGGTTGA |
| Core_Oligo96 | TACAAAAAATATATTTCAGAACCTGTTT |
| Core_Oligo97 | ATTTATCAATGGAATAAATCACATTTCG |
| Core_Oligo98 | GCACTCATTTAGTTCAGTTGACAATGCC |
| Core_Oligo99 | CAAAAATAAATTAAAGCAAAGCCCAATT |
| Core_Oligo100 | AAACGATTTCATTTACCCTGAATTTAGT |
| Core_Oligo101 | AGTTGCTGCTTAGAAGTAAAAAATCATACAGGCAAGTATTGGCTCGTTA |
| Core_Oligo102 | TACACTACAGCGCCATTTTTTCGTTACACT |
| Core_Oligo103 | CTCTCACGGCCTTTGAGCGAGACGTTAATATTTTGGTGTTCACGCACTC |
| Core_Oligo104 | TTGAAGCAGAGTCAAGTTTTGAATTAGC |
| Core_Oligo105 | AGGCTTAAGGTTGGATCATAATATGACCCTGTAATGCTCACTGTTTTTA |
| Core_Oligo106 | ACCAACGATTAATTGTCATAATGGCATCAATTCTAGGC |
| Core_Oligo107 | TAATGCAAGTATAAATGGTTTTCAGGTCATTGCCTACG |
| Core_Oligo108 | AAAAACGACGGCCAGTGCGTGCGGGGGGACGA |
| Core_Oligo109 | GGAATTTCGTTCCGCGTCTGGTCAGCTC |
| Core_Oligo110 | GGAATCAAATCGCAATAACGCCGCAAGG |
| Core_Oligo111 | TCCAGAGTTGCTTCCCCCTCATTGGGGC |
| Core_Oligo112 | TACAATTATCCTTGAGCGTCCTAGTAGC |
| Core_Oligo113 | CTCAGAGAATATAAAGTAATGTAATCATTCAGGCATGTC |
| Core_Oligo114 | TTGCCATTATTCACGAAAGAG |
| Core_Oligo115 | GAAACGCCCCTCAGTAAAGGCCGCTTTTCGGAATCAAT |
| Core_Oligo116 | CGATAGCTACTGGTATCATCGCCTGATATAATA |
| Core_Oligo117 | TGGAGCCAACTGTTTCAGGAA |
| Core_Oligo118 | AATTATCGTATTAAAAGGCAC |
| Core_Oligo119 | GGCAAGTCCTGAACAAGAAAAAGCCGATTTTATGCCGGA |
| Core_Oligo120 | ACCAATGTCCAAGAACGGGTAACCTAAAATTACAGGTAAAGA |
| Core_Oligo121 | CAGCAAAGCCCCCTAACACTCATCTTTGGAA |
| Core_Oligo122 | TAAAGGTGAACCGCTCGCTGA |
| Core_Oligo123 | TAGCGACGCTTTTGTCGAAATCCGCGACGTT |
| Core_Oligo124 | ATCACATGTAGAAACCAAAGGCGTTAGAAAAACACCATC |
| Core_Oligo125 | CAAACGTCCACCCTACGCATA |
| Core_Oligo126 | GGAGGGAGCGGGGTGGAAGTT |
| Core_Oligo127 | CATTACCGTGCCCGGCGATTA |
| Core_Oligo128 | CACGACGCGCCAGCCCGTGCATCTGCCACCGTCGGAAA |
| Core_Oligo129 | GCCTAAGGTAAAGTAATTGCCATATCCTGACGTGAACGG |
| Core_Oligo130 | AGCCGCCCCACCACAATCTTG |
| Core_Oligo131 | GGCGACAGGCGGATGGACTAA |
| Core_Oligo132 | AATAATATCACCAGAGGTGAA |
| Core_Oligo133 | CAATAGAGGAATAGACAGCAT |
| Core_Oligo134 | TAACCTCCATTCGCCAGCTTT |
| Core_Oligo135 | TACCAGCGGGTTGACAACGGC |
| Core_Oligo136 | AGAACTGGTACCGTTGATACC |
| Core_Oligo137 | CGTTTTCGCGCAGTGCGCAGA |
| Core_Oligo138 | TAACGCCTGTGCTGGTGTAGA |
| Core_Oligo139 | CCTTATTGCAAGCCCAATGAC |
| Core_Oligo140 | GGAAATTTCAAGAGAAAATAC |
| Core_Oligo141 | CACCGGAGCATTGATGGCTGA |
| Core_Oligo142 | CAGTATGTTACTTAGCCGGCTCATTTCATAATGTATTAA |
| Core_Oligo143 | AGGATTAAGGTAAATTCTAAGAACGCGATACCTTTACGACGAAATCGGT |
| Core_Oligo144 | AGGGATAACGCAGTCCATATT |
| Core_Oligo145 | AGAGCCAAGAAAATTGCCAGT |
| Core_Oligo146 | TAAATCCCCCCTTACCTGTTTATCAACATGCGTTATCA |
| Core_Oligo147 | TCGGAACTTGGGAACATCGTA |
| Core_Oligo148 | TTAAGGAACCGAACTGACACTTTAATACAAATAACGAAC |
| Core_Oligo149 | AGTACCATTCAACCTAGCGAA |
| Core_Oligo150 | AGTTAATATCACCACAAGCAAGCCGTTTAGAA |
| Core_Oligo151 | ACCCTCAGGCAACAGCGTCTT |
| Core_Oligo152 | CATGAAAACCGTCAGCCCAATAGCAAGCTGT |
| Core_Oligo153 | GTTTTAACGGGGTCAGCAAAGTA |
| Core_Oligo154 | ATAGCCCAAATTCACAAGATT |
| Core_Oligo155 | AACCCATGCATGATAAATAAG |
| Core_Oligo156 | AGGGGGAAGGGTTTAAACTTAAATTTCTTACATCGCGG |
| Core_Oligo157 | AGTTTCGACGGAATTTAACGT |
| Core_Oligo158 | GCTGCGCGCCACGGATAGACTTTCTCCGAAG |
| Core_Oligo159 | TTTACCGGTCAGACTATCCCATCCTAATACC |
| Core_Oligo160 | GGCGATCACCAAGAGCTATGCC |
| Core_Oligo161 | TCAGTTCATTACCTAGGTTGGGACAGTTGGGCTCG |
| Core_Oligo162 | TTGAGGCGGAACCAGACGACGACAATAAAGTAGGGGAG |
| Core_Oligo163 | GTCGAGAGCCAAAGGGAGGTT |
| Core_Oligo164 | TACACTCGGAAACCAGGCAAAGCGCACCGGAACAGTCCC |
| Core_Oligo165 | GCGATCGCAAGCTTGGGATAG |
| Core_Oligo166 | AGGAGTGAGCACCGTCGGCTGTCTTTCCATA |
| Core_Oligo167 | ACCGTACTATTTTGACCCAGC |
| Core_Oligo168 | GACAGATGCTTGAGAGCCAACACATCGC |
| Core_Oligo169 | ACCGATATAAACAGGTGAGTGGAAGGAG |
| Core_Oligo170 | CAACGGACAACATTTTTAATGTCACCTT |
| Core_Oligo171 | GGATGTTTAGTACCGCCAAAAGACAGAATCTT |
| Core_Oligo172 | CAACCTAAATTACGGCTGATGTTATCTA |
| Core_Oligo173 | CCTTCATGGCTTGCTTAACAAGAATGGC |
| Core_Oligo174 | TTTCTTAAGTCAGATTACATTTGATTGT |
| Core_Oligo175 | GATCGCAGTAGCCAACGATGCTAACGGA |
| Core_Oligo176 | CGACAGTTAAATGTAGTGATGTAGAACGTCAGCGTGCG |
| Core_Oligo177 | TCCATTATTACCAGTTAACCTGAGCCGT |
| Core_Oligo178 | AGACTTTCAAAATATAGGTCTTTTGAGG |
| Core_Oligo179 | TGAGCCTCTTCGCTATTATTGTAAAGAGACGC |
| Core_Oligo180 | CGGAACGGGGTAATTTAAGACACTCGTA |
| Core_Oligo181 | CTCAGCAACTGGATAAAACATCGAACGTTATTAATCTT |
| Core_Oligo182 | GGCAAAAGAATCAGATACAGACAAAGTTGAAA |
| Core_Oligo183 | AACAACCATGACCAACAGTACTATTCCT |
| Core_Oligo184 | ACAAGAAAGCTGCTTTAGGCACGTGGCA |
| Core_Oligo185 | TACCAAGCGCGAAATGCCTTGAGTAACAATTAGCAAAGTACC |
| Core_Oligo186 | CCGGCACTAATTCGGCAAACGTCCGGCCAGAGCACCCGGGTCTTTCTTT |
| Core_Oligo187 | GGCTTGCTTGAATCTGTAAATTGCGGAA |
| Core_Oligo188 | CGGTCAACTTATGCTGTTTAGAAAACAG |
| Core_Oligo189 | CCAGGCGCCAGAACCTTAATTGCGCGAACTGATAGCAC |
| Core_Oligo190 | GATAGTTGGTCTTTGAATTACGGCAATT |
| Core_Oligo191 | TACAGAGCAAAAGAATAGTGAGACTTTA |
| Core_Oligo192 | CGGTACGATTGGCCTTGACTTTTCAGTTCAGC |
| Core_Oligo193 | GTAATGCCAACACTGTTATATGCACTAA |
| Core_Oligo194 | TGGGCGCCAAACGGACATAAAAGAATGC |
| Core_Oligo195 | TACACTCGCCATCAAAAACGCTTCTGGTGCTACACT |
| Core_Oligo196 | GGGAAAATAAGGTGCCACGCTGAGACCAGAACGAATTCGT |
| Core_Oligo197 | ATTGCCAAAAACAGGAAGGCAGCCAGCTGGAG |
| Core_Oligo198 | TACCTTTTTCATGGTCAGTTGGCAACCGTTGTAAAGTGT |
| Core_Oligo199 | TTGTTTGAGAGATCTACAGCCAGCAAATCGTC |
| Core_Oligo200 | GATTCATAATTTCAATCAAACCCTCAATGATTAGTCAACATA |
| Core_Oligo201 | TAGTTGGAGCAAACAAGAAGAATGCCACACGA |
| Core_Oligo202 | AACCGTGTGGCAAATGAAAAATCTATCGGCGCTGTTT |
| Core_Oligo203 | TTCAAAAGAAATTGAGAAGAA |
| Core_Oligo204 | TGAGTAATTCCACAAATAACATCACTTG |
| Core_Oligo205 | CAAATGGGTTTGCCAGTGTAGCGGTCACATC |
| Core_Oligo206 | TAATCGTGCACTCTGACATTCTGGCCAATAT |
| Core_Oligo207 | TTTGGTAAGAGCACTACGGAGGCTGAGACTCCATTCATTATATAGA |
| Core_Oligo208 | ATTAACATTTCTTTGAGCACG |
| Core_Oligo209 | ATAAAAAGGGGTGCATCACGC |
| Core_Oligo210 | TCAATCAGGACCTGCTCCAAGCGTCATACATGAGA |
| Core_Oligo211 | TTGACCATCCTGTTGCGGGCG |
| Core_Oligo212 | AATATGACGAGCTCAATATTACCGCCAGTGC |
| Core_Oligo213 | ACTATGAATAACAAGAGTCAGAGCCGCCGCCAACC |
| Core_Oligo214 | AATCATAGCAGTGTAGAACCCTTCTGACAAG |
| Core_Oligo215 | TGTACCACGGGAAATACGCCA |
| Core_Oligo216 | AGCTATATGCAGCAGTAACCA |
| Core_Oligo217 | AAAATTAGCGCGGGGCTAAAC |
| Core_Oligo218 | TAATCAGCCCTGCAATCCCAC |
| Core_Oligo219 | ATTTGAATTACTCATAAGAGCCAGAATGGAAAATC |
| Core_Oligo220 | CTGCGAAAAATCGGGAAGGGA |
| Core_Oligo221 | ACCCGTAACAACCGGATAAGCCGCCACCAGAAACC |
| Core_Oligo222 | AGAGCATTGCATTAAGGGATT |
| Core_Oligo223 | GAGGGTATCGCGTCAATACCTACATTTTAGA |
| Core_Oligo224 | ATTTTTTGTAAAGGACTGTTGCCCTGCGGCTACACT |
| Core_Oligo225 | GCGAGCTACCGCCTCTTAATGCGC |
| | |
| Cy3_Oligo1 | ACCGTCTATCAGGGCGATGAAAAAAAAAAAAAAA |
| Cy3_Oligo2 | GTTCCAGGCAACTAAAATATCAGGCTCCAGTTAGCAAAAAAAAAAAAAAA |
| Cy3_Oligo3 | TCCAACGAATTGCTGAAGCAATTGCGAATAGTAAAAAAAAAAAAAAAAAA |
| Cy3_Oligo4 | CCACTATAGCTCAAGCTTCAACGTTGAATTTTGTCAAAAAAAAAAAAAAA |
| Cy3_Oligo5 | CCCGAGAGAAGTTTAGAGGAAATTGTATCATTCCAAAAAAAAAAAAAAAA |
| Cy3_Oligo6 | AAGGGAGATTATTTTTCAATTAAGCGCAAGTAAGCAAAAAAAAAAAAAAA |
| Cy3_Oligo7 | AAGCACTAATAAAGGCGCAGATAACTGAACCGAAGAAAAAAAAAAAAAAA |
| Cy3_Oligo8 | TTTTGGGTTCAGGTTGAATACGTCAGAGTGAAATAAAAAAAAAAAAAAAA |
| Cy3_Oligo9 | CTTGACGGTTAGAAAGATGATGAGAATAAGTTACCAAAAAAAAAAAAAAA |
| Cy3_Oligo10 | AAAAAAGCCCACTACGTGAACCATCATAT |
| Cy3_Oligo11 | GTAGATTGTCGAGGCGTGGACAAAAAAAAAAAAAAA |
| Cy3_Oligo12 | TGGAAGGGGGAAAGAGAATAGAAAAAAAAAAAAAAA |
| Cy3_Oligo13 | CAGATGAACCCAAAGCGAAAAAAAAAAAAAAAAAAA |
| Cy3_Oligo14 | ACAAACGGATAATATCAAATAATAAAAAAAAAAAAAAA |
| Cy3_Oligo15 | AAACAGAAAATCGGCAAGAGTAAAAAAAAAAAAAAA |
| Cy3_Oligo16 | TATCAAACCCCCGAGAGTGTTAAAAAAAAAAAAAAA |
| Cy3_Oligo17 | AAAAAACATCGGGAGCTTTAATTGCTCAAAAAAAAAAAAAAA |
| Cy3_Oligo18 | GGAGAATGGCGAATAATTCGACATGTTTAAAAAAAAAAAAAAA |
| Cy3_Oligo19 | GAACAAACAAGTTACAGACCGGAATATAAAAAAAAAAAAAAAA |
| Cy3_Oligo20 | TTACAGAGAAACAAAAGATTACATTCCAAAAAAAAAAAAAAAA |
| Cy3_Oligo21 | AACAGGGACCTGAGAGACTTCAAGTACGAAAAAAAAAAAAAAA |
| Cy3_Oligo22 | GAGCGCTTCGCCTGCAGGTCAATGGCTTAAAAAAAAAAAAAAA |
| Cy3_Oligo23 | AAAAAAGAATTGAGTCAACTTTCAACAAAAAAAAAAAAAAAA |
| Cy3_Oligo24 | AGAGCAATCTGTATAACAACTAAAAAAAAAAAAAAA |
| Cy3_Oligo25 | CCCTTTTTCTAAAGAATCTCCAAAAAAAAAAAAAAA |
| Cy3_Oligo26 | GCAATAGCAGACGTTAATAATAAAAAAAAAAAAAAA |
| Cy3_Oligo27 | AGATAGCCCCTCATAAAAGGAAAAAAAAAAAAAAAA |
| Cy3_Oligo28 | AGAAGGACCTGTAGCGGTTTAAAAAAAAAAAAAAAA |
| Cy3_Oligo29 | TGCTAAATAAGCCCGAGAGATAACCCACAAAAAAAAAAAAAAAAA |
| Cy3_Oligo30 | TACAACGAACCGAGGCAGCCTAAAAAAAAAAAAAAA |
| Cy3_Oligo31 | GTAACGATAAGAAATTAGACGAAAAAAAAAAAAAAA |
| Cy3_Oligo32 | TGAATTTGAAACAAGGTAATTAAAAAAAAAAAAAAA |
| Cy3_Oligo33 | CAGACAGCGAACAAACATAAAAAAAAAAAAAAAAAA |
| Cy3_Oligo34 | GTCTTTCCTATCTTACACCCTAAAAAAAAAAAAAAA |
| Cy3_Oligo35 | TTTTTCAAGCGAACCAAAATCAAATTGCAAAAAAAAAAAAAAA |
| Cy3_Oligo36 | AAAAAAAGCGTTTTTATTCATGCACGTAAAAAAAAAAAAAAA |
| Cy3_Oligo37 | AAAGGAAACTCCAAATTGCTTTTAACGTAAAAAAAAAAAAAAA |
| Cy3_Oligo38 | TCAGCTTCATCAAAACATCAATGAATAAAAAAAAAAAAAAAAA |
| Cy3_Oligo39 | GCCTTTAGCCCGAACAAAAGACCTACCAAAAAAAAAAAAAAAA |
| Cy3_Oligo40 | AAAAAAGTTTCAGCGGAGTGAGAATAGA |
| Cy3_Oligo41 | GTACAAACAATGTAACAGTACCTTTTAAAAAAAAAAAAAAAA |
| Cy3_Oligo42 | AAAAAACTTTTGATAAGAGGTCATT |
| Cy3_Oligo43 | ATGCTGTTAAAGAATGCCGTAAAAAAAAAAAAAAAA |
| Cy3_Oligo44 | TATAACAAATCAAACCGGCGAAAAAAAAAAAAAAAA |
| Cy3_Oligo45 | TTTGCGGGGATTAGAGAAAGGGGGATTTAAAAAAAAAAAAAAA |
| Cy3_Oligo46 | GTGTCTGTAGGGTTTTTAGAGAAAAAAAAAAAAAAA |
| Cy3_Oligo47 | TAAATATTTTGGAAAACCCTAAAAAAAAAAAAAAAA |
| Cy3_Oligo48 | AGAGCTTTCAAAGGTCAAGTTAAAAAAAAAAAAAAA |
| Rotor-motor_Conn1 | TCCTTACCGCCCGACTTTTTTATCAAAAAGAGTATTGACTTAAAGTCTAACCTATAGGATACTTACAGCCATCGAGAGGGACACGGGGGATCCTCTAGACTGCAGAAAGTTGTGAAAGACCTTATGAGG |
| Rotor-motor_Conn2 | TTCTAAAATTCCGGCATTTTCTTTCTGCAGTCTAGAGGATCCCCCGTGTCCCTCTCGATGGCTGTAAGTATCCTATAGGTTAGACTTTAAGTCAATACTCTTTTTGATAATTGAGCCATCTATTATAGA |

## 4) Adapters , templates and connections

| Names | Sequence |
|---|---|
| Adapter A | /5Cy5/GTCTAATAGGGTTTTCCCAG/3AzideN/ |
| Adapter B | /5AzideN/ACGACGACGGAAAAAG/3Cy3Sp/ |
| Adapter C | /5Cy5/TGCCGGAATTTTAGAA/3AmMC6T/ |
| Adapter D | /5AmMC6/CCTCATAAGGTCTTTCACCCTCAGA/3Cy3Sp/ |
| Template _AB | TAC CAT CTT TTT CCG TCG TCG TGT TTG GGA AAA CCC TAT T |
| Template _CD | GGC TCT GAG GCT GAA AGA CCT TAT GAG GCT TCT AAA ATT CCG GCA TAC CAT |
| T-AD | GTG AAA GAC CTT ATG AGG AAA AAC TGG GAA AAC CCT ATT |
| T-BC | CTT TTT CCG TCG TCG TAA AAA TTC TAA AAT TCC GGC A |
| A' | CTG GGA AAA CCC TAT T |
| B' | CTT TTT CCG TCG TCG T |
| C | TTC TAA AAT TCC GGC A |
| D' | TCT GAG GGT GAA AGA CCT TAT GAG G |
| ds_Biotin_1_B' | /5Biosg/CCC CCC CCC CAA CGA CGG CCA GTG AAT TGT AGC CAC CAA CTT CTT TTT CCG TCG TCG T |
| ds_Biotin_2_A' | CTG GGA AAA CCC TAT TTT GTT GGT GGC TAC AAT TCA CTG GCC GTC GTT CCC CCC CCC C/3Bio/ |

**References**

1. Li, Q. *et al.* Macroscopic contraction of a gel induced by the integrated motion of light-driven molecular motors. *Nature Nanotechnology 2015 10:2* 10, 161–165 (2015).
2. Helmi, S. & Turberfield, A. J. Template-directed conjugation of heterogeneous oligonucleotides to a homobifunctional molecule for programmable supramolecular assembly. *Nanoscale* 14, 4463–4468 (2022).
3. Stahl, E., Martin, T. G., Praetorius, F. & Dietz, H. Facile and scalable preparation of pure and dense DNA origami solutions. *Angew Chem Int Ed Engl* 53, 12735–40 (2014).